\documentclass[a4paper,11pt]{article}
\pdfoutput=1 

\usepackage{jheppub}
\usepackage{xcolor}

\usepackage[T1]{fontenc} 
\usepackage{bbold}
\usepackage{amsmath}
\usepackage{amssymb}
\usepackage{mathtools}
\usepackage{stmaryrd}
\usepackage{tensor}

\usepackage{subcaption}
\usepackage{graphicx}
\usepackage{stmaryrd}
\let\a=\alpha 
\let\b=\beta 
\let\g=\gamma 
\let\d=\delta 
\let\e=\epsilon

\let\y=\psi

\let\D=\Delta

\let\F=\Phi

\let\W=\Omega   
\let\G=\Gamma

\let\<=\langle
\let\>=\rangle

\newcommand{\CC}{\mathcal{C}}

\newcommand{\GG}{\mathcal{G}}

\newcommand{\NN}{\mathcal{N}}

\newcommand{\RR}{\mathcal{R}}

\def\be#1\ee{\begin{align}#1\end{align}}

\newcommand{\op}[1]{\operatorname{#1}}
\newcommand{\al}[1]{\begin{align}#1\end{align}}
\newcommand{\spl}[1]{\begin{split}#1\end{split}}
\newcommand{\mat}[1]{\begin{pmatrix}#1\end{pmatrix}}
\newcommand{\ga}[1]{\begin{gathered}#1\end{gathered}}

\usepackage{tikz}
\usetikzlibrary{calc,decorations.markings}
\tikzset{every picture/.style={line width=0.8pt}}

\tikzset{graph-1/.style = {
  line cap = round,
   line join = round,
     > = triangle 45,
     x=0.7cm, y=0.7cm,
      every node/.append style = {inner ysep=2mm}
                        }
    }
    
\usetikzlibrary{matrix,decorations.pathreplacing}

\title{
Einstein gravity from a matrix integral - Part II
}

\author[a]{Shota Komatsu}
\author[b]{Adrien Martina}
\author[b]{Joao Penedones}
\author[b]{Antoine Vuignier}
\author[b,c]{Xiang Zhao}
\affiliation[a]{CERN, Theoretical Physics Department,
CH-1211 Geneva 23, Switzerland}
\affiliation[b]{Fields and Strings Laboratory, Institute of Physics, Ecole Polytechnique Federale de Lausanne (EPFL),
CH-1015 Lausanne, Switzerland}
\affiliation[c]{Université Paris-Saclay, CNRS, CEA, Institut de Physique Théorique, 91191, Gif-sur-Yvette, France
}

\abstract{
Using supersymmetric localization, we compute the partition function and some protected correlators of the polarized IKKT matrix model. 
Surprisingly, we find that the original IKKT model is different from polarized IKKT in the limit of vanishing mass deformation. 
We study different regimes of the localization results and recover the electrostatic problem which defines the gravity dual.

\vspace{3cm}
}

\begin{document} 
\maketitle
\flushbottom

\newpage

\section{Introduction}
Holographic duality \cite{tHooft:1993dmi, Susskind:1994vu, Maldacena:1997re} is one of the most -- if not the most -- profound ideas in modern theoretical physics as it provides a framework to describe quantum gravity microscopically using quantum field theories in lower dimensions. It builds on 't Hooft's observation \cite{tHooft:1973alw} that diagrammatic expansions of large 
$N$ gauge theories share structural similarities with two-dimensional Riemann surfaces, and formalizes this into a concrete correspondence between distinct theories, offering insights both into gauge theories and gravity.

Despite its promise, a precise understanding of how classical Einstein gravity emerges from gauge theories and how it is modified by  quantum corrections still remains elusive. The difficulty arises primarily because Einstein gravity is expected to emerge in a strong-coupling regime of gauge theories, a regime for which we currently lack effective analytical tools. A possible way to make progress is to study simpler cases of holography where the gauge theory involved is more tractable. In this paper and its companion work \cite{Komatsu:2024bop}, we study one of the simplest examples of gauge theory, a $0+0$ dimensional gauge theory, also known as a matrix integral. The holographic duality between matrix integrals and string theories has been a subject of active research in the past (see e.g.~\cite{Douglas:1989ve, Brezin:1990rb, Gopakumar:2022djw, Collier:2023cyw}). However, the ``gravity-side'' of these matrix integrals is not conventional Einstein gravity but instead corresponds to non-standard low-dimensional theories with some gravitational features. In contrast the system we will study is, in our view, the simplest model with an emergent description of Einstein gravity.


\paragraph{Matrix integral}  More precisely, we 
study the polarized IKKT matrix model \cite{Ishibashi_1997, Bonelli_2002, Hartnoll:2024csr,Komatsu:2024bop}. Since this model is in $0+0$ dimension,  the partition function is a matrix integral
\begin{equation}
    Z =\int [d X^I] [d \psi_\alpha] e^{-S}\,,
    \label{eq:polarized IKKT partition function}
\end{equation}
and the physical observables are matrix expectation values, expressed as
\begin{equation}
    \langle f(X,\psi) \rangle = \frac{1}{Z} \int [d X^I] [d \psi_\alpha] f(X,\psi) e^{- S}\,,
\end{equation}
where the action reads
\begin{equation}
\begin{split}
    S = \operatorname{Tr} & \Biggl[-\frac{1}{4} [X_I,X_J]^2 - \frac{i}{2} \bar{\psi} \Gamma^{I} [X_I,\psi] \\ & + \frac{3 \Omega^2}{4^3} X_i X_i + \frac{\Omega^2}{4^3} X_p X_p + i \frac{\Omega}{3} \epsilon^{i j k} X_i X_j X_k - \frac{1}{8} \Omega \bar{\psi} \Gamma^{1 2 3} \psi \Biggr] \, . \label{eq:action}
\end{split}
\end{equation}
The first line is the IKKT action, where indices are contracted with Euclidean signature. This ensures that the partition function converges. The second line is a mass-deformation, parameterized by the dimensionless parameter $\Omega$. The matrices $X^I$ are dimensionless $N \times N$ Hermitian matrices. The indices $I,J\in \{1,2,\dots,10\}$ while $i,j,k\in \{1,2,3\}$ and $p\in \{4,5,\dots 10\}$.
The fermions $\psi_\a$ are specified by 16 independent Grassmann-valued Hermitian matrices\footnote{Note that Majorana-Weyl spinors only exist as spinorial representations of $SO(9,1)$ but not of $SO(10)$. However, this issue is resolved using the so-called doubling trick and thinking in terms of Wick rotation from Lorentzian signature \cite{Nicolai:1978vc, Yee:2003ge}.}. 
The gamma matrices are $32\times 32$
and $\Gamma^{123} = \G^1\G^2\G^3$, while the fermion conjugation is defined as $\bar{\psi} = \psi^\top \mathcal C$ where $\mathcal C$ is the charge conjugation matrix.

This model has 16 supercharges and global $SO(3) \times SO(7)$ as well as $SU(N)$ gauge symmetry.\footnote{We use the word ``gauge'' in analogy to higher-dimensional gauge theories. However, in 0+0 dimensions, gauge and global symmetries can be treated on the same footing, with the distinction that only gauge-invariant observables are considered.} 
The matrix integral has classical saddles $X_i = \frac{3 \Omega}{8} L_i$, given by representations of $SU(2)$ generators $L_i$ \cite{Hartnoll:2024csr, Komatsu:2024bop}. A generic reducible representation of $SU(2)$ is parameterized by its $q$ different spin $j_s$ irreducible components $s=1,...,q$ each of which has $n_s$ copies and dimension $N_s = 2j_s + 1$, such that $N = \sum_{s=1}^q n_s N_s$. The number of different saddles is thus given by the number of partitions of $N$.

\begin{figure}[h]
\[ 
   X^i = \frac{3 \Omega}{8}
   \begin{tikzpicture}[decoration={brace,amplitude=2pt},baseline]
   \footnotesize
   \matrix (magic) [matrix of math nodes,left delimiter=(,right delimiter=)] {
      \color{blue}L^i_{N_1}  &  &  \\
     & \color{blue} \ddots  &  \\
     &  &  \color{blue} L^i_{N_1} \\
     & & & \ddots \\
     & & & & \color{violet} L^i_{N_q} \\
     & & & & & \color{violet} \ddots \\
     & & & & & & \color{violet} L^i_{N_q}\\
    };
    \draw[decorate,blue] (magic-1-1.north) -- (magic-3-3.north) node[above=5pt,midway,sloped] {$n_1$};
    \draw[decorate,violet] (magic-5-5.north) -- (magic-7-7.north) node[above=5pt,midway,sloped] {$n_q$};
  \end{tikzpicture}
 \qquad 
 \begin{tikzpicture}[baseline] 
\small
  \draw[->] (0,-2.5) -- (4.2,-2.5) node[right] {$r$};
  \draw (0,-2.5) -- (0,-1) ;
   \draw[loosely dotted] (0,-1) -- (0,0.75) ;
  \draw[->] (0,0.75) -- (0,2.5) node[above] {$z$};
  \draw[color=blue]   (0,-1.5)  node[left]{$z_1 = \frac{3\pi \mu\alpha'}{8} N_1$} -- (2,-1.5) node[right]{$Q_1 = \frac{\pi^4 \mu^6\alpha'^3 g_s}{32} n_1$};
  \draw[color=violet]   (0,1.5) node[left]{$z_q = \frac{3\pi \mu \alpha'}{8} N_q$} -- (1.5,1.5) node[right]{$Q_q = \frac{\pi^4 \mu^6\alpha'^3 g_s}{32} n_q$};

\end{tikzpicture}
\]
  \caption{Correspondence between the fuzzy sphere vacua of the matrix model and the dual geometries. For each spin-$j_s$ $SU(2)$ irreducible representation of dimension $N_s=2j_s+1$ we put a disk at position $z_s \sim N_s$. The number $n_s$ of copies of this representation appearing determines the charge of that disk $Q_s \sim n_s$. }
  \label{fig:vacua}
\end{figure}
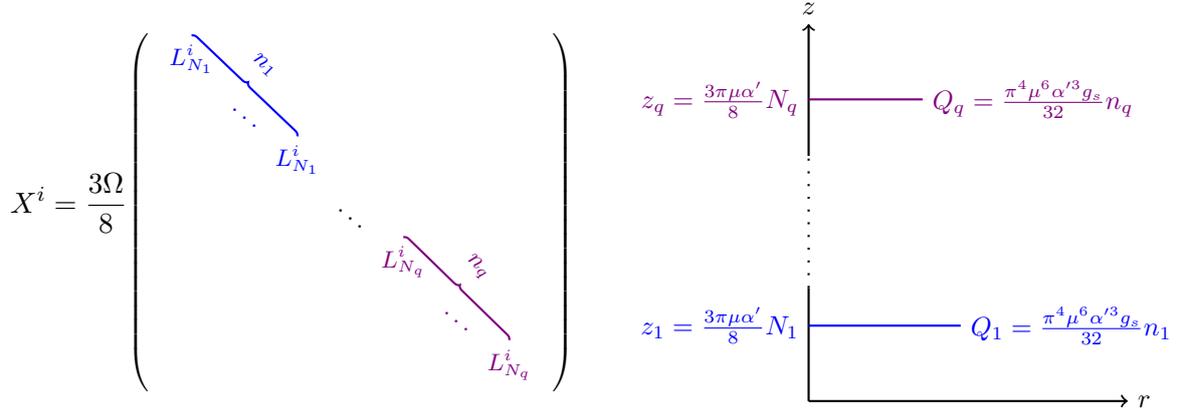

\paragraph{Gravity dual}  In a recent paper \cite{Komatsu:2024bop} we constructed a family of solutions to Euclidean IIB supergravity dual to the polarized IKKT matrix model. Those solutions are warped products of a 2-sphere and a 6-sphere fibered over the two dimensional $(r,z)$ plane, determined entirely by a single function $V(r,z)$ solving the four dimensional axially symmetric Laplace equation 
\begin{equation}
    \frac{\partial^2 V}{\partial z^2}+ \frac{2}{r} \frac{\partial V}{\partial r} + \frac{\partial^2 V}{\partial r^2}=0,
\end{equation}
where $r$ is a radial variable in $\mathbb{R}^3$ and $z$ is the orthogonal coordinate. We can think of $V(r,z)$ as the solution to an electrostatic problem, that is uniquely determined once we specify proper boundary conditions. As explained in \cite{Komatsu:2024bop}, to make the resulting geometry non-singular, we need to consider a background potential
\begin{equation}
    V_\infty(r,z) = V_b \left( z r^2-z^3 \right),
\end{equation}
where $V_b$ is a positive constant. In addition we consider a configuration of 3-dimensional conducting balls, each centered on the $z$-axis, at positions $z_s$ and carrying charges $Q_s$, with their sizes $R_s$ being determined by the condition that the charge density vanishes at the edge. The electrostatic problem is then entirely specified by the parameters $(z_s,Q_s)$ with $s=1,...,q$. This configuration is dual to a saddle of the matrix integral parametrized the degeneracies and irrep dimensions ($n_s$, $N_s$), see figure \ref{fig:vacua}.

\paragraph{Results}
In this paper, we use localization techniques to derive exact results for the partition function and a family of protected correlation functions. As we show in the following sections, we obtain the $U(N)$ partition function\footnote{We use the normalization $[d X] \equiv  \prod_{A = 1}^{N^2} d X^A$ for bosonic matrices and $[d \psi] \equiv \prod_{A = 1}^{N^2} d \psi^A$ for fermionic matrices, where $A$ is an index for the $U(N)$ generators $T^A$, e.g. $X = X^A T^A$, with normalization $\mathrm{Tr} T^A T^B = \delta^{A B}$. More details on this measure convention are given in appendix \ref{app:Normalization_factors}.}
\begin{equation}
\ga{
Z=\sum_\RR Z_\RR\,,\\
    Z_\RR = C_\mathcal{R} e^{ \frac{9 \Omega^4}{2^{15}} \sum_s n_s (N_s^3 - N_s) } \int \left( \prod_s \prod_{i = 1}^{n_s} d m_{s i}\right) Z_{\mathrm{1-loop}} \ \mathrm{exp} \left( - \frac{3 \Omega^4}{2^7} \sum_s \sum_{i = 1}^{n_s} N_s m_{s i}^2 \right)\,,
}
\label{eq:PartitionFunction}
\end{equation}
where
\al{
\ga{
Z_\mathrm{1-loop} = \prod_{(si,tj)}  f_{s t} (m_{s i} - m_{t j}) \label{eq:1-loop-det} \, , \\
f_{s t} (x) \equiv \prod_{J = \frac{|N_s - N_t|}{2} }^{\frac{N_s + N_t}{2}-1} \frac{[(2+3J)^2 + (8x)^2]^{3} [(3J)^2 + (8x)^2]}{[(1+3J)^2 + (8x)^2]^{3} [(3+3J)^2 + (8x)^2]} \, ,
}
}
\begin{equation}
    C_\mathcal{R} = \frac{ (2 \pi)^
    {5 N^2 +  N/2 }}{\prod_{k=1}^{N-1} k! \prod_s n_s !} \left(
    \frac{32}{3\pi } 
    \right)^{\sum_s n_s} \prod_{s } \left(\frac{1}{N_s} \prod_{J = 1}^{N_s -1} \frac{(2 + 3 J)^3}{(1 + 3 J)^3}\right)^{n_s} \,.
    \label{eq:normalization}
\end{equation}
The sum $\sum_\mathcal{R}$ is taken over all the inequivalent $N$-dimensional representations $\mathcal{R}$ of $SU(2)$. The parameters $s,t$ are labels for the different irreducible representations (irreps) appearing in the representation $\mathcal{R}$, and $i=1, ..., n_s$ is an index going over the multiplicity of the irrep $s$.
The product $\displaystyle{\prod_{(si,tj)}}$ in $  Z_\mathrm{1-loop}$ runs over all pairs $(si,tj)$, excluding the identical terms $(si,si)$.
The constant $C_\mathcal{R}$ is a normalization constant which depends on the saddle. Similarly, correlators of gauge invariant functions of the protected matrix $\phi = X_3 - i X_{10}$ can be computed as shown in \eqref{eq:correlators}. We provide analyses in various regimes of the localization results \eqref{eq:PartitionFunction} and \eqref{eq:correlators}, matching them with direct computations.

As surprising as it may sound, the partition function \eqref{eq:PartitionFunction} diverges as $\Omega \to 0$, even though the IKKT partition function has been proven to converge \cite{Austing:2001pk}, namely
\begin{equation}
    \lim_{\Omega \to 0} Z(\Omega) = \infty \neq Z_\mathrm{IKKT} \, .
\end{equation}
This behaviour is due to the fermion mass terms as we illustrate for $N=2$ in appendix \ref{app:fermion_masses_ConvergenceIKKT}. We also obtain this divergence for any $N$ from integrating out the off-diagonal matrices as shown in section \ref{app:strong_limit}. As we show there, the leading divergence can be interpreted as arising from the trivial vacuum, where each particle is very separated from the others and each contributes $Z_{U(1)} \sim \frac{1}{\Omega^2}$. 
Nevertheless, we conjecture that the IKKT partition function can be obtained from the $\Omega \to 0$ limit of \eqref{eq:PartitionFunction} by closing the contour of integration in the complex plane.

Finally, we recover the electrostatic problem describing the backreacted geometries from the density of eigenvalues associated to \eqref{eq:PartitionFunction}. This analysis is very similar to the one relating the BMN model to Lin-Maldacena geometries \cite{Asano:2014vba, Asano:2014eca}, as well as in other setups \cite{Okuda:2008px,He:2024djr}.

\paragraph{Structure of the paper}  In section \ref{sec:review} we review the basics of the IKKT model and its mass deformation. Section \ref{sec:localization} includes the setup and the details of applying supersymmetric localization techniques to compute the polarized IKKT partition function and protected correlation functions. It is intended to be pedagogical. Section \ref{sec: analysis of the result} presents the analysis of the localization result, including strong and weak coupling limits as well as a conjectured prescription to connect the polarized and the original IKKT partition function. In section \ref{sec:gravity from matrix model} we demonstrate the explicit agreement between the polarized IKKT matrix model and the dual geometry constructed in \cite{Komatsu:2024bop}. This is done by showing that in the large $N$ limit the eigenvalue densities from the matrix integral satisfy the same equation as that of the charge densities in the electrostatic system describing the dual geometry (upon an explicit identification of parameters on both sides). We discuss various open problems in section \ref{sec:discussion}. Various technical details are collected in the appendices.

\section{Review of the matrix models}
\label{sec:review}
\subsection{IKKT}
The (Euclidean) IKKT integral \cite{Ishibashi_1997,Aoki:1998bq} is 
\begin{equation}
    Z_{\rm IKKT} = \int \prod_{A=1}^{N^2-1} \Biggl[\prod_{I=1}^{10} dX_I^A \prod_{\alpha=1}^{16} d\psi_\alpha^A \Biggr] e^{-S_{\rm IKKT}},
    \label{eq:IKKT partition function}
\end{equation}
where\footnote{The action has the same convention as \cite{Green:1997tn,Moore:1998et,Green:1998yf}, which is natural from the perspective of dimensionally reducing 10D $\NN=1$ SYM to a point. This convention is related to that in \cite{Hartnoll:2024csr} through a simple redefinition of the charge conjugation matrix: $i\CC_{\rm here} = \CC_{\rm there}$. After Weyl projection to the suitable 16-component spinors, the actions are the same.}
\al{\spl{
    S_{\rm IKKT} &= \operatorname{Tr} \left[-\frac{1}{4} [X_I,X_J]^2 - \frac{i}{2} \sum_{\alpha,\beta=1}^{32} \sum_{I=1}^{10} \psi_\alpha (\mathcal{C}\G^{I})_{\alpha \beta} [X_I,\psi_\beta] \right].
\label{eq:IKKT action}
}}
The bosonic matrices $X_I$ and the fermionic matrices $\psi_\alpha$ are traceless Hermitian matrices, expressed in the basis of $SU(N)$ generators $T^A$, where we choose $\mathrm{Tr} T^A T^B = \delta^{A B}$ as normalization. The fermions $\psi_\alpha$ consist of 16 Grassmann-valued Hermitian matrices. 

One might expect that the partition function \eqref{eq:IKKT partition function} is divergent, because the action has infinitely long valleys (flat directions), corresponding to diagonal (commuting) matrices, along which $S_\mathrm{IKKT}=0$. However, it has been shown in \cite{Austing:2001bd,Austing:2001pk} that 
the width of the valleys are in fact highly suppressed so that their volume remains finite. To understand this point, consider the following toy integral
\begin{equation}
    \int dx dy e^{-y^2 - x^4 y^2} = \sqrt{\pi} \int dx \frac{1}{\sqrt{1 + x^4}} < \infty \,.
\end{equation}
Even though the left-hand-side has flat directions corresponding to $y = 0$, the valley at large $x$ has width $\sim \frac{1}{x^2}$ so that the volume integral over this infinite valley remains finite.

A closed form of the partition function \eqref{eq:IKKT partition function}, first conjectured in \cite{Green:1997tn} and later computed using localization techniques \cite{Moore:1998et} takes the form
\begin{equation}
    Z_\mathrm{IKKT} = \frac{(2 \pi)^{(10 N + 11)(N-1)/2}}{ \sqrt{N} \prod_{k = 1}^{N-1}(k!)} \sum_{m | N} \frac{1}{m^2} \,, \label{eq:IKKTZresult}
\end{equation}
where the sum runs over all divisors of $N$ and the prefactor, determined in \cite{Krauth:1998xh}, has been adapted to our conventions. It can also be expressed in terms of a divisor function $\sigma_2(N)=\sum_{m | N} m^2$ as
\begin{align}
Z_\mathrm{IKKT} = \frac{(2 \pi)^{(10 N + 11)(N-1)/2}}{ N^{5/2} \prod_{k = 1}^{N-1}(k!)} \sigma_2 (N)\,,
\end{align}
whose asymptotic behavior and variance have been studied in the literature \cite{hardy1979introduction}.

\subsection{Polarized IKKT}
The model we are interested in is a mass deformation of IKKT which preserves all sixteen dynamical supersymmetries. Its action, written in \eqref{eq:action}, is invariant under
\al{
\spl{
\delta_\epsilon X^I & = \bar{\epsilon} \Gamma^I \psi \,, \\
\delta_\epsilon \psi & = \frac{i}{2} \Gamma^{I J} \epsilon [X_I, X_J] + \frac{3 \Omega}{8} \Gamma^{1 2 3} \Gamma^i \epsilon X_i + \frac{\Omega}{8} \Gamma^{1 2 3} \Gamma^p \epsilon X_p \,, \label{susy_mass_deform}
}
}
where $\epsilon$ is deemed as Grassmann-even and $\d_\e$ is Grassman-odd susy generator.

The saddle points of the polarized IKKT integral are minima of the bosonic potential
\begin{equation}
    V_B = \operatorname{Tr} \left[-\frac{1}{4} [X_I,X_J]^2 + \frac{3 \Omega^2}{4^3} X_i X_i + \frac{ \Omega^2}{4^3} X_p X_p + i \frac{\Omega}{3} \epsilon_{ijk} X_i X_j X_k\right].
\end{equation}
The minima are given by \cite{Komatsu:2024bop,Hartnoll:2024csr}
\begin{equation}
    X_p = 0\,,\qquad 
    X_i = \frac{3}{8} \Omega  L_i\,, \qquad [L_i,L_j] = i \epsilon_{ijk} L_k\,,
\label{eq:saddle ansatz}
\end{equation}
with $L_i$ being the $N\times N$ matrix representations of $SU(2)$ Lie algebra (not necessarily irreducible). 
At the minima the potential takes values 
\begin{equation}
    V_B =
        -\frac{9}{2^{13}}\Omega^4 \op{Tr}L_i L_i \,. 
\end{equation}
For a general reducible representation, $L_i$ is block diagonal with $q$ different irreps, each having multiplicity $n_s$ and size $N_s \times N_s$. Thus, we have
\begin{equation}
\begin{split}
\operatorname{Tr}L_i L_i &=\sum_{s=1}^q n_s N_s j_s(j_s+1)=
   \sum_{s=1}^q n_s N_s \frac{N_s^2-1}{4}\,,
   \qquad N=\sum_{s=1}^q n_s N_s\,.
\end{split}
\end{equation} 
All these minima preserve supersymmetry \cite{Komatsu:2024bop}.



\section{Localization in polarized IKKT}
\label{sec:localization}
We will now use supersymmetric localization to reduce the polarized IKKT partition function, as well as SUSY-protected correlators, to integrals over the moduli space. Given that our model is very similar to the plane-wave matrix model, also called BMN model \cite{Berenstein:2002jq}, we will closely follow the formalism developed in \cite{Asano:2012zt}. 

Let us review the main idea \cite{cremonesi_introduction_2013,Pestun:2007rz}. Given a supersymmetric theory, one can choose one supersymmetry $\delta_s$ and define a deformed partition function,
\begin{equation}
    Z_t \equiv \int [dX^I] [d \psi_\alpha] \mathrm{exp}(- (S + t \delta_s V)) \,,
\label{eq:localization action}
\end{equation}
where $V$ is a Grassmann-odd polynomial of the matrices. If the purely bosonic part of $\delta_s V$ is non-negative, this deformed quantity is converging for any $t \geq 0$.
Let us now assume that $\delta_s^2 V = 0$. Then,
\begin{equation}
    \frac{d}{dt} Z_t =  \int [d X^I] [d \psi_\alpha] \delta_s \Biggl( V \exp(- \left( S + t \delta_s V \right)) \Biggr) = 0 \,,
\end{equation}
where in the last step we used that $\delta_s$ can be written as a sum of derivatives\footnote{This follows from the supersymmetries \eqref{eq:SUSY_16}. We can schematically write $$\delta_s \sim \psi \frac{\delta}{\delta X} + ([X,X]+\Omega X + K) \frac{\delta}{\delta \psi} + ([X, \psi] + \Omega \psi) \frac{\delta}{\delta K}$$ Note also that all derivatives commute with what's in front of them, so they can be moved to the left.}. The assumption that $\delta_s^2 V = 0$ typically restricts $V$ to be invariant under some subgroup of $SU(N) \times SO(3) \times SO(7)$. The fact that $Z_t$ is independent of $t$ has a very useful consequence: The saddle point approximation corresponding to $t\to \infty$ is exact, and is equal to the partition function $Z = Z_{t=0}$. This reduces the integral to the moduli space obeying $\delta_s V = 0$. 

Note that this argument can also be applied to correlators of protected operators $\mathcal{O}$ such that $\delta_s \mathcal{O} = 0$, namely
\begin{equation}
    \langle \mathcal O \rangle = \langle \mathcal{O} \rangle_t \equiv \frac{1}{Z} \int [d X^I] [d \psi_\alpha] \mathcal{O} \exp(- \left( S + t \delta_s V \right)) \,, \qquad \forall t \geq 0 \,.
\end{equation}

\subsection{Closure of the SUSY algebra}
The requirement that $\delta_s^2$ reduces to some combination of gauge and global symmetry generators, under which $V$ should be invariant, requires the SUSY algebra to close on this symmetry group. In the formulation \eqref{susy_mass_deform}, this is only true on-shell. 

An off-shell formulation is obtained by including seven auxiliary fields $K_{a = 1,..., 7}$, following \cite{Berkovits:1993hx,Asano:2012zt}. Namely, we will now consider the action
\al{
\spl{
S = \operatorname{Tr} \Biggl[-\frac{1}{4} & [X_I,X_J]^2 - \frac{1}{2} \psi^\top \bar \gamma^{I} [X_I,\psi] + \frac{1}{2} K_a K_a \\ & + \frac{3 \Omega^2}{4^3} X_i X_i + \frac{\W^2}{4^3} X_p X_p + i \frac{\W}{3} \epsilon^{i j k} X_i X_j X_k + \frac{i}{8} \W \psi^\top \bar \gamma^{1 2 3} \psi \Biggr], \label{eq:S_with_auxiliaries}
}
}
where we wrote the Majorana-Weyl fermions in 16-component notation, i.e. $\psi_{\alpha=1,...,16}$ are Hermitian matrices of left-handed spinors, and $\bar \gamma^{I}$ are $16 \times 16$ Weyl matrices defined in \eqref{eq:Weyl_matrices}. Note that the effect of auxiliary fields to the partition function is trivial, which is just a multiplication by $\int [d K_a] e^{- \frac{1}{2} \mathrm{Tr} K_a K_a} = (2\pi)^{7N^2/2}$. 
The supercharge, leaving the action \eqref{eq:S_with_auxiliaries} invariant, is now acting as
\al{
\spl{
\delta_\epsilon X^I & = - i \epsilon^\top \bar{\gamma}^I \psi \,, \\
\delta_\epsilon \psi & = \frac{i}{2} \bar{\gamma}^{I J} \epsilon [X_I, X_J] + \frac{3 \Omega}{8} \bar \gamma^{1 2 3} \bar{\gamma}^i \epsilon X_i + \frac{\Omega}{8} \bar \gamma^{1 2 3} \bar{\gamma}^p \epsilon X_p + i \nu_a K_a \,, \\ \delta_\epsilon K_a & = i \nu_a^\top \bar{\gamma}^I [X_I, \psi] +  \frac{\Omega}{4} \nu_a^\top \bar \gamma^{1 2 3} \psi \,,\label{eq:SUSY_16}
}
}
where $\epsilon$ is deemed as Grassmann-even and $\d_\e$ is Grassman-odd. The parameters $\nu_{a=1,2,\ldots,7}$ are 7 arbitrary Grassmann-even Majorana-Weyl spinors at this point.  However, we require the SUSY algebra to close, namely,
\begin{equation}
    \{ \delta_{\epsilon_1}, \delta_{\epsilon_2} \} (\mathrm{fields}) = \delta_B (\mathrm{fields}) \, ,
\end{equation}
where $\delta_B$ is a bosonic symmetry generator (i.e. some combination of $SU(N)$ and $SO(3) \times SO(7)$). This imposes constraints on $\nu_a$, as we show in  \ref{App_subsec:susy algebra closure}, which read 
\al{\ga{
    \epsilon^\top \bar{\gamma}^I \nu_a = 0\,, \\
    \frac{1}{2} (\epsilon_1^\top \bar{\gamma}^I \epsilon_2) \gamma^I_{\alpha \beta}= (\epsilon_1)_\alpha (\epsilon_2)_\beta + (\nu_a^1)_\alpha (\nu_a^2)_{\beta}\,,
    \\
    (\nu_a^1)^\top \bar{\gamma}^I \nu_b^2 = \delta_{a b} \epsilon^\top_1 \bar{\gamma}^I \epsilon_2\,. 
    \label{eq:nu_constraints}
}}
We solve these constraints for our choice of $\epsilon$, as discussed in the next section and shown explicitly in appendix \ref{app:constraintsFromAuxiliaries}.

\subsection{Localization setup}
Throughout this and the following sections, it will prove convenient to use the following notation for the indices
\al{\ga{
I,J,... = 1,..., 10 \, , \quad i,j,... = 1,...,3 \, , \quad p,q,... = 4,...,10 \, , \quad a = 1,...,7 \, ,  \\
I',J',... = 1,..., 9 \, , \quad i',j',... = 1,2 \,, \quad p',q',... = 4,...,9 \,, \quad a' = 1,...,6 \,.
\label{eq:Notations_indices}
}}
We now choose a particular supersymmetry transformation $\delta_s$ by specifying $\e$. Following \cite{Asano:2012zt}, we find it natural\footnote{While the final result of the integral \eqref{eq:polarized IKKT partition function} is independent of the choice of the deformation $\d_s V$, the partition function written as a reduced (moduli space) integral, e.g. \eqref{eq:PartitionFunction}, can be different. In fact, in \cite{Hartnoll:2024csr} a different deformation of the polarized IKKT action was used and it leads to different saddle points and different moduli space integral. 

In \cite{Asano:2014eca}, it was observed that a similar choice of BPS sector (or equivalently $\phi$) to ours leads to a direct match with the gravity dual of the BMN model. 
In section \ref{sec:gravity from matrix model}, we will see that our choice of $\phi$ also leads to the matching with the gravity dual to the polarized IKKT model \cite{Komatsu:2024bop}.} to choose $\epsilon$ such that it preserves a specific linear combination of $X^I$,
\begin{equation}
    \delta_s\phi = 0\,, \qquad
    \phi \equiv X_3 - i X_{10}\,. 
\end{equation}
Choosing a gamma matrix representation such that $\bar \gamma^3$ and $\bar \gamma^{10}$ are diagonal, taking the same value in the first 8 components, as well as in the last 8 components, as given explicitly in \eqref{eq:SO9Matrices}, \eqref{eq:Weyl_matrices}, we obtain that
\begin{equation}
    \epsilon = \begin{pmatrix}
        \eta \\ 0
    \end{pmatrix} \,,
\end{equation}
where $\eta$ is a 8-component spinor. We also choose the normalization $\epsilon^\top \epsilon = \eta^\top \eta = 1$. The constraints on $\nu_a$ \eqref{eq:nu_constraints} are solved in appendix \ref{app:constraintsFromAuxiliaries}, where we find that $\nu_a$ have to live on the same 8-component space as $\epsilon$, have to be orthogonal to $\epsilon$ and have to be orthogonal to each other. This justifies why $a$ has to run from 1 to 7. From now on, we will choose $\eta = (1,0,...,0)$. Correspondingly, we will choose $(\nu_a)_\alpha = \delta_{\alpha, a+1}$.\footnote{This is fixed up to $SO(7)$ rotation.}

Let us also package the matrices in a convenient way. As explained in appendix \ref{app:constraintsFromAuxiliaries}, the spinors $\{ \bar{\gamma}^{I'} \epsilon, \nu_{a }\}$ form an orthonormal basis of 16-component spinors.
We thus decompose the fermionic matrices as
\begin{equation}
    \psi = v^{I'} \psi_{I'} + \nu^a \chi_a\,, \qquad v^{I'} \equiv \bar{\gamma}^{I'} \epsilon\,.
    \label{eq:spinor_basis}
\end{equation}
This allows us to package matrices into
\begin{equation}
    Y \equiv \begin{pmatrix}
        X_{I'} \\ \chi_a
    \end{pmatrix} \,,
    \qquad Y' \equiv \begin{pmatrix}
        - i \psi_{I'} \\ i H_a
    \end{pmatrix}\,,
\label{eq:localization_variables}
\end{equation}
where
\begin{equation}
    H_a \equiv K_a + \frac{\Omega}{8} \nu_a^\top \tilde{\epsilon} X_{10} - i s_a\,,
\end{equation}
with
\begin{equation}
    \tilde{\epsilon} \equiv \bar \gamma^{1 2 3} \epsilon\,, \qquad 
    s_a \equiv \frac{3 \Omega}{8} \nu_a^\top \gamma^i \tilde{\epsilon} X_i - \frac{\Omega}{8} \nu_a^\top \gamma^{p'} \tilde{\epsilon} X_{p'} + \frac{i}{2} \nu_a^\top \bar{\gamma}^{I' J'} \epsilon [X_{I'}, X_{J'}]\,.
\end{equation}
From now on, we will consider the matrices in $Y$ and $Y'$ as our physical variables. This rewriting is very similar to \cite{Pestun:2007rz,Asano:2012zt}, and it has the advantage of simplifying the SUSY transformations to
\begin{equation}
    \delta_s Y = Y' \,, \quad \quad \delta_s Y' = - i (\delta_\phi + \delta_{U(1)}) Y \,, \label{eq:SUSY_spinorbasis}
\end{equation}
where $\delta_\phi$ is a gauge transformation with parameter $\phi $,
\begin{equation}
    \delta_{\phi} = i [\phi, \cdot] \,,
\end{equation}
and $\delta_{U(1)}$ is a $U(1)$ subgroup of $SO(3) \times SO(7)$ which acts as
\begin{align}
    &\delta_{U(1)} (X_1 \pm i X_2)= \mp i\frac{3}{8} (X_1 \pm i X_2) \,, \qquad 
    \delta_{U(1)} (\chi_1\pm i\chi_6) = \pm i\frac{1}{4} (\chi_1\pm i \chi_6) \,,\nonumber
\\
    &\delta_{U(1)} (X_4 \pm i X_5) = \mp i \frac{1}{8} (X_4 \pm i X_5) \,,\qquad  
    \delta_{U(1)} (\chi_2\pm i\chi_5) = \mp i\frac{1}{4} (\chi_2\pm i \chi_5) \,,
     \nonumber
\\
 & \delta_{U(1)} (X_6 \pm i X_7) = \mp i \frac{1}{8} (X_6 \pm i X_7) \,,\qquad 
 \delta_{U(1)} (\chi_3\pm i\chi_4) = \pm i\frac{1}{4} (\chi_3\pm i \chi_4) \,,\nonumber
 \\&\delta_{U(1)} (X_8 \pm i X_9) = \mp i \frac{1}{8} (X_8 \pm i X_9) \,, \qquad
    \delta_{U(1)} X_3 = 0 \,,
    \qquad \delta_{U(1)} \chi_7 = 0 \,.\nonumber
\end{align}
This notation will prove very useful in the following sections.

Finally we need to choose a Grassmann-odd potential $V_0$ to deform the action. We take the conventional choice in supersymmetric localization \cite{Pestun:2007rz,Cremonesi:2013twh}, namely
\begin{equation}
    V_0 =  \mathrm{Tr} \ \psi_\alpha (\delta_s \psi_\alpha)^\dagger \, .
\end{equation}
This ensures that the purely bosonic part of $ \delta_s V_0$ is non-negative, since
\begin{equation}
    t \delta_s V_0 = t \ \mathrm{Tr} \left( |\delta_s \psi_\alpha|^2 - \psi_\alpha \delta_s \Bigl[ (\delta_s \psi_\alpha)^\dagger \Big]\right) \,. \label{eq:deformation_new}
\end{equation}
The first term is purely bosonic and positive, whereas the second term is a fermionic bilinear. 
\subsection{Saddle points}
To do the saddle point approximation at $t \to \infty$, we first need to identify the loci of the saddle points in \eqref{eq:deformation_new}. This is done by writing explicitly
\al{
\spl{
\mathrm{Tr} |\delta_s \psi|^2 = \mathrm{Tr} & \Biggl\{ \left(K_7 + \frac{\Omega}{8} X_{10}\right)^2 + K_{a'}^2 + \left( \frac{3 \Omega}{8} X_i + \frac{i}{2} \epsilon_{i j k} [X_j, X_k] \right)^2 \\ & + \left( \frac{\Omega}{8} X_4 - \frac{i}{2} [X_5,X_3] \right)^2 + \left( \frac{\Omega}{8} X_5 - \frac{i}{2} [X_3,X_4] \right)^2 \\ &  + \left( \frac{\Omega}{8} X_6 - \frac{i}{2} [X_7,X_3] \right)^2 + \left( \frac{\Omega}{8} X_7 - \frac{i}{2} [X_3,X_6] \right)^2 \\ & + \left( \frac{\Omega}{8} X_8 - \frac{i}{2} [X_9,X_3] \right)^2 + \left( \frac{\Omega}{8} X_9 - \frac{i}{2} [X_3,X_8] \right)^2 \\ & - \frac{3}{4} [X_3, X_{p'}]^2 -  [X_{i'}, X_{p'}]^2 - \frac{1}{2} [X_{p'}, X_{q'}]^2 - [X_I, X_{10}]^2 \Biggr\}\,,
} 
\label{eq:purely bosonic part of S_loc}
}
\al{\spl{
\mathrm{Tr}  \left( \psi^\top \delta_s \Bigl[ (\delta_s \psi)^* \Big]\right) = \mathrm{Tr} \Biggl\{ & - \frac{3 i \Omega}{4} \psi_1^\top \psi_2 - \frac{i \Omega}{4}\psi_4^\top \psi_5 - \frac{i \Omega}{4} \psi_6^\top \psi_7 - \frac{i \Omega}{4} \psi^\top_8 \psi_9 \\ & - \frac{i \Omega}{2} \chi_1^\top \chi_6 + \frac{i \Omega}{2} \chi_2^\top \chi_5 - \frac{i \Omega}{2} \chi_3^\top \chi_4 + \frac{3 i \Omega}{4} \chi_7^\top \psi_3 + \dots \Biggr\} \,,
}
\label{eq:fermionic part of S_loc}
}
where $\dots$ denote fermionic bilinears involving bosonic matrices and the notation for the indices follows \eqref{eq:Notations_indices}.
Since the matrices are Hermitian, all terms in the purely bosonic part are positive. Thus, the saddle point loci are determined by imposing that each term in \eqref{eq:purely bosonic part of S_loc} vanishes, leading to\footnote{The reason why fermions are set to zero can be seen through a simple example 
\al{
\lim_{t\to\infty}
\int d^{n}\y\, d^{n}\chi\, e^{-t\, \chi^\top A \y} f(\chi,\y) = t^n/n! \int d^{n}\y\, d^{n}\chi\,   (\chi^\top A \y)^n f(0,0) + O(t^{n-1})\,,
}
where $\chi$, $\y$ are $n$-component fermions and $A$ is a matrix (operator). Also note that from \eqref{eq:fermionic part of S_loc} we see there is no fermonic zero modes in $\d_s V$.}
\begin{equation}
    X_i = \frac{3 \Omega}{8} L_i \,, \quad X_{10} = M \,, \quad K_7 = - \frac{\Omega}{8} M \,, \quad X_{p'} = 0\,, \quad K_{a'} = 0 \,, \quad \psi_{I'} = 0 \,, \quad \chi_a = 0 \,,
\label{eq:localization saddle point}
\end{equation}
where 
\begin{equation}
    [L_i, L_j] = i \epsilon_{i j k} L_k\,, \quad [M, L_i] = 0\,.
\end{equation}
The matrices $L_i$ are $SU(2)$ generators, and the requirement $[M,L_i] = 0$ implies, by Schur's lemma, that $M$ is block-diagonal in the irrep types, and takes constant values in each irrep. For concreteness, $L_i$ and $M$ take the form (recall that $N_s=2j_s+1$) 
\begin{equation}
    L_i = \begin{pmatrix}
        \mathbb{1}_{n_1} \otimes L_i^{[j_1]} & & & & \\ & \ddots & & & \\ & & \mathbb{1}_{n_s} \otimes L_i^{[j_s]} & & \\ & & & \ddots & \\ & & & & \mathbb{1}_{n_q} \otimes L_i^{[j_q]}
    \end{pmatrix} \, , \label{eq:L_i}
\end{equation}
\bigskip
\begin{equation}
    M = \begin{pmatrix}
        M_{1} \otimes \mathbb{1}_{N_1} & & & & \\ & \ddots & & & \\ & & M_s \otimes \mathbb{1}_{N_s} & & \\ & & & \ddots & \\ & & & & M_{q} \otimes \mathbb{1}_{N_q}
    \end{pmatrix} \, . \label{eq:M}
\end{equation}
As we will see later, using the $SU(N)$ gauge invariance we can diagonalize $M$, meaning that each block $M_s$ will be written as 
\begin{equation}
    M_s = \mathrm{diag}(m_{s i})_{i=1,...,n_s}\,, \label{eq:M-diagonalized}
\end{equation}
which is convenient for  localization.

\subsection{Gauge fixing}
The saddle points $X_i \propto L_i$,
for non-trivial $L_i$ have some gauge redundancy $L_i \to U L_i U^\dagger$ (and correspondingly $M\to U M U^\dagger$).
It is not easy to gauge fix $L_i$ since this is representation-dependent. For example, the gauge redundancy of the $N$-dimensional $SU(2)$ irrep is $SU(N)$ 
whereas the $SU(2)$ trivial representation $L_i = 0$ has no gauge redundancy.
More generally, the non-trivial gauge redundancy of \eqref{eq:L_i} is given by the action of the group\footnote{The action of $U(n_s)$ treats each of the $n_s$ $SU(2)$ irrep blocks in \eqref{eq:L_i}, i.e. $L_j^{[j_s]}$, as a matrix element of an $n_s\times n_s$ matrix. Concretely, elements of $U(n_s)$ have the same matrix form as $M$ in \eqref{eq:M} and they commute with $L_j$.}
\begin{equation}
    \mathcal{G}_\mathcal{R} =
    \frac{SU(N)}{\left(\otimes_{s = 1}^{q} U(n_s)\right)/U(1)}
    =\frac{U(N)}{\otimes_{s = 1}^{q} U(n_s) }\,,
    \label{eq:gauge_redundancy}
\end{equation}
where $\RR$ denotes a saddle point in \eqref{eq:localization saddle point} or equivalently an $N$-dimensional representation of $SU(2)$ generators.

The convenient way to gauge fix is to introduce Faddeev-Popov ghosts. This allows us to gauge fix the non-trivial unitary rotations of $L_i$, in a way that treats all the $SU(2)$ representations on the same footing. Before elaboration, let us first review the main idea. 
The partition function will be written as a sum over the saddle points $Z= \sum_\RR Z_\RR$, and for each $Z_\RR$ we introduce
\al{
1 = \int [d \alpha] \mathrm{det}'\left(\frac{\delta F}{\delta \alpha}\right) \delta'(F[\Phi])\,,
\label{eq:FP trick}
}
where $F[\Phi]$ is a gauge-fixing condition on the fields collectively denoted by $\Phi$, $\alpha$ are coordinates of a group element \eqref{eq:gauge_redundancy}. Due the to slightly unusual gauge group \eqref{eq:gauge_redundancy}, the gauge fixing requires extra care for treating the zero modes of the ghosts. The end result is captured by the primes ($\det^{\prime}$ and $\delta^{\prime}$) above, which indicate the exclusion of zero modes such that the determinant and the Dirac delta function are well-defined (see section \ref{subsec:ghost action}). Using gauge invariance, the integral over $\alpha$ factors out, giving the gauge group volume, whereas the determinant and the delta function are written in terms of a ghost action $S_\mathrm{gh}$. In our localization setup, this would give\footnote{Note that the inclusion of the parameter $t$ in front of the ghost action has no incidence on the analysis, as one can rescale fields in a measure invariant way to make this factor appear.} 
\begin{equation}
    Z_\mathcal{R} = \int [d \Phi] e^{- (S + t \delta_s V_0)} = \mathrm{Vol}(\mathcal{G}_\mathcal{R}) \int [d \Phi] [d\F_\mathrm{gh}] e^{- (S + t \delta_s V_0 + t S_\mathrm{gh})} \,,\label{eq:FaddeevPopovTrick}
\end{equation}
where $\F_{\rm gh}$ denotes ghost fields collectively. 

Written in this form, the exponent in the integrand is no longer of the form $S + t \delta V$ with $\delta^2 V=0$, which was the form we wanted to localize the integral. However, we can get back to this form by formulating the ghost action in the BRST formalism and combine it with supersymmetry, as was done in \cite{Pestun:2007rz,Asano:2012zt}. This will allow us to treat $\delta_s V_0$ and $S_\mathrm{gh}$ on the same footing and write the exponent as $S+t Q V$, where the operator $Q=\delta_s+\delta_b$ combines both the supersymmetry $\delta_s$ and the BRST $\delta_b$ transformations and $Q^2 V=0$. Let us describe this in more detail.

\subsubsection{The ghost action}
\label{subsec:ghost action}
To begin with, we choose the following gauge fixing condition\footnote{Schematically we can write $X_j=L_j + \d X_j^{\text{(phys)}} + \d X_j^{\text{(gauge)}}$, and the condition \eqref{eq:gauge fixing} indicates that $[L_j,\d X_j^{\text{(phys)}}]=0$. See \cite[Sec.3.2]{Hartnoll:2024csr} for an explanation on why fluctuations satisfying this constraint are orthogonal to the gauge orbit. One can also check that \eqref{eq:gauge fixing} gives $N^2-\sum_s n_s^2$ independent constraints on $X_j$, which equals exactly the number of gauge transformation directions in the gauge group $\GG_\RR$.
}
\begin{equation}
    F[X]\equiv - i [L_j, X_j]\overset{!}{=}0\,.
\label{eq:gauge fixing}
\end{equation}
This is a zero-dimensional version of the background gauge, which was employed e.g.~in the analysis of scatterng amplitudes on the Coulomb branch of $\mathcal{N}=4$ super Yang-Mills \cite{Alday:2009zm}.

To fix the gauge through the ghosts they should live in the space of the gauge transformation directions, and their fluctuations living in the same space as the physical degrees of freedom should be removed. This will be implemented by introducing additional zero-mode ghosts $\Phi_{\rm gh,0}$ satisfying $[L_j,\Phi_{\rm gh,0}]=0$, as will be explained in the remainder of this section.
Explicitly, we add the ghost fields
\begin{equation}
    \{ \Phi_\mathrm{gh} \} = \{ C, C_0, \tilde{C}, \tilde{C}_0, b, b_0, a_0, \tilde{a}_0 \} \,,
\end{equation}
\begin{equation}
    \{ \Phi_{\mathrm{gh},0} \} = \{ C_0, \tilde{C}_0, b_0, a_0, \tilde{a}_0 \} \,,
\end{equation}
where the ghosts $C, C_0, \tilde{C}, \Tilde{C}_0$ are fermionic whereas $b$, $b_0$, $a_0$, $\tilde a_0$ are bosonic. The ghosts $C$, $\tilde C$ and $b$ are Hermitian $N$ by $N$ matrices. Since $\Phi_{\mathrm{gh},0}$ commute with $L_i$, they all take the same form as $M$, \eqref{eq:M}, parameterized by Hermitian $n_s \times n_s$ matrices. The full ghost action reads
\al{
\spl{
S_\mathrm{gh} & = \mathrm{Tr} \Bigl( i b F[X] + i \tilde{C} [L_j, [X_j, C]] + i b b_0 + i \tilde{C} C_0 + i C \tilde{C}_0 
\\ 
&\quad + i \left(a_0 + i \frac{3 \Omega}{8} L_3 - i \phi + i C^2\right) \tilde{a}_0  + \tilde{C} [L_j, \psi_j] \Bigr)\,.
}
\label{eq:ghost action}
}
The first two terms $i b F$ and $i\tilde{C} [...,C]$ are the conventional ones, giving respectively the Dirac $\delta$ function for the gauge fixing and the Faddeev-Popov determinant. The terms $i b b_0$, $i \tilde{C} C_0$ and $i C \tilde{C}_0$  remove the components of $C, \tilde C$ and $b$ which are zero-modes under $F$ and $\delta F / \delta \alpha$. The second line, which wouldn't be necessary if our purpose was only to gauge fix, comes from the constraint of BRST symmetry as well as consistency with supersymmetry as shown in section \ref{sec:BRSTplusSUSY} and in appendix \ref{app:BRST_cohomology}. Let us describe in more details the construction of the zero-modes and why the second line doesn't affect the discussion.

The ghost $b$ has zero modes under ${\rm Tr}(F[X]\, \cdot)$, i.e.,
\al{
0 \equiv {\rm Tr}(F[X] b^{\text{(zero)}}) = 
i{\rm Tr}(X_j [L_j,b^{\text{(zero)}}])\,.
}
In order for the Dirac delta function in \eqref{eq:FP trick} to be well-defined, we introduce $b_0$ and impose  
\al{
[L_j, b_0]=0 \,,
\label{eq:condition for b_0}}
such that it indeed lives in the same space as the zero modes of $b$. Then  the integral of $\exp{(ibb_0)}$ gives a Dirac delta on the zero modes of $b$ and eliminates them. Similarly, the ghosts $C$ and $\tilde{C}$ have zero modes under the operator $[L_j,[X_j,\,\cdot] \equiv \frac{\d F}{\d \a}$, i.e.,\footnote{One can similarly analyze the zero modes of $\tilde{C}$ using the identity
\al{
{\rm Tr}(\tilde{C}[L_j, [X_j,C]]) = - {\rm Tr}(C[X_j,[L_j,\tilde{C}]])\,.
}
}
\al{
0\equiv[L_j,[X_j,C^{\text{(zero)}}]]
=
[X_j,[L_j,C^{\text{(zero)}}]]\,.
}
In the last equality we have used the Jacobi identity as well as the gauge fixing condition \eqref{eq:gauge fixing}. This allows to identify the zero-mode ghosts as\footnote{Here we only show that \eqref{eq:condition for C_0} is sufficient. Using \eqref{eq:condition for b_0} and \eqref{eq:BRST_algebra}, one can also show $[L_j,C_0]=0$.}
\al{
[L_j,C_0]=[L_j,\tilde{C}_0]=0\,.
\label{eq:condition for C_0}
}
Then just as for $b$, the terms $i\tilde{C} C_0$ and $i C \tilde{C}_0$ will remove the zero modes of $C$, $\tilde{C}$ after integrating out $\tilde{C}_0$, $C_0$. 

Up to this point we have justified the first line of \eqref{eq:ghost action}. The second line comes from the constraint of BRST symmetry as well as consistency with supersymmetry as previously mentioned. Note that the term $i(a_0 + ...)\tilde{a}_0$ does not affect the gauge fixing since the integral over $a_0, \tilde a_0$ is a constant. Finally, note the last term $\mathrm{Tr} \tilde C [L_j, \psi_j]$ never contributes to physical expectation values without extra ghost insertions.



\subsubsection{The BRST+SUSY formalism}
\label{sec:BRSTplusSUSY}
To implement the BRST formalism we have introduced a family of \textit{eight} ghosts, denoted by $(C, C_0, \tilde{C}, \tilde{C}_0, b, b_0, a_0, \tilde{a}_0)$, in \eqref{eq:ghost action}. 
As shown in appendix \ref{app:BRST_cohomology}, it is natural to define the BRST anti-commuting operator as
\al{
\spl{
& \delta_b Y = [Y, C] \,, \\
& \delta_b C = i\bigg( a_0 + i \frac{3 \Omega}{8} L_3\bigg) - C^2 \,, \\
& \delta_b \tilde{C} = i b \,, \\
& \delta_b \tilde{a}_0 = - i \tilde{C}_0 \,, \\
& \delta_b b_0 = - i C_0 \,, \\
& \delta_b a_0 = 0 \,,
}
\hspace{-2cm}
\spl{
& \delta_b Y' = [Y', C] \,, \\
& \delta_b \phi = [\phi, C] \,, \\
& \delta_b b = [\tilde{C}, a_0 + i \frac{3 \Omega}{8} L_3] \,, \\
& \delta_b \tilde{C}_0 = - [\tilde{a}_0, a_0 + i \frac{3 \Omega}{8} L_3] \,, \\
& \delta_b C_0 = - [b_0, a_0 + i \frac{3 \Omega}{8} L_3] \,,
}
\label{eq:BRST_algebra}
}
where by the notation $[\cdot, C]$ we mean the commutator for the bosonic components of  $Y,Y'$  and minus the anticommutator, $-\{ \cdot , C\}$ for the fermionic components of  $Y, Y'$. It obeys the property,
\begin{equation}
    \delta_b^2 = i \left[\cdot, a_0 + i \frac{3 \Omega}{8} L_3\right]\,, \quad \quad \delta_b \mathrm{Tr} [\mathrm{physical \ fields}] = 0 \implies \delta_b V_0 = 0\,.
\end{equation}

Since both BRST symmetry and supersymmetry have a cohomology structure, it is convenient to combine them for the supersymmetric localization \cite{Pestun:2007rz}. Let us define the combined operator
\begin{equation}
    Q \equiv \delta_s + \delta_{b}\,.
\end{equation}
We also define the action of supersymmetry on the ghosts as
\begin{equation}
    \delta_s C = \phi\,, \qquad \delta_s(\mathrm{other \ ghosts}) = 0\,.
\end{equation}
Then it is straightforward to check that
\al{
\spl{
& Q Y = Y' + [Y, C] \,, \\
& Q C = \left(\phi - \frac{3 \Omega}{8} L_3 \right) + i a_0 - C^2 \,, \\
& Q \tilde{C} = i b \,, \\
& Q \tilde{a}_0 = - i \tilde{C}_0 \,, \\
& Q b_0 = - i C_0 \,, \\
& Q a_0 = 0 \,,
}
\spl{
& Q Y' = - i (\delta_\phi + \Omega \delta_{U(1)}) Y + [Y', C] \,, \\
& Q \phi = [\phi, C] \,, \\
& Q b = [\tilde{C}, a_0 + i \frac{3 \Omega}{8} L_3] \,, \\
& Q \tilde{C}_0 = - [\tilde{a}_0, a_0 + i \frac{3 \Omega}{8} L_3] \,, \\
& Q C_0 = - [b_0, a_0 + i \frac{3 \Omega}{8} L_3] \,.
}
\label{eq:action of Q}
}
The transformation $Q$ satisfies the property\footnote{We also choose $\d_{U(1)}\F_{\rm gh}=0$ for all ghost fields.}
\al{
\spl{
Q^2 = R \equiv - i \Omega \delta_{U(1)} + i \left[\cdot, a_0 + i \frac{3 \Omega}{8} L_3\right]\,.
}
}

The BRST+SUSY formalism can now be used to write $S_\mathrm{gh}$ as a $Q$-exact quantity, namely
\begin{equation}
    S_\mathrm{gh} = Q V_\mathrm{gh} \,, \qquad V_\mathrm{gh} = \mathrm{Tr} (\tilde{C} (F + b_0) + C \tilde{a}_0)\,.
\end{equation}
Using $\delta_s V_0 = Q V_0$, we can define
\begin{equation}
    V \equiv V_0 + V_\mathrm{gh}\,,
\end{equation}
and write the partition function \eqref{eq:FaddeevPopovTrick} 
as a $Q$-exact quantity
\begin{equation}
    Z_\mathcal{R} = \mathrm{Vol}(\mathcal{G}_\mathcal{R}) \int [d \Phi] [d\F_\mathrm{gh}] e^{- (S + t Q V)} \,.
    \label{eq:FaddeevPopovTrick2}
\end{equation}
We will use this as our localizing partition function, in view of \eqref{eq:localization action}.\footnote{
One might worry that $QV$ does not have the same saddle points as $\d_s V_0$. As explained in e.g. \cite[Sec.3.1]{Kapustin:2009kz}, the added ghost action does not modify the saddle point loci.}

\subsection{The saddle point approximation}
We now perform the saddle point approximation of \eqref{eq:FaddeevPopovTrick2} around each saddle in the $t \to \infty$ limit\footnote{Note that the ghost variables are all set to 0 at the saddle point, except for $a_0 \to M$.}, in order to compute the partition function. First note that at the saddle $\mathcal{R}$, the action localizes to
\begin{equation}
    S_0 = - \frac{9\Omega^4}{2^{13}} \mathrm{Tr} L_i^2 + \frac{3 \Omega^2}{2^7} \mathrm{Tr} M^2 = - \frac{9 \Omega^4}{2^{15}} \sum_s n_s (N_s^3 - N_s) + \frac{3 \Omega^2}{2^7} \sum_{s} \sum_{i = 1}^{n_s} m_{s i}^2 \,, \label{eq:Action_SaddlePoint}
\end{equation}
where $m_{s i}$ are the eigenvalues of the block $s$, $M_s$ part of $M$.
The remaining quantity to be computed is the 1-loop determinant. Before computing it, it is useful to do a change of variables of \eqref{eq:localization_variables},
\al{
\ga{
- i \tilde{\psi}_{I'} \equiv - i \psi_{I'} + [X_{I'}, C] \,, \\
i \tilde{H}_a \equiv i H_a - \{ \chi_a, C \} \,,  \\
 \tilde{\phi} \equiv a_0 + i \frac{3 \Omega}{8} L_3 - i \phi + i C^2 \,. \label{eq:change_of_var}
}
}
The first two are simply translations of $\psi_{I'}$ and $H_a$, whereas the last can be thought as a translation both on $a_0$, which commutes with $L_i$ and the components of $X_{10}$ which do not commute with $L_i$. Note that at the saddle point,
\begin{equation}
    \tilde{H}_a = 0 \,, \quad (\mathrm{all \ fermions}) = 0 \,, \quad \tilde{\phi} = 0 \implies a_0 = M \,.
\end{equation}
These new variables are useful for the following reason. Defining,
\al{
\ga{
Z_b \equiv (X_{I'}, \tilde{a}_0, b_0) \,, \quad Z_f \equiv (\chi_a, C, \tilde{C}) \,, \\
Z_f' \equiv (\tilde \psi_{I'}, \tilde{C}_0, C_0) \,, \quad Z_b' \equiv (\tilde{H}_a, \tilde{\phi}, b) \,,
}
}
they transform nicely under $Q$,
\begin{equation}
    Q Z_b = - i Z'_f \,, \quad Q Z_f = i Z'_b \,, \quad \quad Q Z_f' = i R Z_b \,, \quad Q Z_b' = - i R Z_f \,,
\end{equation}
where
\begin{equation}
    R \equiv - i \Omega \delta_{U(1)} + i \left[\cdot, a_0 + i \frac{3 \Omega}{8} L_3\right] \,.
\end{equation}
To compute the 1-loop determinant, we expand each $Z$ around the saddle point as $Z = \hat{Z} + \tilde{Z}$ where $\hat{Z}$ is the saddle and $\tilde{Z}$ is the fluctuation, obtaining schematically
\begin{equation}
    V_0 + V_\mathrm{gh} = \begin{pmatrix}
        \tilde Z'_f && \tilde Z_f
    \end{pmatrix} \begin{pmatrix}
        D_{0 0} && D_{0 1} \\ D_{1 0} && D_{1 1}
    \end{pmatrix} \begin{pmatrix}
        \tilde Z_b \\ \tilde Z'_b
    \end{pmatrix} + \mathcal{O}\left(\tilde{Z}^3\right) \, .
\end{equation}
Indeed, one can check that there is no linear piece in $\tilde Z$, as shown in appendix \ref{app:explicit_formulae}. Then, noting that $Q Z$ is always at least linear in the fluctuation,
\al{
\spl{
Q (V_0 + V_\mathrm{gh}) = &  i \begin{pmatrix}
        R_0 \tilde Z_b && \tilde Z'_b
    \end{pmatrix} \begin{pmatrix}
        D_{0 0} && D_{0 1} \\ D_{1 0} && D_{1 1}
    \end{pmatrix} \begin{pmatrix}
        \tilde Z_b \\ \tilde Z'_b
    \end{pmatrix} \\ & + i \begin{pmatrix}
        \tilde Z'_f && \tilde Z_f
    \end{pmatrix} \begin{pmatrix}
        D_{0 0} && D_{0 1} \\ D_{1 0} && D_{1 1}
    \end{pmatrix} \begin{pmatrix}
        \tilde Z'_f \\ R_0 \tilde Z_f
    \end{pmatrix}+ \mathcal{O}\left(\tilde{Z}^3\right) \,, \label{eq:Deformation_NiceForm}
}
}
where
\begin{equation}
    R_0 \equiv - i \Omega \delta_{U(1)} - i \left[ M + i \frac{3 \Omega}{8} L_3 , \cdot \right] \, .
\end{equation}
Indeed, it is straight-forward to check that $R Z = R_0 \tilde{Z} + \dots$ where the additional terms are higher order in $\tilde Z$ as shown in \eqref{eq:RZequalsRZ0}.
This is a nice structure because the determinants of the $D_{i j}$ matrices cancel between the fermions and the bosons. We are thus left with
\begin{equation}
    Z_\mathrm{1-loop} \propto \left(\frac{\det_{\tilde Z_f} R_0}{\det_{\tilde Z_b} R_0}\right)^{1/2} \,.
\end{equation}
We obtain these determinants by computing the eigenvalues of $R_0$ on the relevant matrices. 
To do it, we first define
\al{
\ga{
B_1 \equiv X_1 - i X_2 \,, \quad \quad \quad B_2 \equiv X_4 - i X_5 \,, \\  B_3 \equiv X_6 - i X_7 \,, \quad \quad \quad B_4 \equiv X_8 - i X_9 \,, \\
\xi_1 \equiv \chi_1 + i \chi_6 \,, \quad \quad \xi_2 \equiv  \chi_2 - i \chi_5 \,, \quad \quad \xi_3 \equiv \chi_3 + i \chi_4 \,.
}
}
These definitions allow to make $-i \Omega \delta_{U(1)}$ diagonal,
\al{
\ga{
- i \Omega \delta_{U (1)} B_1 = \frac{3 \Omega}{8} B_1 \,, \quad - i \Omega \delta_{U (1)} B_{i = 2, 3, 4} = \frac{\Omega}{8} B_i \,, \quad - i \Omega \delta_{U(1)} \xi_i = \frac{2\Omega}{8} \xi_i \,, \\
- i \Omega \delta_{U(1)} (\mathrm{other \ matrices}) = 0 \,.
}
}
We also diagonalize $M$ using the remaining $\otimes_s U(n_s)$ gauge symmetry. We denote the diagonal values of $M$ by $m_{s i}$, where $s$ labels the irrep and $i$ goes over its multiplicity. Thus, the operator $[M, \cdot]$ acts diagonally on the matrices. Regarding $[L_3, \cdot]$, the eigenvalues are easily obtained using fuzzy spherical harmonics.

This procedure is detailed in appendix \ref{app:eigenvalues}. The eigenvalues of $R_0$ and the Vandermonde determinant associated to the diagonalization of $M$ combine to the 1-loop determinant\footnote{We also performed a rescaling $M \to \Omega M$ to get rid of the $\Omega$ dependence in the 1-loop determinant.}
\al{
\spl{
& Z_{\mathrm{1-loop}}  = \prod_{s,t = 1}^{q} \prod_{\substack{i, j = 1\\ i \neq j}}^{n_s, n_t} \prod_{J = |j_s - j_t|}^{j_s + j_t}  \\ &  \quad \quad \quad \quad \left[ \frac{\left[\left( \frac{1}{8} \right)^2 (2 + 3 J)^2 + (m_{s i} - m_{t j})^2 \right]^3 \left[ \left( \frac{1}{8} \right)^2 (3 J)^2 + (m_{s i} - m_{t j})^2  \right]}{\left[\left( \frac{1}{8} \right)^2 (1 + 3 J)^2 + (m_{s i} - m_{t j})^2 \right]^3 \left[ \left( \frac{1}{8} \right)^2 (3 + 3 J)^2 + (m_{s i} - m_{t j})^2  \right]} \right]^{1/2} \,, \label{eq:1_loop_det_2}
}
}
which can also be written in the form \eqref{eq:1-loop-det}.

\subsection{The partition function and protected observables}
At the saddle point, $S = S_0$ in \eqref{eq:Action_SaddlePoint}, and the 1-loop determinant is \eqref{eq:1_loop_det_2}. Since this approximation is exact and independent of $t$, this is an exact result for the partition function, reduced to a sum over the saddle points, i.e. $SU(2)$-representations, of an integral over the moduli space parameterized by $m_{si}$. Combining \eqref{eq:1_loop_det_2}, \eqref{eq:Action_SaddlePoint} we obtain, with the rescaling $m_{s i} \to \Omega m_{s i}$,
\begin{equation}
    Z = \sum_{\mathcal{R}} C_\mathcal{R} e^{ \frac{9 \Omega^4}{2^{15}} \sum_s n_s (N_s^3 - N_s)} \int \prod_s \prod_{i = 1}^{n_s} d m_{s i} Z_{\mathrm{1-loop}} \ \mathrm{exp} \left( - \frac{3 \Omega^4}{2^7} \sum_s \sum_{i = 1}^{n_s} N_s m_{s i}^2 \right)  \,,
\label{eq:LocalizationResult}
\end{equation}
\al{
\spl{
C_\mathcal{R} & = \frac{ (2 \pi)^
    {5 N^2 +  N/2 }}{\prod_{k=1}^{N-1} k! \prod_s n_s !} \left(
    \frac{32}{3\pi } 
    \right)^{\sum_s n_s} \prod_{s} \left(\frac{1}{N_s} \prod_{J = 1}^{N_s -1} \frac{(2 + 3 J)^3}{(1 + 3 J)^3}\right)^{n_s} \,,
}
}
where the normalization factor $C_\mathcal{R}$ is computed in appendix \ref{app:Normalization_factors}. 

This analysis also applies to correlators of gauge invariant combinations of the protected operators $\phi = X_3 - i X_{10}$, namely
\al{
\spl{
\langle f(\phi) \rangle = \frac{1}{Z} & \sum_{\mathcal{R}} C_\mathcal{R} e^{\frac{9 \Omega^4}{2^{15}} \sum_s n_s (N_s^3 - N_s)} \\ & \int \prod_s \prod_{i = 1}^{n_s} d m_{s i} Z_{\mathrm{1-loop}} f\left(\frac{3\Omega}{8}L_3 - i \Omega M \right) \ \mathrm{exp} \left( - \frac{3 \Omega^4}{2^7} \sum_s \sum_{i = 1}^{n_s} N_s m_{s i}^2 \right) \,, \label{eq:correlators}
}
}
where $f$ is a gauge invariant function of $\phi$. Note that in this equation, $L_3$ and $M$ take the form \eqref{eq:L_i}, \eqref{eq:M}, where we diagonalized each block $M_s$.

So far, we have discussed the partition function of  the $U(N)$ theory because we  did not impose the matrices to be traceless. The $SU(N)$ partition function is obtained by integrating out the traces of $X^I$ and $\psi_\alpha$ which are purely Gaussian. Writing the matrices in generator components $X_I = X_I^A T^A$, $\psi_\alpha = \psi_\alpha^A T^A$, we define the $SU(N)$ partition function as
\begin{equation}
    Z_{SU(N)} = \int \prod_{A =1}^{N^2-1} \left(\prod_{I=1}^{10} d X^A_I \prod_{\alpha =1 }^{16}  d \psi^A_\alpha\right) e^{- S} \, ,
\end{equation}
where we also use the normalization 
\begin{equation}
    \mathrm{Tr} T^A T^B =  \delta^{A B} \, . \label{eq:GeneratorsNormalization}
\end{equation}
In this case we find that the partition function $Z$ and $Z_{SU(N)}$ are simply related by
\begin{equation}
    Z = Z_{U(1)}Z_{SU(N)} \,, \qquad Z_{U(1)} = \frac{2^{14} \pi^5}{3 \sqrt{3} \Omega^2} \,. \label{eq:relationToTraceless}
\end{equation}
Similarly, we can relate $U(N)$ and $SU(N)$ correlators. 
Expanding $\phi$ in its traceless part and its trace, we obtain
\al{
\spl{
\langle \mathrm{Tr} \phi^{2 n} \rangle  
= \sum_{k=0}^n {2n \choose 2k} \left( \frac{-2^7}{3 N \Omega^2} \right)^{n-k} \frac{\Gamma(n-k + \frac{1}{2})}{\sqrt{\pi}} \langle \mathrm{Tr} \phi^{2k} \rangle_{SU(N)} \,. \label{eq:CorrelatorsToTraceless}
}
} 
where $\phi$ in the right-hand-side is traceless, and on both sides we use the same normalization \eqref{eq:GeneratorsNormalization} for both the $N^2$ generators of $U(N)$ or the $N^2-1$ generators of $SU(N)$.

\section{Analysis of the localization result}
\label{sec: analysis of the result}
In this section, we perform various analyses of the localization results \eqref{eq:LocalizationResult} and \eqref{eq:correlators}.

Firstly, we write what our general results predict for $N=2$. In this simple case, the integrals can be simplified explicitly in terms of special functions, and we provide  plots of some quantities.

Secondly, we discuss the limit of vanishing mass deformation $\Omega \to 0$. Naively, one might expect that the limit $\Omega \to 0$ would allow to obtain the value of the IKKT partition function based on solid grounds. However, \eqref{eq:LocalizationResult} predicts that it diverges,
\begin{equation}
    \lim_{\Omega \to 0} Z(\Omega) = \infty \neq Z_\mathrm{IKKT} \, .
\end{equation}
The reason for this  is the presence of massive fermions in the polarized IKKT model. In order to establish the convergence of the IKKT partition function, one crucial factor was that the integral over fermions  produces a power-law decay along the commuting modes valley \cite{Austing:2001bd,Aoki_1998}. When giving a tiny mass to the fermions, this power-law decay is modified and the partition function with traceless matrices diverges as $1/\Omega^{2(N-1)}$, as predicted by our localization result. This analysis is carefully carried in the  section \ref{app:strong_limit}. In appendix \ref{app:fermion_masses_ConvergenceIKKT}, we present a direct argument following the original reasoning for the convergence of the IKKT partition function.

Thirdly, we study the limit $\Omega \to \infty$. In this case, it is interesting to compare our result to the one obtained in \cite{Hartnoll:2024csr}. We find precise agreement as shown in section \ref{app:free_limit}.

Finally, we show in section \ref{sec:RecoverIKKT} that we can recover the IKKT partition function for $N=2$ and $N=3$ by assuming a contour prescription to avoid the $\Omega \to 0$ singularity, which we conjecture to be valid for arbitrary $N$.

\subsection{Explicit results for $N = 2$}
\begin{figure}
    \centering
    \includegraphics[scale=0.8]{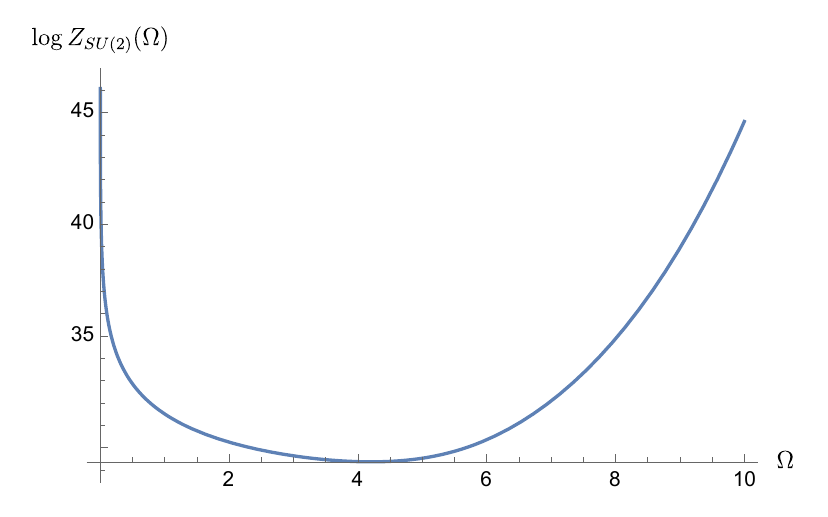}
    \caption{The free energy $\log Z(\Omega)$ of $SU(2)$ polarized IKKT. It has a minimum at $\Omega \approx 4.2046(5)$}
    \label{fig:ZSU2}
\end{figure}

\begin{figure}
    \centering
    \includegraphics[scale=0.75]{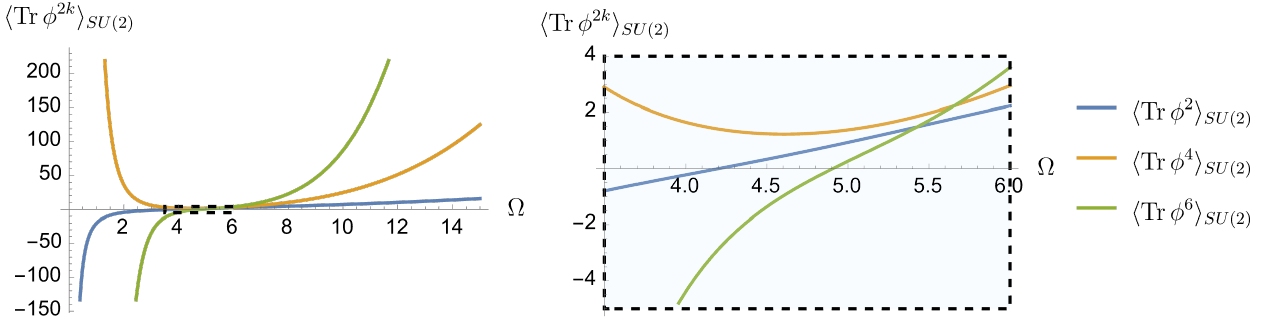}
    \caption{ The correlation function $\langle \mathrm{Tr}\phi^{2n} \rangle = \langle \mathrm{Tr} (X_3 - i X_{10})^{2n} \rangle$ of $SU(2)$ polarized IKKT.
    In the zoomed figure on the right, we can see that $\langle \mathrm{Tr}\phi^{4} \rangle$ is always positive while $\langle \mathrm{Tr}\phi^{2} \rangle$ and $\langle \mathrm{Tr}\phi^{6} \rangle$ change sign at different values of $\Omega$.
    }
    \label{fig:TrPhiKSU2}
\end{figure}

It is enlightening to consider the case $N=2$ where we can explicitly write the integrals in terms of special functions. There are only two saddles, namely the trivial and the irreducible representations. The trivial saddle involves a double integral, but the integral over the center of mass $m_1 + m_2$ decouples, leaving only one non-trivial integral. The irreducible saddle can be evaluated exactly for any $N$, as shown in \eqref{eq:Z_largeOmegaIrrep}. Thus the case $N=2$ trivially follows. Altogether we obtain
\begin{equation}
    Z_{SU(2)} = \frac{2^{16} \pi^{29/2}}{3} \int_{- \infty}^{\infty} dx \, \left( \frac{(4+x^2)^3 x^2}{(1+x^2)^3 (9+x^2)} \right) e^{-\frac{3 \Omega^4}{2^{14}} x^2 } + 2^8 5^3 \pi^{31/2} e^{\frac{27}{2^{14}} \Omega^4} \,. \label{eq:ZSU2}
\end{equation}
This can be simplified in terms of the complementary error function $\mathrm{erfc}(x)$, namely
\al{
\spl{
Z_{SU(2)} & = \frac{2^{23} \pi^{15}}{3 \sqrt{3} \Omega^2} - 5 \cdot 3^2 \cdot 2^3 \sqrt{3} \pi^{15} \Omega^2 + \frac{3^3 \sqrt{3} \pi^{15} \Omega^6}{2^{9}}  \\ & \quad \quad + \pi^{31/2} e^{\frac{3 \Omega^4}{2^{14}}} \mathrm{erfc} \left( \frac{\sqrt{3} \Omega^2}{2^7} \right) \left( 195 \cdot 2^7 + \frac{3^4}{2^{4}} \Omega^4 - \frac{3^4}{2^{16}} \Omega^8 \right) \\ & \quad \quad + 5^3 2^8 \pi^{31/2} e^{\frac{27 \Omega^4}{2^{14}}} \left(1 - \frac{1}{2} \mathrm{erfc}\left( \frac{3\sqrt{3} \Omega^2}{2^7} \right)\right) \,.
}
}
An explicit plot is provided in figure \ref{fig:ZSU2}.
It is also interesting to consider $SU(2)$ correlators.
The $U(2)$ localization result precisely takes the form \eqref{eq:CorrelatorsToTraceless}, namely,
\begin{equation}
    \langle \mathrm{Tr} \phi^{2n} \rangle_{U(2)} = \sum_{k=0}^n {2n \choose 2k} \left( \frac{-2^6}{3 \Omega^2} \right)^{n-k} \frac{\Gamma(n-k + \frac{1}{2})}{\sqrt{\pi}} \langle \mathrm{Tr} \phi^{2k} \rangle_{SU(2)} \,,
\end{equation}
where we identify
\al{
\spl{
\langle \mathrm{Tr} \phi^{2k} \rangle_{SU(2)} & = \frac{\Omega^{2k}}{Z_{SU(2)}} \Biggl\{ \frac{2^{17} \pi^{29/2}}{3} \left( \frac{-1}{2^8} \right)^k \int_{- \infty}^{\infty} dx \, \left( \frac{(4+x^2)^3 x^2}{(1+x^2)^3 (9+x^2)} \right) x^{2k} e^{-\frac{3 \Omega^4}{2^{14}} x^2 } \\ & \qquad \qquad \qquad + 5^3 2^9 \pi^{31/2} \left( \frac{3}{16} \right)^{2k} e^{\frac{27}{2^{14}} \Omega^4} \Biggr\} \, .
}
}
This can be written in terms of error functions as well. The corresponding results are plotted in figure \ref{fig:TrPhiKSU2}.
Notice that in the limit $\Omega \to \infty$, the irreducible saddle dominates and we find
\al{
\langle \mathrm{Tr} \phi^{2k} \rangle_{SU(2)} & = 2\left( \frac{3\Omega}{16} \right)^{2k} +O\left(e^{-\frac{27}{2^{14}} \Omega^4}\right)\, ,
}
as expected from the leading irreducible saddle contribution \eqref{eq:saddle ansatz}, and discussed in more generality in \eqref{eq:CorrelatorsLargeOmega}.

\subsection{The strong coupling limit $\Omega \to 0$}
\label{app:strong_limit}
We now study the limit of vanishing mass-deformation $\Omega \to 0$. Written in terms of dimensionful quantities, the polarized IKKT action takes the form
\begin{equation}
\begin{split}
    S = \frac{1}{g_\mathrm{YM}^2}\operatorname{Tr} & \Biggl[-\frac{1}{4} [X'_I,X'_J]^2 - \frac{i}{2} \bar{\psi}' \Gamma^{I} [X'_I,\psi'] \\ & + \frac{3 \mu^2}{4^3} X'_i X'_i + \frac{\mu^2}{4^3} X'_p X'_p + i \frac{\mu}{3} \epsilon^{i j k} X'_i X'_j X'_k - \frac{1}{8} \mu \bar{\psi}' \Gamma^{1 2 3} \psi' \Biggr] \, .
\end{split}
\end{equation}
where $g_\mathrm{YM}$ has dimension two in 0+0 dimensions and $\mu$ has dimension one. This is related to \eqref{eq:action} by rescaling $X'_I = \sqrt{g_\mathrm{YM}} X_I$,  $\psi' = g_\mathrm{YM}^{3/2} \psi$ and $\Omega = \mu / \sqrt{g_\mathrm{YM}}$. We thus interpret $\Omega \to 0$ as a strong coupling limit $g_\mathrm{YM} \to \infty$.

\paragraph{From the localization result: }
To take the $\Omega \to 0$ limit of the localization result \eqref{eq:LocalizationResult}, it is useful to rescale the moduli space integration by $m_{si} \to m_{s i} / \Omega^2$. In the limit $\Omega \to 0$, we obtain
\al{
\spl{
Z & = \sum_{\mathcal{R}} \underbrace{C_\mathcal{R} \left(\frac{1}{\Omega^2} \right)^{\sum_s n_s} \int \prod_{s = 1}^{q} \prod_{i = 1}^{n_s} d m_{s i}  \mathrm{exp} \left( - \frac{3}{2^7} \sum_{s = 1}^{q} \sum_{i = 1}^{n_s} N_s m_{s i}^2 \right)}_{\equiv Z_{U(N),\mathcal{R}}^{(\Omega \to 0)}} (1 + \mathcal{O}(\Omega)) \,.
}
}
Each saddle contribution in this limit can be written as 
\begin{align}
\frac{Z_{U(N),\mathcal{R}}^{(\Omega \to 0)}}{A_N} = \prod_s \frac{1}{n_s!} \left( \frac{Z_{U(N_s), \text{irrep}}^{(\Omega \to 0)}}{A_{N_s}} \right)^{n_s}  \,,
\label{eq:MasslessDecouplingForm}
\end{align}
where
\begin{equation}
    A_N = \frac{(2 \pi)^{5 N^2 + N/2}}{\prod_{k=1}^{N-1} k!} \,,
\end{equation}
and $Z_{U(N_s),\text{irrep}}$ is the irreducible vacuum contribution \eqref{eq:ZUN_largeOmegaIrrep} when $\Omega \to 0$
\begin{equation}
Z_{U(N),\text{irrep}}^{(\Omega \to 0)} = \frac{(2 \pi)^{5 N^2 + (N-1)/2}}{\prod_{k=1}^{N-1} k!} \frac{2^9}{3 \sqrt{3} N \sqrt{N} \Omega^2} \prod_{J=1}^{N-1} \left( \frac{3J + 2}{3 J + 1} \right)^3 \,.
\end{equation}
The notation \eqref{eq:MasslessDecouplingForm} suggests that each irreducible block inside the reducible representation $\mathcal{R}$ decouples from the other blocks in the $\Omega \to 0$ limit.
This is very similar to the picture of free particles and their bound states in an harmonic trap that characterizes the strong coupling limit of the BMN model \cite{Komatsu:2024vnb}.
As we explain in the second part of this section, this can be understood by integrating out the off-diagonal matrix elements that couple the blocks to each other, without using localization.

The result \eqref{eq:MasslessDecouplingForm} predicts that each saddle in the $U(N)$ case diverges as $1 / \Omega^{2 \sum_s n_s}$. In particular, the leading divergence is governed by the trivial vacuum $n_s = N$, $N_s = 1$
\begin{equation}
    Z_{U(N)}^{(\Omega \to 0)} = \frac{(2 \pi)^{5 N^2}}{\prod_{k=1}^N k!} \left( \frac{2^9}{3 \sqrt{3} \Omega^2}\right)^N \,. \label{eq:Localization_massless}
\end{equation}
Also note that in the $SU(N)$ case, the trivial saddle $n_s = N$ has the leading divergence $1/ \Omega^{2(N-1)}$ whereas the irreducible saddle $n_s = 1$ has a convergent contribution. This convergent behavior of the irreducible saddle was also captured in \cite{Hartnoll:2024csr}, but the above analysis shows that this is a special feature. All other saddles diverge.

Let us now consider the protected correlators $\langle \mathrm{Tr} \phi^{2n} \rangle$. With the rescaling $M \to \frac{1}{\Omega^2} M$, the dominant contribution also comes from the trivial vacuum and we obtain
\al{
\spl{
\langle \mathrm{Tr} \phi^{2 n} \rangle_{U(N),\Omega \to 0} & = \frac{\int \prod_{i=1}^N dm_i \sum_j \left(\frac{-i m_j}{\Omega}\right)^{2n} e^{-\frac{3}{2^7} \sum_i m_i^2}}{\int \prod_{i=1}^N dm_i e^{-\frac{3}{2^7} \sum_i m_i^2}} \\ & = N \left( \frac{-2^7}{3 \Omega^2} \right)^n \frac{\Gamma(n+ \frac{1}{2})}{\sqrt{\pi}} \,. \label{eq:LocalizationResultMasslessCorrelators}
}
}
Note that this can be expressed as
\al{
\spl{
\langle \mathrm{Tr} \phi^{2 n} \rangle_{U(N),\Omega \to 0} & = \sum_{k=0}^n {2n \choose 2k} \left( \frac{-2^7}{3 N \Omega^2} \right)^{n-k} \frac{\Gamma(n-k + \frac{1}{2})}{\sqrt{\pi}} \left( \frac{-2^7 (N-1)}{3 N \Omega^2}\right)^k \frac{\Gamma(k+\frac{1}{2})}{\sqrt{\pi}} \,.
}
}
We thus identify thanks to \eqref{eq:CorrelatorsToTraceless},
\begin{equation}
    \langle \mathrm{Tr} \phi^{2 n} \rangle_{SU(N),\Omega \to 0} = N\left( \frac{-2^7 (N-1)}{3 N \Omega^2}\right)^n \frac{\Gamma(n+\frac{1}{2})}{\sqrt{\pi}} \, .
\end{equation}
\paragraph{From integrating out off-diagonal modes: }
First note that we can rescale coordinates such that
\begin{equation}
    Z(\Omega) = \left(\frac{1}{\Omega^2}\right)^{N^2} \tilde Z(\Omega) \,, \label{eq:ToStrongCouplingParameters}
\end{equation}
where $\tilde Z = \int [dX^I] [d \psi_\alpha] e^{- \tilde S}$ with
\al{
\spl{
\tilde S & = \operatorname{Tr} \Biggl[-\frac{1}{4 \Omega^4} [X_I,X_J]^2 - \frac{i}{2 \Omega^2} \bar{\psi} \Gamma^{I} [X_I,\psi] + i \frac{1}{3 \Omega^2} \epsilon^{i j k} X_i X_j X_k \\ & \qquad \qquad \qquad + \frac{3}{4^3} X_i X_i + \frac{1}{4^3} X_p X_p  - \frac{1}{8}  \bar{\psi} \Gamma^{1 2 3} \psi \Biggr] \,. \label{eq:S_strongLimitExpansion}
}
}
This is better suited for the strong coupling limit $g =\frac{1}{\Omega^2} \gg 1$.\footnote{Remember that this definition of $g$ is related to the introduction of the dimension-full quantities $\Omega = \mu / \sqrt{g_\mathrm{YM}}$. Then, the dimensionless gauge coupling is $g = g_\mathrm{YM} / \mu^2 = 1/\Omega^2$.} Then, we separate the matrices as
\begin{equation}
    X^I = r^I + q^I \,, \quad \quad \psi_\alpha = \theta_\alpha + \Theta_\alpha \,,
\end{equation}
where $r^I$, $\theta_\alpha$ are diagonal and $q^I$, $\Theta_\alpha$ are off-diagonal. We will use the notation
\begin{equation}
    r_{ab}^I\equiv r_a^I-r_b^I \,, \quad \hat{r}_{a b}^I \equiv r_{a b}^I/|r_{a b}| \,, \quad |r_{a b}| \equiv \sqrt{\sum_I r_{a b}^I r_{a b}^I} \,.
\end{equation}
We then use a $SO(10) \supset SO(3) \times SO(7)$ preserving gauge fixing
\begin{equation}
    Z(\Omega) = c_N \left(\frac{1}{\Omega^2}\right)^{N^2} \int [dr^I][d \theta_\alpha] [dq^I] [d \Theta_\alpha] \delta(\hat r_{a b} \cdot q_{a b}) (\Delta(r) + \mathcal{O}(\Omega)) e^{- \tilde S} \,,
\end{equation}
where $\Delta(r) = \prod_{a \neq b} |r_{a b}|$ is the Vandermonde determinant, which is the leading term in the change of variables, and $c_N = \frac{1}{N!} \frac{\mathrm{Vol} \ U(N)}{\mathrm{Vol} \  U(1)^N} = (2 \pi)^{(N^2-N)/2} / \prod_{k=1}^N (k!)$ is a volume factor. Details on this gauge fixing method are given in \cite{Komatsu:2024vnb,Lin:2014wka}. 
Then, we rescale 
\begin{equation}
    q^I = \Omega^2 y^I \,, \quad \quad \Theta_\alpha = \Omega \chi_\alpha \,.
\end{equation}
This reduces the partition function to
\begin{equation}
    Z(\Omega) = c_N \left( \frac{1}{\Omega^2} \right)^N \int [dr^I][d \theta_\alpha] [dy^I] [d \chi_\alpha] \delta(\hat r_{a b} \cdot y_{a b}) \Delta(r) e^{-S_0 - S_{y, \chi}} + \text{subleading} \,,
\end{equation} 
where
\begin{equation}
    S_{y, \chi} = \frac{1}{2} |r_{a b}|^2 y_{a b}^{I} y_{b a}^{I} - \frac{1}{2} r_{a b}^I \chi_{b a}^\top \bar{\gamma}^I \chi_{a b} + \mathcal{O}(\Omega) \,,
\end{equation}
\begin{equation}
    S_{0} = \frac{3}{4^3} r^i_a r^i_a + \frac{1}{4^3} r^p_a r^p_a + \frac{i}{8} \theta_a \gamma^{1 2 3} \theta_a \,.
\end{equation}
In the limit $\Omega \to 0$, these are Gaussian integrals,
\begin{equation}
    \int [dy^I] \delta(\hat r_{a b} \cdot y_{a b}) e^{- \frac{1}{2} |r_{a b}|^2 y_{a b}^I y_{b a}^I} = (2 \pi)^{9(N^2-N)/2} \prod_{a \neq b} |r_{a b}|^{- 9} \,,
\end{equation}
\begin{equation}
    \int [d \chi_\alpha] e^{\frac{1}{2} r_{a b}^I \chi_{b a} \bar \gamma^I \chi_{a b}} = \prod_{a \neq b} |r_{a b}|^8 \,.
\end{equation}
Combined with $\Delta(r)$, all these products cancel. We thus obtain
\al{
\spl{
Z(\Omega) & = c_N (2 \pi)^{\frac{9(N^2-N)}{2}} \left( \frac{1}{\Omega^2} \right)^N \int [dr^I] [d \theta_\alpha] e^{- S_0}(1 + \mathcal{O}(\Omega)) \\ & = \frac{(2 \pi)^{5 N^2}}{\prod_{k=1}^N k!} \left( \frac{2^9}{3 \sqrt{3} \Omega^2}\right)^N + \text{subleading} \,. \label{eq:DirectMasslessLimit}
}
}
This method can also be applied to compute correlators at small $\Omega$. The original correlator is related to the rescaled coordinates \eqref{eq:S_strongLimitExpansion} by
\begin{equation}
    \langle \mathrm{Tr} \phi^{2 n} \rangle = \frac{1}{\Omega^{2 n}} \langle \mathrm{Tr} \phi^{2 n} \rangle_{\tilde S} \,, \qquad \qquad \langle \mathcal{O} \rangle_{\tilde S} \equiv \frac{1}{\tilde Z} \int [dX^I] [d \psi_\alpha] \mathcal{O} e^{- \tilde S} \,.
\end{equation}
To compute $\langle \mathrm{Tr} \phi^{2 n} \rangle_{\tilde S}$ at small $\Omega$'s, we insert $\phi = r^3 - i r^{10} + \mathcal{O}(\Omega)$ in the previous derivation, thus obtaining 
\al{
\spl{
\langle \mathrm{Tr} \phi^{2 n} \rangle_{\Omega \to 0} & = \left(\frac{1}{\Omega}\right)^{2 n} \frac{\int [dr^3] [dr^{10}] (r^3 - i r^{10})^{2 n} e^{- \frac{3}{4^3} r_a^3 r_a^3 - \frac{1}{4^3} r_a^{10} r_a^{10}}}{\int [dr^3] [dr^{10}] e^{- \frac{3}{4^3} r_a^3 r_a^3 - \frac{1}{4^3} r_a^{10} r_a^{10}}} \\ & = N \left( \frac{-2^7}{3 \Omega^2} \right)^n \frac{\Gamma(n+ \frac{1}{2})}{\sqrt{\pi}} \,.\label{eq:DirectSmallCorrelators}
}
}

Both \eqref{eq:DirectMasslessLimit} and \eqref{eq:DirectSmallCorrelators} agree exactly with the localization results when $\Omega \to 0$ \eqref{eq:Localization_massless}, \eqref{eq:LocalizationResultMasslessCorrelators}.
Note that the same reasoning for $Z$ can be applied to the pure IKKT partition function deep in the commuting modes valley once removing the Gaussian deformation ($S_0 \to 0$), setting $\Omega \to 1$,
and replacing the expansion in $\Omega$ by an expansion in $|r_{a b}| \gg 1$. The equation \eqref{eq:DirectMasslessLimit} would then reduce to an undetermined expression $\int [dr^I] [d \theta_\alpha] 1 = \infty \cdot 0$. In fact, a careful treatment of obtaining an effective action for large commuting modes separation $|r_{a b}| \gg 1$ has been carried in \cite{Aoki_1998}, and the integral over the fermions $\theta_\alpha$ in this case would reduce to a power-law $1/r^\#$ suppressed enough to ensure the convergence of the IKKT partition function. We see explicitly in \eqref{eq:DirectMasslessLimit} that this power-law decay is subleading, with the leading piece being due to the fermion mass term present in polarized IKKT. This leading piece is the cause of the divergence in the $\Omega \to 0$ limit of polarized IKKT.\footnote{A similar analysis can be found in another attempt to deform IKKT in 4d \cite[Sec.4.1]{Austing:2001ib}.} This discussion is consistent with appendix \ref{app:fermion_masses_ConvergenceIKKT} where we see the divergence coming from the Pfaffian mass terms, the original IKKT Pfaffian being subleading.

\subsection{The free limit $\Omega \to \infty$}
\label{app:free_limit}
\paragraph{From the localization result: } Rescaling $m_{si} \to m_{si}/ \Omega^2$ and taking $\Omega \to \infty$ in the localization result \eqref{eq:LocalizationResult}, we obtain

\begin{equation}
    Z_\mathcal{R}^{(\Omega \to \infty)} = \mathcal{N}_\mathcal{R} \left( \frac{1}{\Omega^2} \right)^{\sum_s n_s^2} e^{\frac{9 \Omega^4}{2^{15}} \sum_s n_s (N_s^3 - N_s)} \int \prod_s \prod_{i = 1}^{n_s} \Delta(m) \exp\left(- \frac{3}{2^7} \sum_s \sum_{i=1}^{n_s} N_s m_{s i}^2\right) \,,
\end{equation}
where
\begin{equation}
    \Delta(m) = \left(\prod_s \prod_{i \neq j} (m_{s i} - m_{s j})^2 \right)^{1/2} \,,
\end{equation}
\begin{equation}
    \mathcal{N}_\mathcal{R}=  \frac{ (2 \pi)^
    {5 N^2 +  N/2 - \sum_s n_s/2}}{\prod_{k=1}^{N-1} k! \prod_s n_s !}  \left( \frac{64}{3} \right)^{\sum_s n_s^2}\prod_{s,t} \prod_{\substack{J = |j_s - j_t|\\ J \neq 0}}^{j_s + j_t} \left[ \frac{\left[ (2 + 3 J)^2  \right]^3 \left[(3 J)^2  \right]}{\left[(1 + 3 J)^2 \right]^3 \left[  (3 + 3 J)^2  \right]} \right]^{n_s n_t} \,. \label{eq:largeOmegaConstant}
\end{equation} 
This can be computed for any saddle $\mathcal{R}$, since we can write it as a matrix integral. We obtain
\al{
\spl{
Z_\mathcal{R}^{(\Omega \to \infty)} & = \frac{(2 \pi)^{5 N^2 + N/2}}{\prod_{k=1}^{N - 1} k!} \prod_s \left[\prod_{k=1}^{n_s - 1} (k!) \left( \frac{1}{\sqrt{2 \pi}} \right)^{\sum_s n_s} \left( \frac{2^9}{3 \sqrt{3} \sqrt{N} \Omega^2} \right)^{\sum_s n_s^2} \right] \\ & \qquad \qquad \cdot \prod_{s,t} \prod_{J = |N_s - N_t|/2}^{(N_s + N_t - 2)/2} \left[ \frac{\left[ (2 + 3 J)^2  \right]^3 \left[(3 J)^2  \right]}{\left[(1 + 3 J)^2 \right]^3 \left[  (3 + 3 J)^2  \right]} \right]^{n_s n_t} e^{\frac{9 \Omega^4}{2^{15}} \sum_s n_s (N_s^3 - N_s)} \,.
}
}

In particular we obtain that
\begin{equation}
    Z^{(\Omega \to \infty)}_{U(N), \mathrm{trivial}} = \left( \frac{2^{14} \pi^5}{3 \sqrt{3} \Omega^2} \right)^{N^2} \,, \label{eq:trivial_infiniteMass}
\end{equation}
\begin{equation}
    Z_{U(N),\text{irrep}} = \frac{(2 \pi)^{5 N^2 + (N-1)/2}}{\prod_{k=1}^{N-1} k!} \frac{2^9}{3 \sqrt{3} N \sqrt{N} \Omega^2} \prod_{J=1}^{N-1} \left( \frac{3J + 2}{3 J + 1} \right)^3 e^{\frac{9 \Omega^4}{2^{15}} (N^3-N)} \,, \label{eq:ZUN_largeOmegaIrrep}
\end{equation}
\begin{equation}
Z_{SU(N), \mathrm{irrep}} =  \frac{(2 \pi)^{(10 N + 11)(N-1)/2}}{N \sqrt{N} \prod_{k = 1}^{N-1}(k!)} \prod_{J=1}^{N-1} \left( \frac{3J+2}{3J+1} \right)^3 e^{\frac{9 \Omega^4}{2^{15}} (N^3-N)} \, . \label{eq:Z_largeOmegaIrrep}
\end{equation}
Note that we wrote the irreducible result for the $SU(N)$ partition function using \eqref{eq:relationToTraceless}. We also dropped the $(\Omega \to \infty)$ superscript from $Z_{U(N),\mathrm{irrep}} $
because this is the exact formula for any $\Omega$ (in this case $Z_\mathrm{1-loop} = 1$).

The correlators $\langle \mathrm{Tr} \phi^{2n} \rangle$ can also be easily computed since, after rescaling $M \to \frac{1}{\Omega^2} M$, the dominant term in \eqref{eq:correlators} comes from the irreducible fuzzy sphere $\frac{3 \Omega}{8} L_i$. We thus obtain
\begin{equation}
    \langle \mathrm{Tr} \phi^{2n} \rangle_{\Omega \to \infty} = \langle \mathrm{Tr} \phi^{2n} \rangle_{SU(N), \Omega \to \infty} = \left( \frac{3 \Omega}{8}\right)^{2 n} \mathrm{Tr} L_3^{2 n} = \left( \frac{3 \Omega}{8}\right)^{2 n} \sum_{m = - \frac{N-1}{2}}^{\frac{N-1}{2}} m^{2 n} \,, \label{eq:CorrelatorsLargeOmega}
\end{equation}
where the equality between $SU(N)$ and $U(N)$ correlators follows from using \eqref{eq:CorrelatorsToTraceless} at large $\Omega$.

\paragraph{From direct methods: } The limit $\Omega \to \infty$ around the trivial vacuum can be directly obtained from looking at the mass-deformation of the polarized IKKT action. In view of \eqref{eq:ToStrongCouplingParameters}, we obtain
\al{
\spl{
Z_{\mathrm{trivial}}^{(\Omega \to \infty)} & = \left(\frac{1}{\Omega^2}\right)^{N^2} \int [d X^I] [d \psi_\alpha] e^{- \mathrm{Tr}\left(\frac{3}{4^3} X_i X_i +\frac{1}{4^3} X_p X_p - \frac{1}{8} \bar \psi \Gamma^{1 2 3} \psi \right)} = \left( \frac{2^{14} \pi^5}{3 \sqrt{3} \Omega^2} \right)^{N^2} \,,
}
}
which agrees with \eqref{eq:trivial_infiniteMass}.

The limit $\Omega \to \infty$ around the irreducible vacuum was carefully treated in \cite{Hartnoll:2024csr}. Their result is the same as \eqref{eq:Z_largeOmegaIrrep} once changing the measure convention from ours ($\mathrm{Tr} T^{A} T^B = \delta^{A B}$) to theirs ($\mathrm{Tr} T^A T^B = 2 \delta^{A B}$).

\subsection{Recovering IKKT}
\label{sec:RecoverIKKT}
In this section, we give evidence for the conjecture that the IKKT partition function can be recovered from \eqref{eq:PartitionFunction} by not only taking $\Omega \to 0$, but also closing the contour of the moduli integrals on the upper complex plane, thus picking poles and making the result finite as $\Omega \to 0$. Let us first recall that
\begin{equation}
    \lim_{\Omega \to 0} Z_{SU(N)} = \sum_\mathcal{R} \lim_{\Omega \to 0} \frac{Z_\mathcal{R}}{Z_{U(1)}} \,,
\end{equation}
\begin{equation}
     \lim_{\Omega \to 0} Z_\mathcal{R} =  \lim_{\Omega \to 0} C_\mathcal{R} \int \prod_{s, i} dm_{s i} Z_\mathrm{1-loop} e^{- \frac{3 \Omega^4}{2^7} \sum_{s i} N_s m_{s i}^2} \,.
\end{equation}
We will give a prescription to make the ratio $Z_\mathcal{R}/ Z_{U(1)}$ finite, such that
\begin{equation}
    Z_\mathcal{R}^\mathrm{IKKT} \equiv \lim_{\Omega \to 0} \left(\frac{Z_\mathcal{R}}{Z_{U(1)}}\right)_\mathrm{regulated} < \infty \,.
\end{equation}
The recipe we provide to define $Z_\mathcal{R}^\mathrm{IKKT}$ is the following.
\begin{enumerate}
    \item Parameterize the $(\sum_s n_s)$-dimensional integral $Z_\mathcal{R}$ in terms of its trace $m_\mathrm{com} \equiv \sum_{s,i} m_{s i}$ and in terms of $(\sum_s n_s - 1)$ relative variables $\D_{i j} \equiv m_{s i} - m_{t j}$ (for some $s,t$) such that all the  variables $\{ \D_{i j} \}$ fully determine all differences $m_{s i} - m_{t j}$.
    \item Integrate out the trace $m_\mathrm{com} = \sum_{s i} m_{s i}$ which does not appear in $Z_\mathrm{1-loop}$. This trace will simplify with $Z_{U(1)}$ up to a constant.
    \item The remaining $(\sum_s n_s - 1)$-dimensional integral over $\{ \Delta_{i j} \}$ is divergent when $\Omega \to 0$ if integrated on the real axis. Here is the prescription: \textbf{Close the contour on the upper-half plane} as shown in figure \ref{fig:ContourIKKT}. The integral over $\Delta_{i j}$ will pick contributions from poles at constant positions, but also from poles due to the other variables.
    \item Iterate the pole-picking procedure up to having exhausted all integrations. 
\end{enumerate}
We conjecture that
\begin{equation}
    Z_\mathrm{IKKT} = \sum_\mathcal{R} Z^\mathrm{IKKT}_\mathcal{R} \, .
\end{equation}
In the following, we prove it for $N=2$ and $N=3$.

\subsubsection{IKKT for $N=2$}
Let us consider the toy example $N=2$. After removing the center of mass moduli integral, we're left with \eqref{eq:ZSU2} which we repeat here for convenience
\begin{equation}
    Z_{SU(2)} = \frac{2^{16} \pi^{29/2}}{3} \int_{- \infty}^{\infty} dx \, \left( \frac{(4+x^2)^3 x^2}{(1+x^2)^3 (9+x^2)} \right) e^{-\frac{3 \Omega^4}{2^{14}} x^2 } + 2^8 5^3 \pi^{31/2} e^{\frac{27}{2^{14}} \Omega^4} \,.
\end{equation}
Setting $\Omega \to 0$ and closing the $x$ integral on the upper complex plane, we pick the residues at $x_1=i$ and $x_2=3i$, giving
\al{
\spl{
Z_\mathrm{trivial}^\mathrm{IKKT} = \frac{2^{16} \pi^{29/2}}{3} 2 \pi i \sum_{m=1}^2 \mathrm{Res} \left( \frac{(4+x^2)^3 x^2}{(1+x^2)^3 (9+x^2)}, x= x_m \right) = 2^8 35 \pi^{31/2} \,,
}
}
\begin{equation}
Z_\mathrm{irrep}^\mathrm{IKKT} = 2^8 5^3 \pi^{31/2} \,.
\end{equation}
They indeed sum to the $N=2$ IKKT partition function \eqref{eq:IKKTZresult} 
\begin{equation}
    Z_\mathrm{trivial}^\mathrm{IKKT}+ Z_\mathrm{irrep}^\mathrm{IKKT} = 2^{13} 5 \pi^{31/2} = Z_\mathrm{IKKT} \,.
\end{equation}
\subsubsection{IKKT for $N=3$} 
The number 3 has three partitions, $3 = 1 + 1 + 1$, the trivial saddle, which we will denote by $\mathcal{R} = (1,1,1)$, $3 = 3$, the irreducible saddle which we denote by $\mathcal{R} = (3)$ and $3 = 1 + 2$, corresponding to $n_s = 1, N_s = 1$, $n_t = 1$, $N_t = 2$, which we will denote by $\mathcal{R} = (1,2)$. Let us start by computing the trivial saddle contribution $Z_{(1,1,1)}^\mathrm{IKKT}$. The first step is performed by changing the $m_1, m_2, m_3$ variables to $m_\mathrm{com} \equiv m_1 + m_2 + m_3$ and $\Delta_{1 2} \equiv m_1 - m_2$, $\Delta_{1 3} \equiv m_1 - m_3$. Note that $m_2 - m_3 = \Delta_{1 3} - \Delta_{1 2}$ is expressed in terms of them.
\al{
\spl{
Z_{(1,1,1)} & = C_{(1,1,1)} \frac{1}{3} \int d \Delta_{1 2} d \Delta_{1 3} d m_\mathrm{com} Z_\mathrm{1-loop} \\ & \qquad \qquad \qquad \qquad \exp \left(-\frac{3 \Omega^4}{2^7} \frac{2}{3} (\Delta_{1 2}^2 + \Delta_{1 3}^2 - \Delta_{12} \Delta_{13}) - \frac{\Omega^4}{2^7} m_\mathrm{com}^2\right) \,.
}
}
Integrating out $m_\mathrm{com}$, dividing by $Z_{U(1)}$ and rescaling $\Delta_{i j} \to \Delta_{i j} / 8$, we obtain
\al{
\spl{
Z_{(1,1,1)}^\mathrm{IKKT} = C_{(1,1,1)} \frac{1}{2^{16} \pi^{9/2}} \sqrt{\frac{3}{2}} \oint_C d \Delta_{1 2} d \Delta_{1 3} A(\Delta_{12}, \Delta_{13}) \,,
}
}
where
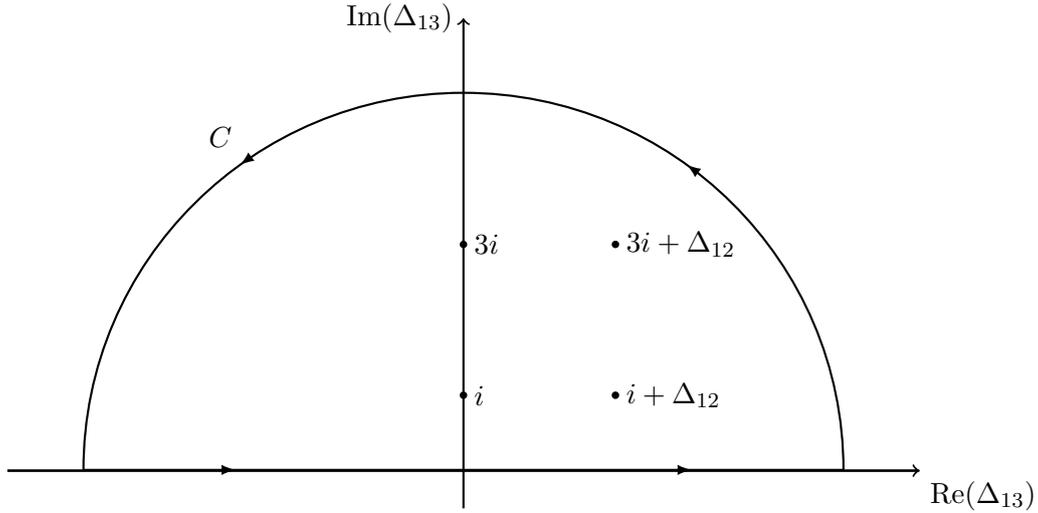
\begin{figure}
\centering
\begin{tikzpicture}
    \draw[->, thick] (-6, 0) -- (6, 0) node[below right] {\(\operatorname{Re}(\Delta_{13})\)};
    \draw[->] (0, -0.5) -- (0, 6) node[left] {\(\operatorname{Im}(\Delta_{13})\)};
    \filldraw[black] (0, 1) circle (1pt) node[right] {\(i\)};
    \filldraw[black] (0, 3) circle (1pt) node[right] {\(3i\)};
    \filldraw[black] (2, 1) circle (1pt) node[right] {\(i + \Delta_{12}\)};
    \filldraw[black] (2, 3) circle (1pt) node[right] {\(3i + \Delta_{12}\)};
    \draw[thick] (-6, 0) -- (6, 0);
    \draw[thick,
    decoration={ markings,  
      mark=at position 0.3 with {\arrow{latex}}, 
      mark=at position 0.7 with {\arrow{latex}}}, 
      postaction={decorate}]
  (5, 0.01) arc[start angle=0, end angle=180, radius=5] node[label={[yshift=4cm,xshift=1.8cm,black]above: $C$}] {};
    \draw[thick, decoration={ markings,  
      mark=at position 0.2 with {\arrow{latex}}, 
      mark=at position 0.8 with {\arrow{latex}}}, 
      postaction={decorate}] (-5, 0.01) -- (5, 0.01) ;
\end{tikzpicture}
\caption{Contour integral displayed for the variable $\Delta_{13}$ in the trivial saddle of $N=3$. The original polarized IKKT integral was taken on the real axis, and we propose to close it with a semi-circular arc in the upper half plane to regulate the divergence as $\Omega \to 0$.}
\label{fig:ContourIKKT}
\end{figure}
\begin{equation}
    A(x,y) = \frac{(4+x^2)^3 x^2}{(1+x^2)^3 (9+x^2)} \frac{(4+y^2)^3 y^2}{(1+y^2)^3 (9+y^2)} \frac{(4+(x-y)^2)^3 (x-y)^2}{(1+(x-y)^2)^3 (9+(x-y)^2)} \,.
\end{equation}
Let us consider the integral over $\Delta_{1 3}$ first. Our contour prescription, illustrated in figure \ref{fig:ContourIKKT}, is such that it encircle the poles at $\Delta_{1 3}^{(1)} = i, \Delta_{1 3}^{(2)}=3i, \Delta_{1 3}^{(3)}= \Delta_{1 2} + i$ and $\Delta_{1 3}^{(4)} = \Delta_{1 2} + 3i$, yielding
\al{
\spl{
B(\Delta_{12}) \equiv \oint_C d \Delta_{1 3} & A(\Delta_{1 2}, \Delta_{13}) = \sum_{m = 1}^4 2 \pi i \mathrm{Res} (A(\Delta_{12}, \Delta_{13}), \Delta_{13} = \Delta_{13}^{(m)}) \\ & = \frac{\pi}{128} \frac{P(\Delta_{1 2})}{(\Delta_{1 2}^2 + 1)^3 (\Delta_{1 2}^2 + 4)^2 (\Delta_{1 2}^2 + 9) (\Delta_{1 2}^2 + 16)^3 (\Delta_{1 2}^2 + 36)} \,,\label{eq:SU3-1loopIntegral}
}
}
\al{
\spl{
P(x) & \equiv 105 x^{20} + 10920 x^{18} + 457170 x^{16} + 10423560 x^{14} + 142736685 x^{12} + 1175466936 x^{10} \\ & \qquad + 5477149824 x^8 + 12450112896 x^6 + 11571604224 x^4 + 22451613696 x^2 \,.
}
}
We now iterate by integrating $B(\Delta_{1 2})$ over $\Delta_{1 2}$ whose contour encircles the poles at $\Delta_{1 2}^{(1)} = i$,$\Delta_{1 2}^{(2)} = 2i$,$\Delta_{1 2}^{(3)} = 3i$,$\Delta_{1 2}^{(4)} = 4i$ and $\Delta_{1 2}^{(5)} = 6i$. We thus obtain
\begin{equation}
    Z_{(1,1,1)}^\mathrm{IKKT} = C_{(1,1,1)} \frac{1}{2^{16} \pi^{9/2}} \sqrt{\frac{3}{2}} 2 \pi i \sum_{m=1}^5 \mathrm{Res}(B(\Delta_{1 2}), \Delta_{1 2} = \Delta_{1 2}^{(m)}) = \frac{2^{39} \cdot 83 \cdot 241 \pi^{41}}{3^2 5^2 7^3 \sqrt{3}} \,.
\end{equation}

The only other vacuum which is non-trivial to compute is $(1,2)$. Integrating out the center of mass and rescaling variables we obtain
\begin{equation}
    Z_{(1,2)}^\mathrm{IKKT} = C_{(1,2)} \frac{2^{35} 5^3 \pi^{40}}{3 \sqrt{3}} \oint_C d x \frac{(\frac{49}{3}+x^2)^3 (\frac{9}{4}+x^2)}{(\frac{25}{4}+x^2)^3 (\frac{81}{4}+x^2)} \,.
\end{equation}
The contour integral picks the poles at $x = \frac{5}{2} i$ and $x = \frac{9}{4} i$ thus giving
\begin{equation}
    Z_{(1,2)}^\mathrm{IKKT} = \frac{2^{39} 499 \pi^{41}}{3 \cdot 5^2 7^3 \sqrt{3}} \,.
\end{equation}

Finally, $Z_{(3)}^\mathrm{IKKT}$ trivially follows from \eqref{eq:Z_largeOmegaIrrep},
\begin{equation}
    Z_{(3)}^\mathrm{IKKT} = \frac{2^{43} 5^3 \pi^{41}}{3 \cdot 7^3 \sqrt{3}} \,.
\end{equation}
The sum precisely agrees with the $N=3$ IKKT partition function \eqref{eq:IKKTZresult}
\begin{equation}
    Z_{(1,1,1)}^\mathrm{IKKT} + Z_{(1,2)}^\mathrm{IKKT} + Z_{(3)}^\mathrm{IKKT} = \frac{2^{41}5 \pi^{41}}{9 \sqrt{3}} = Z_\mathrm{IKKT} \,.
\end{equation}

\section{Gravity from the matrix model}
\label{sec:gravity from matrix model}
In this section we will show that, in the large $N$ limit, the localization equations are equivalent to the electrostatic equations defining the gravity dual described in \cite{Komatsu:2024bop}. A similar analysis was carried out in \cite{Asano:2014vba,Asano:2014eca} for the case of the BMN model, 
whose vacua are dual to Lin-Maldacena geometries \cite{Lin:2005nh}.

\subsection{The electrostatic problem}
The dual geometry is determined by the  4-dimensional electrostatic potential of a configuration of conducting 3d balls and a background potential \cite{Komatsu:2024bop}, see the right of figure \ref{fig:vacua}. We will write an equation for the charge density of those conductors.

\begin{figure}[h]
    \centering
\begin{tikzpicture}
\draw (2,2) circle (3cm);
\draw[dashed,fill=blue!20] (3.5,2) ellipse (0.15cm and 2.6cm);
\draw[->] (2,2)--(2.75,2) node[above] {$u$}--(3.5,2);
\draw[->] (3.8,2)--(3.8,3) node[right] {$ y$};
\draw[->] (2,2)--(2.75,3.3) node[left]{$R$}--(3.5,4.6);
\end{tikzpicture}
    \caption{Slicing of a 3d ball into 2d disks (blue). The center of the disk is at a distance $u$ from the center of the ball, and $y$ is a radial coordinate on the disk. The linear charge density $f(u)$ measures the charge on the blue slice.}
    \label{fig:slice_ball}
\end{figure}
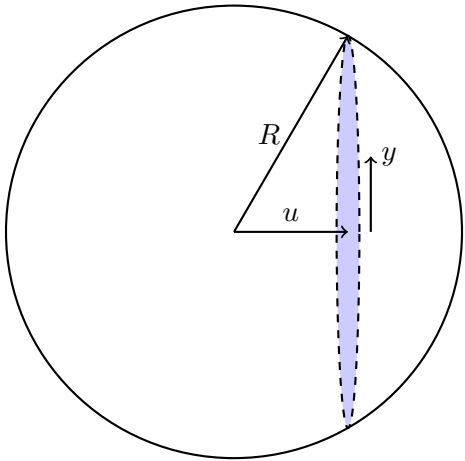

Consider a 3d ball  of radius $R$  with a spherically symmetric charge density $\sigma(r)$, with $0<r<R$.
Let us slice this ball
\footnote{At this point it might be surprising to consider the charge density of the slices instead of the one of the ball itself. In \cite{Asano:2014vba,Asano:2014eca}, it was the charge density of such slices that reproduced the density of eigenvalues of the matrix side.} in thin disks, and compute the linear charge density $f(u)$ of such a slice, at a distance $u$ from the center of the ball, measured on the equatorial plane, see figure \ref{fig:slice_ball}. 
We obtain 
\begin{equation}
    f(u) =2\pi \int_0^{\sqrt{R^2-u^2}} dy \,y \,\sigma\left(r= \sqrt{u^2+y^2} \right) = \pi \int_{u^2}^{R^2} dz \sigma(\sqrt{z}),
\end{equation}
where $y$ is a coordinate perpendicular to the equatorial plane, and the second equality comes from changing $y$ to the variable $z=u^2+y^2$. Taking a derivative with respect to $u$ we get 
\begin{equation}
    f'(u) = -2\pi u \sigma(u).
\end{equation}
Since this is the charge of a slice, we can simply compute the total charge of the ball by integrating over slices\footnote{Alternatively one can compute \begin{equation}
    Q = 4\pi \int_0^R  du u^2\sigma(u) = -2 \int_0^R  du u f'(u) = -\int_{-R}^R  du uf'(u) = \int_{-R}^R  du f(u).
\end{equation}}
\begin{equation}
    Q= \int_{-R}^R du f(u),
\end{equation}
where we used $f(-u)=f(u)$.

The 4d electrostatic potential due to a ball of radius $R_s$, placed at position $z_s$ and with charge density $\sigma_s(r)$, reads\footnote{This can be obtained simply by integrating the 4d electrostatic potential $V(x)$ of a single charge at position $y$ ($V(x)=\frac{1}{4\pi^2}\frac{1}{|x-y|^2}$) over a ball with charge density $\sigma_s$.}
\begin{equation}
V_{\mathrm{ball},s}(r,z) = \frac{1}{4\pi r} \int_0^{R_s} du \, u \sigma_s(u) \operatorname{log}\left(1+ \frac{4 r u}{(r-u)^2+(z-z_s)^2} \right).
\end{equation}
Writing it in terms of $f(u)$ and integrating by parts we obtain 
\begin{equation}
V_{\mathrm{ball},s}(r,z) = \frac{1}{8\pi^2 r} \int_{-R_s}^{R_s} du f_s(u) \left( \frac{r + u}{(r+u)^2 + (z-z_s)^2}+\frac{r - u}{(r-u)^2 + (z-z_s)^2} \right)\,.
\end{equation}
Using $f(u) = f(-u)$, we can write 
\begin{equation}
V_{\mathrm{ball},s}(r,z) = \frac{1}{4\pi^2 r} \int_{-R_s}^{R_s} du f_s(u)   \frac{r - u}{(r-u)^2 + (z-z_s)^2}.
\end{equation}
One may worry about evaluating this function at $z=z_s$ when $r < R_s$. In this case, we should take the limit\footnote{This is equivalent to taking the Cauchy principal value of the integral with $\epsilon=0$.}
\begin{equation}
    V_{\mathrm{ball},s}(r,z_s) = \frac{1}{4\pi^2 r} \lim_{\epsilon\to 0} \int_{-R_s}^{R_s} du f_s(u)   \frac{r-u}{(r-u)^2 + \epsilon^2}.
\end{equation}
The total potential $V(r,z)$ is constant on the conducting ball $s$ and takes the value $V(r,z_s) = V_s$ when $r < R_s$. Thus
\begin{equation}
    V_s = V_b\left(r^2 z_s-z_s^3 \right) + \sum_{t+ \mathrm{images}} V_{\mathrm{ball},t}(r,z_s),
\end{equation}
where the sum runs over all the balls, including images\footnote{By inclusion of images, we mean that for each $t$, we also sum over $\tilde t$ where $f_{\tilde t} = - f_t$ and $z_{\tilde t} = - z_t$. This is the method of image charges in electrostatics, which ensures that $V=0$ at $z=0$.}, and the first term is the background potential. 
Multiplying by $r$ and taking a derivative with respect to $r$, we obtain 
\begin{equation}
    - V_s + V_b(3 r^2 z_s - z_s^3) + \sum_{t + \mathrm{images}} \frac{1}{(2\pi)^2} \mathcal{R} \hspace{-.4cm}\int_{- R_t}^{R_t} dr' \frac{(z_s - z_t)^2 - (r-r')^2}{((z_s - z_t)^2 + (r-r')^2)^2} f_t(r') = 0 \label{eq:LinearCharge}
\end{equation}
where we changed  integration variable $u \to r'$ for later convenience and defined 
\begin{equation}
    \mathcal{R} \hspace{-.4cm}\int dy \frac{1}{y^2} f(y) \equiv \lim_{\epsilon \to 0}   \int dy \frac{y^2-\epsilon^2}{(y^2+\epsilon^2)^2} f(y) =\lim_{\epsilon \to 0} \left[ \int_{|y|>\epsilon} \frac{dy}{y^2} f(y) - \frac{2}{\epsilon} f(0) \right]\,,\label{eq:regulator}
\end{equation}
which regulates the integral over the ball $s$.
The linear charge densities $f_s(r)$, the size of the balls $R_s$ and the potentials $V_s$ are determined by \eqref{eq:LinearCharge} and by the conditions
\begin{align}
    Q_s = \int_{-R_s}^{R_s} dr f_s(r)\,,\qquad\qquad 
    f_s(R_s)=f_s'(R_s)=0\,. \label{eq:ConditionsLinearDensity}
\end{align}

The integral equations \eqref{eq:LinearCharge} and \eqref{eq:ConditionsLinearDensity} are equivalent to the Laplace equation with boundary conditions set by the conducting balls. Hence those equations determine the 10 dimensional Euclidean geometry of the gravity dual \cite{Komatsu:2024bop}. In the next subsection we will show that the large $N$ limit of the localization result gives exactly the same equations.


\subsection{The particle density}
We now study the localization result \eqref{eq:PartitionFunction} in the large $n_s$  limit in which the particles form a continuum which can be described by continuous densities.

The localization result  \eqref{eq:PartitionFunction} can be written as 
\begin{equation}
    Z_\mathcal{R} = C_\mathcal{R} \int \prod_{s, i} dm_{s i} e^{- S_\mathrm{eff}} \,,
\end{equation}
where 
\begin{equation}
    S_\mathrm{eff} = - \frac{9 \Omega^4}{2^{15}} \sum_s n_s (N_s^3 - N_s) + \frac{3 \Omega^4}{128} \sum_s N_s \sum_i m_{s i}^2 + 
    \sum_{(si,tj)}
g_{s t}(m_{s i}- m_{t j})\,,
\end{equation}
with 
\al{
\spl{
 g_{s t}(x) = \sum_{J = |N_s - N_t|/2}^{(N_s + N_t -2)/2} h_J(x)\,,
}
}
and
\al{
\spl{
 h_J(x)=- \log \frac{\left[ (2 + 3 J)^2 + (8 x)^2 \right]^3 \left[ (3 J)^2 + (8 x)^2  \right]}{\left[(1 + 3 J)^2 + (8 x)^2 \right]^3 \left[  (3 + 3 J)^2 + (8 x)^2  \right]}   \,.
}
}
It is useful to think of this as a statistical mechanics system with  $n_s$ particles of charge $N_s$ in a 1D harmonic background potential and with two-body interactions described by $g_{st}(m_{si}-m_{tj})$.
The two-body term depends on the charges $N_s$ and $N_t$ of the two particles and on the distance between them.

\begin{figure}[]
    \centering
\includegraphics[scale=0.7]{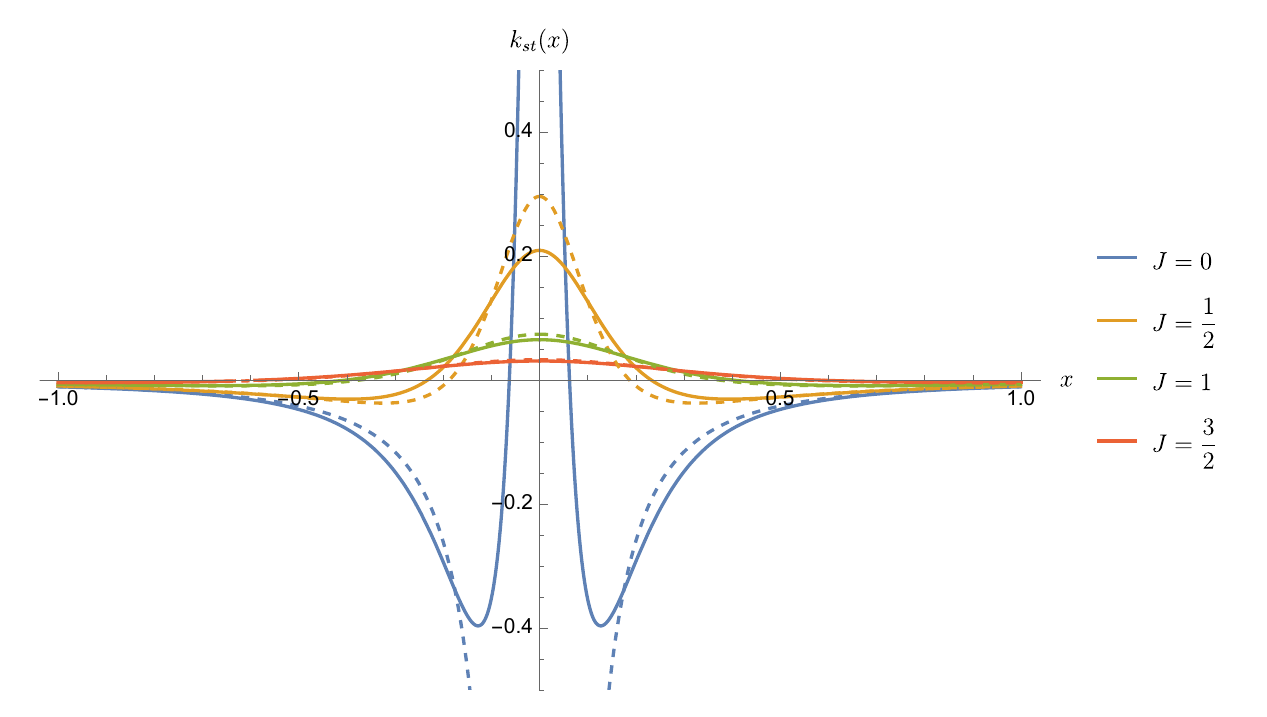}
    \caption{The two-body interaction potential $k_{st}(x)$ for $J=0,\frac{1}{2},1,\frac{3}{2}$ given in \eqref{exactkst}. The dashed lines are the approximation in \eqref{approxkst}. }
    \label{fig:2bodypotential}
\end{figure}

In order to connect with the electrostatic problem described in the previous subsection, 
it is convenient to write 
\begin{align}
  g_{s t}(x)  = k_{st}(x) +
 k_{s\tilde{t}}(x)
 \,,\qquad    
\end{align}
with
\begin{align}
    k_{st}(x)=\sum_{J=|N_s - N_t|/2}^{\infty} h_J(x)\,,
 \qquad \qquad
 k_{s\tilde{t}}(x)=-\sum_{J=|N_s + N_t|/2}^{\infty} h_J(x)\,.
\end{align}
As we shall see, the term $k_{s\tilde{t}}$ is similar to the interaction between the ball $s$ and the image ball $\tilde{t}$ in the electrostatic problem.
We can then write
\begin{equation}
    S_\mathrm{eff} = - \frac{9 \Omega^4}{2^{15}} \sum_s n_s (N_s^3 - N_s) + \frac{3 \Omega^4}{128} \sum_s N_s \sum_i m_{s i}^2 + \sum_{(si,tj)\atop +\text{images}} 
k_{s t}(m_{s i}- m_{t j})\,,
\end{equation}
where for each pair $(si,tj)$ we also sum over its image $(si,\tilde{t}j)$.
The potential between particles can be written in closed form,
\begin{align}
    k_{st}(x)&= 3 \log \frac{\Gamma\left(\frac{2+3J+8ix}{3}\right)\Gamma\left(\frac{2+3J-8ix}{3}\right)}{\Gamma\left(\frac{1+3J+8ix}{3}\right)\Gamma\left(\frac{1+3J-8ix}{3}\right)} -\log\left(J^2+\frac{(8x)^2}{3^2}\right)
    \label{exactkst}
    \\
    &=\frac{1}{3}\left(\frac{1}{(3J+8ix)^2}+
    \frac{1}{(3J-8ix)^2}\right) + O\left(\frac{1}{(3J\pm8ix)^4}\right)\,,
    \label{approxkst}
\end{align}
with 
$J=|N_s - N_t|/2$.
In this formulation, each particle $(si)$ interacts with every other particle $(tj)$ and its  image $(\tilde{t} j)$.
From the plots of figure \ref{fig:2bodypotential}, we observe that the interaction is repulsive at short distances and attractive at large distances.
This is not surprising because  the basic building block $h_J(x)$ shows a similar behavior (in fact, it obeys $\int^\infty_{-\infty} dxh_J(x)=0 $ for any $J\ge 0$).
Particles with the same charge have a short distance repulsion of infinite 
strength. At long distances, the two-body potentials decay to zero as the inverse of the fourth power of distance.

Let us now introduce the eigenvalue densities
\begin{equation}
    \rho^{(s)} (x) = \sum_{i=1}^{n_s} \delta(x - m_{s i})\,.
\end{equation}
In the limit of large $n_s$, we expect that we can replace this exact eigenvalue density by a smooth function. We shall comment in more detail later on the validity of this approximation.
This allows us to write
\begin{align}
    S_\mathrm{eff} =& - \frac{9 \Omega^4}{2^{15}} \sum_s n_s (N_s^3 - N_s)  + \frac{3 \Omega^4}{2^7} \sum_s N_s \int dx x^2\rho^{(s)}(x) 
   \label{eq:Effective_Action} 
    \\&+ \frac{1}{2}\sum_{(s,t)\atop +\text{images}} 
    \int dx dy \rho^{(s)}(x) \rho^{(t)}(y)
k_{s t}(x- y)
- \sum_s \mu_s \left( \int dx \rho^{(s)}(x) - n_s \right)\,, \nonumber
\end{align}
where we introduce $\mu_s$ as a Lagrange multiplier to enforce the constraint $\int dx \rho^{(s)}(x) = n_s$.
The saddle point equation for $\rho^{(s)}(x)$ is then
\begin{equation}
    \frac{3 \Omega^4}{2^7} N_s x^2 + \sum_{t+\text{images}}  \int dy k_{s t}(x-y) \rho^{(t)}(y) - \mu_s = 0\,.
\end{equation}
Finally, we approximate the exact two-body potential $k_{st}(x)$ by its large $J \sim x$ behavior given in \eqref{approxkst},
\begin{equation}
    \frac{3 \Omega^4}{2^7} N_s x^2 - \mu_s + \frac{2}{3} \sum_{t + \mathrm{images}} \mathcal{R} \hspace{-.4cm}\int_{-x^{(t)}}^{x^{(t)}} dy \frac{\left(\frac{3}{2} |N_s - N_t|\right)^2 - (8 (x-y))^2}{\left(\left(\frac{3}{2} |N_s - N_t|\right)^2 + (8 (x-y))^2\right)^2} \rho^{(t)}(y) = 0 \, , \label{eq:SaddlePointEigenvalues}
\end{equation}
where we defined images $\tilde t$ for each $t$ as
\begin{equation}
    \rho^{(\tilde t)} (x) = - \rho^{(t)} (x)\,, \qquad \qquad N_{\tilde t} = - N_t\,.
\end{equation}
In addition, the term $t=s$ is defined by the limit $N_t \to N_s$ outside the integral. This is the meaning of the regulated integral symbol. Notice that this preserves the property $\int dx k_{ss}(x)=0$ of the exact two-body potential.
We also assumed that the densities $\rho^{(s)}(x)$ have support in an interval $[-x^{(s)},x^{(s)}]$. Therefore, we have
\begin{equation}
    n_s = \int_{-x^{(s)}}^{x^{(s)}} dx\ \rho^{(s)}(x) \, , \qquad \qquad \rho^{(s)}(x^{(s)}) = 0 \, . 
\end{equation}
Notice that the kernel in \eqref{eq:SaddlePointEigenvalues} can be written as a derivative of a simpler function that also decays at large distance, 
\begin{align}
    \frac{\left(\frac{3}{2} |N_s - N_t|\right)^2 - (8 (x-y))^2}{\left(\left(\frac{3}{2} |N_s - N_t|\right)^2 + (8 (x-y))^2\right)^2} = -
    \frac{d}{dy} \frac{   (x-y)}{\left(\frac{3}{2} |N_s - N_t|\right)^2 + (8 (x-y))^2 }\,.
\end{align}
Using this property and integrating by parts in \eqref{eq:SaddlePointEigenvalues} we find 
\begin{equation}
    \frac{3 \Omega^4}{2^7} N_s x^2 - \mu_s + \frac{2}{3} \sum_{t + \mathrm{images}} \mathcal{R} \hspace{-.4cm}\int_{-x^{(t)}}^{x^{(t)}} dy \frac{   (x-y)}{\left(\frac{3}{2} |N_s - N_t|\right)^2 + (8 (x-y))^2 }
   \frac{d\rho^{(t)}}{dy} (y) = 0 \, .
\end{equation}
The new kernel for $t=s$ is proportional to the principal value of $\frac{1}{x-y}$ which is the usual kernel in Gaussian matrix models. Therefore, we expect the behavior close to $x^{(s)}$ to be of square root type, 
\begin{align}
    \frac{d\rho^{(t)}}{dy} (y) \sim \sqrt{x^{(s)}-y}\qquad \Rightarrow \qquad
   \rho^{(t)}  (y) \sim \left(x^{(s)}-y\right)^\frac{3}{2}\,.
\end{align}
Thus, we conclude that
\begin{equation}
    n_s = \int_{-x^{(s)}}^{x^{(s)}} dx\ \rho^{(s)}(x) \, , \qquad \qquad \rho^{(s)}(\pm x^{(s)}) =\frac{d\rho^{(t)}}{dx}(\pm x^{(s)}) = 0 \, . \label{eq:ConstraintsEigenDensities}
\end{equation}


\paragraph{Validity of the approximations} We made two approximations to arrive at \eqref{eq:SaddlePointEigenvalues}. Firstly, we replaced discrete sums by integrals of continuous functions. Secondly, we replaced the two-particle potential $k_{st}(x)$ by its large $J\sim x$ behavior.
In order to estimate the error in the first approximation, it is convenient to write
\begin{align}
    \rho^{(s)}(x) = \frac{n_s}{x^{(s)}} b\left(\frac{x}{x^{(s)}}\right)
\end{align}
with $b(w)$ an order one function with support in the interval $[-1,1]$ and normalized by $\int_{-1}^1 dw b(w)=1$.
In appendix \ref{app:approxsollutions}, we study the electrostatic problem with just one ball (and its image). There we show that 
\begin{align}
    b(w)=\begin{cases*}
      \frac{15}{16}(1-w^2)^2 & if \  $x^{(s)} \gg N_s $ \\
       \frac{8}{3\pi^2}(1-w^2)^\frac{3}{2}     & if \   $x^{(s)} \ll N_s $ 
    \end{cases*}\,.
\end{align}
In addition, we found that 
\begin{align}
    \frac{x^{(s)}}{N_s}\sim \begin{cases*}
      \xi^\frac{1}{5} & if \  $\xi \gg 1 $ \\
       \xi^\frac{1}{4}    & if \   $\xi \ll 1 $ 
    \end{cases*}\,,\qquad\qquad
    \xi \equiv \frac{n_s}{\Omega^4 N_s^5}\,.
\end{align}
We can then estimate the order of magnitude of the first 3 terms in the effective action \eqref{eq:Effective_Action}, namely $\Omega^4 n_s N_s^3$, $\Omega^4 n_s N_s (x^{(s)})^2$ and
$n_s^2 /(x^{(s)})^2 $.
Notice that these 3 terms are of order $n_s^2/N_s^2$  if we keep $\xi$ fixed.

Doing a Taylor expansion of $\rho^{(t)}(y)$ around $x$ and referring to figure \ref{fig:2bodypotential}, the error introduced by approximating $k_{st}(x)$ can be estimated by
\begin{align} 
    \int dx dy \rho^{(s)}(x) \rho^{(t)}(y)
\left[ k_{s t}(x- y)-k_{s t}^{\text{approx}}(x- y)\right]
\sim \int dx  \rho^{(s)}(x) \frac{d^2}{dx^2}\rho^{(t)}(x) \sim \frac{n_s^2}{(x^{(s)})^3}\,.\nonumber
\end{align}
This is suppressed by a factor of $x^{(s)}$ relative to the third term in \eqref{eq:Effective_Action}. Therefore, our approximation is valid as long as $x^{(s)}\gg 1$. 
The error introduced by the approximation of a continuous density of particles goes to zero as long as the density $\rho^{(s)} \sim n_s/x^{(s)}$ goes to infinity. 
Finally, in order to neglect the entropy term\footnote{The entropy term comes from converting the measure $\int \prod_i d m_{s i} \to \int [d \rho^{(s)}(x)]$. For a review, refer to e.g. \cite[App.C]{Bourgine:2013ipa}.}
\begin{align}
    \sum_s \int dx\rho^{(s)}(x)\log \frac{\rho^{(s)}(x)}{n_s} \sim n_s \log x^{(s)}\,,
\end{align}
relative to the other terms in \eqref{eq:Effective_Action}, we need 
$n_s/ (x^{(s)})^2 \gg \log x^{(s)} \gg 1$. Putting everything together, we need
\begin{align}
    n_s \gg (x^{(s)})^2\log (x^{(s)}) \gg 1  \,.
\end{align}
This can be achieved in many ways. For example, we can keep $\xi$ fixed (and therefore $x^{(s)}/N_s$  is also fixed) and take $n_s \gg N_s^2 \log N_s$.
Alternatively, we can keep $N_s$ and $\Omega$ fixed and send $n_s \to \infty$.

\subsection{Matching with gravity}

The localization equations \eqref{eq:SaddlePointEigenvalues} and \eqref{eq:ConstraintsEigenDensities} are exactly the same as the electrostatic equations \eqref{eq:LinearCharge} and \eqref{eq:ConditionsLinearDensity} up to an identification of parameters. To find the correct identification we start by writing
\begin{equation}
    x,y = a r, a r' \,, \quad x^{(t)} = a R_t \,, \quad \rho^{(t)}(x = a r) = b f_t(r) \,,
\end{equation}
where we denote by $x,y$ the variables of the densities of eigenvalues and $r, r'$ the variables of the linear charge densities.
Then, matching the kernels in \eqref{eq:SaddlePointEigenvalues} and \eqref{eq:LinearCharge}, we find
\begin{equation}
    \frac{3}{16 a} N_s = z_s.
    \label{eq Ns z}
\end{equation}
We can now rescale one of the equations so that the terms with the integral are identical. Then we match the $r$ independent terms and the terms proportional to $r^2$, which gives 
\begin{equation}
\label{eq Vs}
    V_s+z_s^3 V_b = \frac{24 a}{b \pi^2} \mu_s, \qquad  3 V_b z_s = \frac{9 a^3 N_s \Omega^4}{16 b \pi^2}.
\end{equation}
Finally we match the equations for the total charge \eqref{eq:ConstraintsEigenDensities} and \eqref{eq:ConditionsLinearDensity} to get 
\begin{equation}
 \label{eq Qs ns}
    a b Q_s = n_s.
\end{equation}
As explained in \cite{Komatsu:2024bop}, quantizing the different fluxes in the gravity side leads to quantization of the heights of the disks $z_s$ and their charges $Q_s$ as 
\begin{equation}
    z_s = \frac{3  \pi \mu \alpha'}{8}N_s, \qquad Q_s = \frac{\pi^4 \mu^6 \alpha'^3 g_s}{32} n_s,
    \label{eq:zs and Qs in terms of Ns and ns}
\end{equation}
where the constant $\mu$ is related to $V_b$ as $V_b = \mu^5/2^7$.
We can then solve the equations \eqref{eq Ns z}, \eqref{eq Qs ns} and the first one in \eqref{eq Vs} for the parameters $a,b,\mu_s$ and we find 
\begin{equation}
    a = \frac{1}{2 \pi \mu \alpha'} \, , \quad b = \frac{64}{\pi^3 \mu^5 \alpha'^2 g_s} \, , \quad \mu_s = \frac{16}{3 g_s \alpha' \mu^4}V_s + \frac{9}{2^{12} g_s }N_s^3 \pi^3 \alpha'^2 \mu^4 \, .
\end{equation}
In addition, we still have the second equation in \eqref{eq Vs} that we did not use. We can solve it for $\mu$ and we find
\begin{equation}
    \mu = \sqrt{g_{YM}} \Omega,
\end{equation}
which exactly agrees with the matching derived in \cite{Komatsu:2024bop}.
To summarize, the identification of localization parameters with gravity parameters is
\begin{equation}
    x = \frac{1}{2 \pi \mu \alpha'} r \,, \quad \quad \rho^{(s)}(x) = \frac{64}{\pi^3  \mu^5 \alpha'^2 g_s} f_s(r), \qquad \mu_s = \frac{16}{3 g_s \alpha' \mu^4}V_s + \frac{9}{2^{12} g_s }N_s^3 \pi^3 \alpha'^2 \mu^4 \, .
    \label{eq:quantities_matching}
\end{equation}

\subsection{The on-shell action}
The action \eqref{eq:Effective_Action} evaluated on-shell reduces to
\begin{equation}
    S_\mathrm{on-shell} = - \frac{9 \Omega^4}{2^{15}} \sum_s n_s N_s^3 + \frac{1}{2} \frac{3 \Omega^4}{2^7} \sum_s N_s \int dx x^2 \rho^{(s)}(x) + \frac{1}{2} \sum_s \mu_s n_s.
    \label{eq:on shell}
\end{equation}
Note that for a single ball, writing it in terms of the fixed quantity $\xi =n_s \Omega^{-4} N_s^{-5}$, we get 
\begin{equation}
    S_\mathrm{on-shell} = -\frac{N^2}{\lambda^{2/3}} H(\xi),\qquad H(\xi) \approx \begin{cases*}
    -\frac{5}{7\, 2^7} (15 \pi/8)^{2/5}  \xi^{1/15} & if \  $\xi \gg 1 $ \\
       \frac{9}{2^{15}} \xi^{-1/3}    & if \   $\xi \ll 1 $
    \end{cases*}\,,
    \label{eq:Hxi}
\end{equation}
where $N= n_s N_s$ and  $\lambda \equiv N/\Omega^4 = N g^2_{\text{YM}}/\mu^4$ is the dimensionless 't Hooft coupling. This 
aligns with the scaling similarity arguments in \cite{Bobev:2019bvq,Biggs:2023sqw,Bobev:2024gqg} and our scaling analysis from the supergravity perspective \cite{Komatsu:2024bop}.

Using matching of parameters \eqref{eq:quantities_matching}, \eqref{eq:on shell} becomes
\begin{equation}
    S_\mathrm{on-shell} = \sum_s \left(  - \frac{9 \pi^3 \alpha'^2 \mu^4}{2^{13}} n_s N_s^3 + \frac{2^{8}}{3 \pi^4 g_s^2 \alpha'^4 \mu^{10}}  Q_s V_s + \frac{3 }{4 g_s^2 \pi^3 \alpha'^3 \mu^4 } N_s \int dr r^2 f_s(r) \right).
    \label{eq:S on shell}
\end{equation}
Remarkably, the first and the last term combine exactly to give the octopole\footnote{For more details on the octopole, refer to \cite{Komatsu:2024bop}.} of the electrostatic configuration expressed as
\begin{equation}
    q_{3,s} = 2z_s^3 Q_s -6z_s \int_{-R_s}^{R_s} dr r^2 f_s(r) = \frac{27 g_s \alpha'^6 \mu^9 \pi^7}{2^{13}}  n_s N_s^3- \frac{9}{4} \pi \mu \alpha' N_s  \int dr r^2 f_s(r).
\end{equation}
We can then rewrite \eqref{eq:S on shell} as
\begin{equation}
    S_\mathrm{on-shell} = \sum_s \left(  \frac{2^{8}}{3 \pi^4 g_s^2 \alpha'^4 \mu^{10}}  Q_s V_s - \frac{1}{3 g_s^2 \pi^4 \alpha'^4 \mu^5} q_{3,s} \right).
    \label{eq:S on shell final}
\end{equation}
In the limit $z_s \gg R_s$, where we can compute $V_s \approx - \frac{\mu^5}{2^7}z_s^3$ and $q_3 \approx 2 z_s^3 Q_s$ and we obtain
\begin{equation}
    \frac{2^{8}}{3 \pi^4 g_s^2 \alpha'^4 \mu^{10}}  Q_s V_s \underset{N_i \to \infty}{\simeq}  - \frac{9}{2^{16}} n_s N_s^3 \Omega^4, \qquad - \frac{1}{3 g_s^2 \pi^4 \alpha'^4 \mu^5} q_{3,s} \underset{N_i \to \infty}{\simeq} - \frac{9}{2^{16}} n_s N_s^3 \Omega^4,
\end{equation}
which correctly reproduces the value of the potential at the minima.
 
It is interesting to compare equation \eqref{eq:S on shell final} with the on-shell action found in the gravity dual \cite{Komatsu:2024bop}. There, we found a term proportional to $\sum_s Q_s V_s$ which matches the first term in \eqref{eq:S on shell final} up to a factor of 2. We also found that boundary terms generically give something proportional to $\sum_s q_{3,s}$.
It would be nice to carefully finish the gravitational computation and match the result \eqref{eq:S on shell final}.

\section{Discussion}
\label{sec:discussion}
\paragraph{Summary and main results} In this paper, we studied the polarized IKKT matrix model and reduced the partition function, as well as a family of protected observables, to a sum over saddles where each term is parameterized by one block-diagonal matrix integral. Surprisingly, we found that the limit of vanishing mass-deformation did not reduce to the IKKT model, but instead led to a divergence. We made sense of the divergence by interpreting this limit as a strong coupling limit where the off-diagonal entries of the matrices can be integrated out perturbatively. 
In fact, we obtained that the structure exhibited by each saddle contribution is very similar to the picture in BMN, where each irreducible block forms a bound state which decouples from the others in the strong coupling limit. We also considered the limit where the mass-deformation dominates and obtained results consistent with \cite{Hartnoll:2024csr}. Given that the limit of vanishing mass-deformation of the partition function diverges, one can ask whether there exists some natural regularization which would give rise to IKKT. We proposed an answer to this question by conjecturing a contour prescription of the moduli integrals and we verified that it works for $N=2,3$.

Then, we gave evidence that each saddle of the localization result was in one to one correspondence with the geometries we constructed in \cite{Komatsu:2024bop}. This was derived by analyzing the dynamics of the eigenvalues of the localization result in the limit where they form a continuum. In this limit, the partition function is dominated by some configurations of eigenvalues densities which obey a saddle point integral equation. In the gravity side, the metric is characterized by a potential which is obtained after solving an electrostatic problem. This problem can be reduced to an integral equation on the charge densities. We found exact agreement between the two sides once parameters were identified. We also obtained that the on-shell action of the eigenvalues densities is consistent with the bosonic potential evaluated at its minima, and that the structure agrees with the gravitational on-shell action we derived in \cite{Komatsu:2024bop}.
\paragraph{Future directions}
\begin{itemize}
\item It would be instructive to pursue the small $\W$ expansion of the effective action to subleading orders. The leading term derived in section \ref{app:strong_limit} corresponds to free particles in an harmonic potential. The next terms will give rise to interactions between these particles. Can we understand how the distribution of particles changes when there are many interacting particles ? The dual gravitational description should emerge from this density field of interacting particles.

\item It should be possible to obtain the $S^{6}$ partition function of $k$ NS5-branes from a careful scaling limit of our result.

\item Our conjectured prescription to obtain IKKT from polarized IKKT predicts that the sum over $SU(2)$ representations of the residues of the 1-loop determinant should somehow reduce to a sum over divisors of $N$, $\sum_{m | N} \frac{1}{m^2}$. 
Proving this conjecture might involve some non-trivial number theoretic 
relations. 

\item The (undeformed) IKKT partition function, which is given in terms of the divisor function $\sigma_2(N)$, does not have a uniform large $N$ limit and instead behaves ``erratically" as a function of $N$. The asymptotic behavior of $\sigma_2$ has been studied in the literature and it would be interesting to see if and how they can be reproduced from the gravity side. See \cite{Schlenker:2022dyo} for discussions on the gravitational path integral and quantities that depend erratically on $N$.
\end{itemize}

\paragraph{Acknowledgements}
We thank N. Bobev, P. Bomans, F. Cachazo, F. F. Gautason,  S. Hartnoll, A. Kidambi, J. Liu, J. Matos, N. Nekrasov, S. Pufu, A. Rossboth and J. Vilas Boas for discussions. This work was supported by the Simons Foundation grant 488649 (Simons Collaboration on the Nonperturbative Bootstrap) and by the Swiss National Science Foundation through the project 200020\_197160 and through the National Centre of Competence in Research SwissMAP.

\appendix

\addtocontents{toc}{\protect\setcounter{tocdepth}{1}}

\section{Convention for the gamma matrices}
\label{app:gamma_conventions}
To define the Euclidean $SO(10)$ gamma matrices, we first define $SO(9)$ Gamma matrices, inspired from \cite[App.3.A.5]{Freedman:2012zz},
\al{
\spl{
\gamma^1 & \equiv \sigma^2 \otimes \sigma^2 \otimes \sigma^1 \otimes \mathbb{1} \,, \\ 
\gamma^2 & \equiv \sigma^2 \otimes \sigma^3 \otimes \mathbb{1} \otimes \sigma^2 \,, \\ 
\gamma^3 & \equiv - \sigma^3 \otimes \mathbb{1} \otimes \mathbb{1} \otimes \mathbb{1} \,, \\ 
\gamma^4 & \equiv -\sigma^2 \otimes \sigma^1 \otimes \mathbb{1} \otimes \sigma^2 \,, \\ 
\gamma^5 & \equiv \sigma^2 \otimes \mathbb{1} \otimes \sigma^2 \otimes \sigma^3 \,, \\ 
\gamma^6 & \equiv \sigma^1 \otimes \mathbb{1} \otimes \mathbb{1} \otimes \mathbb{1} \,, \\ 
\gamma^7 & \equiv \sigma^2 \otimes \sigma^2 \otimes \sigma^2 \otimes \sigma^2 \,, \\ 
\gamma^8 & \equiv -\sigma^2 \otimes \mathbb{1} \otimes \sigma^2 \otimes \sigma^1 \,, \\ 
\gamma^9 & \equiv \sigma^2 \otimes \sigma^2 \otimes \sigma^3 \otimes \mathbb{1} \,, \\ \label{eq:SO9Matrices}
}
}
where $\sigma^i$ are Pauli matrices, $\mathbb{1} = \mathbb{1}_2$ and we have reshuffled the indices and signs for convenience in our localization computations. The features of this representation are that
\begin{itemize}
    \item The complex valued matrix $\sigma^2$ always appears an even number of times, so that $\gamma^{I'}$ are real and symmetric.
    \item We chose $\gamma^3$ diagonal.
    \item The matrices $\gamma^{I'=1,...,2}$ and $\gamma^{I'=4,...,9}$ take a block off-diagonal form, where the off-diagonal blocks are 8 by 8.
\end{itemize}
We then define the $SO(10)$ ``Weyl'' matrices, as
\begin{equation}
    \gamma^I = (\gamma^{I'}, - i \mathbb{1}_{16})\,, \qquad \bar \gamma^I = (\gamma^{I'}, i \mathbb{1}_{16}) \,,
    \quad (I=1,2,\ldots 10, \ I'=1,2,\ldots,9) \,. \label{eq:Weyl_matrices}
\end{equation}
This allows to define the $SO(10)$ gamma matrices as
\al{\spl{
& \G^I = 
\mat{0 & \g^I \\ \bar\g^I & 0}\,,
\\
& \left\{\Gamma^I, \Gamma^J\right\} 
=\mat{2\g^{(I}\bar\g^{J)} & 0 \\ 0 & 2\bar\g^{(I}\g^{J)}}
=2 \d^{IJ}\mathbb{1}_{32}\,,
\\
&\G^{IJ} 
\equiv
\G^{[I}\G^{J]}
=\mat{\g^{[I}\bar\g^{J]} & 0 \\ 0 & \bar\g^{[I}\g^{J]}}
\equiv
\mat{\bar\g^{IJ} & 0 \\ 0 & \g^{IJ}}\,,
\\
&\G^{IJK} 
\equiv
\G^{[I}\G^{J}\G^{K]}
=\mat{0 & \g^{[I}\bar\g^{J}\g^{K]} \\ \bar\g^{[I}\g^{J}\bar\g^{K]} & 0}
\equiv
\mat{0 & \g^{IJK} \\  \bar\g^{IJK} & 0}\,,
\\
&\G^{IJKL}
=\mat{\g^{[I}\bar\g^{J}\g^{K}\bar\g^{L]} & 0 \\ 0 &  \bar\g^{[I}\g^{J}\bar\g^{K}\g^{L]} }
\equiv
\mat{\bar\g^{IJKL} & 0 \\ 0 & \g^{IJKL}}\,,
\\
& \Gamma_*  \equiv (-i)^{10/2}\G^1 \G^2 \ldots \G^{10}  = \left(\begin{array}{cc}
\mathbb{1}_{16} & 0 \\
0 & -\mathbb{1}_{16}
\end{array}\right)\,,
}\label{eq:gamma matrices Euclidean}}
and higher rank gamma matrices also follow the same pattern as above. Our choice for the conjugation matrix $\mathcal{C}$ is
\begin{equation}
    \mathcal{C} = \Gamma^{10} = \begin{pmatrix}
        0 & - i \mathbb{1}_{16} \\ i \mathbb{1}_{16} & 0
    \end{pmatrix} \,.
\end{equation}
In this convention, the Majorana-Weyl condition is trivially satisfied by real left-handed spinors of the form $\psi = (\psi_{\alpha=1,...,16},0,...,0)$ where $\psi_\alpha$ are real Grassmann numbers. For such spinors,
\al{
\ga{
\mathcal{C} \Gamma^I \psi = - i \bar \gamma^I \psi \,, \qquad \Gamma^{I J} \psi = \bar \gamma^{I J} \psi \,, \qquad \mathcal{C} \Gamma^{1 2 3} \psi = - i \bar \gamma^{1 2 3} \psi \,, \\ \Gamma^{1 2 3} \Gamma^I \psi = \bar \gamma^{1 2 3} \bar \gamma^I \psi \,, \qquad \dots
}
}
where it is clear that we use the notation $\psi = (\psi,0)$ implicitly both for the 32-component and 16-component spinor $\psi$. 

\section{The IKKT convergence argument, broken by fermion masses}
\label{app:fermion_masses_ConvergenceIKKT}
In this appendix, we discuss convergence of the pure $SU(2)$ IKKT partition function, and show that the presence of massive fermions affects the convergence argument, leading to a divergence when $\Omega \to 0$. Let us start by looking at the bosonic part of the pure IKKT model 
\begin{equation}
    S_\mathrm{bos}^\mathrm{IKKT} = - \frac{1}{4} \mathrm{Tr} [X_I, X_J]^2 \,.
\end{equation} 
We define $X^I = R x^I$ where $\sum_{I} \mathrm{Tr} x^I x^I = 1$. The divergent behaviour may arise from the region $R \gg 1$. However, it has been shown in \cite{Austing:2001pk,Austing:2001ib,Austing:2001bd} that correlators behave at large $R$ as
\begin{equation}
    \int [d X^I] \mathrm{Tr} X^p e^{- S_\mathrm{bos}^\mathrm{IKKT}} \sim \int_{R \gg 1} \frac{d R}{R} R^{p-6} \label{eq:LargeRBosonicIKKT}\,.
\end{equation}
In particular the partition function of $SU(2)$ bosonic IKKT converges, as well as correlators with degree $p < 6$. Adding a bosonic mass deformation only gives an $e^{- \Omega^2 R^2}$ factor. Namely, with
\begin{equation}
\begin{split}
    S_\mathrm{bos} = \operatorname{Tr} & \Biggl[-\frac{1}{4} [X_I,X_J]^2 + \frac{3 \Omega^2}{4^3} X_i X_i + \frac{\Omega^2}{4^3} X_p X_p + i \frac{\Omega}{3} \epsilon^{i j k} X_i X_j X_k \Biggr] \, ,
\end{split}
\end{equation}
correlators far in the commuting modes valley behave as
\begin{equation}
    \int [d X^I] \mathrm{Tr} X^p e^{- S_\mathrm{bos}} \sim \int_{R \gg 1} \frac{d R}{R^6} R^p e^{- \Omega^2 R^2} \sim \Omega^{6-p} \label{eq:LargeRBosonicPolarizedIKKT} \,.
\end{equation}
There are no surprises yet. Bosonic polarized IKKT has the same divergences when $\Omega \to 0$ as bosonic IKKT. The surprise comes when adding fermions. 

When we add fermions to bosonic pure IKKT, we basically add
\begin{equation}
    \int [d \psi_\alpha] e^{\frac{1}{2} \psi^\top \bar \gamma^{I} [X_I,\psi]} = \mathrm{Pf}(X^I)
\end{equation}
in the bosonic integral. One can integrate out the $\psi$'s, giving a Pfaffian $\mathrm{Pf}(X^I)$ which is a polynomial in the $X$ matrices. For the case $N=2$ that we consider here, it has been shown in \cite{Krauth:1998xh} that
\begin{equation}
    \mathrm{Pf}(X^I) \propto \left( \mathrm{Tr} [X_I, X_J] [X_J, X_K] [X_K,X_I] \right)^4 \,.
\end{equation}
To discuss the convergence behaviour of the $SU(2)$ IKKT partition function with fermions, we thus simply insert the Pfaffian in the left-hand-side of \eqref{eq:LargeRBosonicIKKT}. Since $\mathrm{Pf}(X^I)$ only involves commutators, it can only \textit{help} the convergence of the partition function since it vanishes in the commuting modes valleys. In fact, the careful analysis of \cite{Austing:2001pk,Austing:2001ib} shows that correlators now converge up to $p < 14$. Let us now discuss what happens when we perform the mass deformation. In this case, we obtain the Pfaffian
\begin{equation}
    \int [d \psi_\alpha] e^{\frac{1}{2} \bar \psi^\top \bar \gamma^I [X_I, \psi] - \frac{i}{8} \Omega \psi^\top \bar \gamma^{1 2 3} \psi} = \mathrm{Pf}(X^I, \Omega)
\end{equation}
which reduces to the previous one when $\Omega = 0$. One can check numerically that when $X^I \sim R \ \sigma^3$ where $\sigma^3 = \mathrm{diag}(1,-1)$, i.e. when $X^I$ are commuting matrices, with magnitude $\sim R$, the mass terms of the fermions create new contributions such that the Pfaffian goes as
\begin{equation}
    \mathrm{Pf}(X^I, \Omega) \sim \Omega^{24} + \Omega^{22} R^2 + \Omega^{20} R^4 + \cdots + \Omega^{8} R^{16} \,.
\end{equation}
Inserting this Pfaffian in \eqref{eq:LargeRBosonicPolarizedIKKT}, one obtains
\al{
\spl{
\int [d X^I] [d \psi_\alpha] e^{- S} & = \int [d X^I] \mathrm{Pf}(X^I, \Omega) e^{- S_\mathrm{bos}} \\ & \sim \int_{R \gg 1} \frac{d R}{R} R^{-6} (\Omega^{8} R^{16} + \dots) e^{- \Omega^2 R^2} \sim \frac{1}{ \Omega^2} + \dots
}
}
Therefore, we see a $1 / \Omega^2$ divergence appearing, due to the massive terms of the Pfaffian. Formally, we see in this last integral why $\lim_{\Omega \to 0}$ and $\int [d X^I] [d \psi_\alpha]$ don't commute, i.e.
\begin{equation}
    0 = \int_{R \gg 1} \frac{d R}{R} \lim_{\Omega \to 0}   \Omega^8 R^{10} e^{- \Omega^2 R^2} \neq \lim_{\Omega \to 0} \int_{R \gg 1} \frac{d R}{R}  \Omega^8 R^{10} e^{- \Omega^2 R^2} = \infty \,.
\end{equation}

\section{Technical details for localization}
\subsection{Constraints from SUSY closure}
\label{App_subsec:susy algebra closure}
To impose the SUSY closure on gauge and global symmetry generators, we will need to use gamma matrices identities which are easier to express in 32-component notation. We thus begin by expressing the supersymmetries \eqref{eq:SUSY_16} as
\al{
\spl{
\delta_\epsilon X^I & = \bar{\epsilon} \Gamma^I \psi \,, \\
\delta_\epsilon \psi & = \frac{i}{2} \Gamma^{I J} \epsilon [X_I, X_J] + \frac{3 \Omega}{8} \Gamma^{1 2 3} \Gamma^i \epsilon X_i + \frac{\Omega}{8} \Gamma^{1 2 3} \Gamma^p \epsilon X_p + i \nu_a K_a \,, \\ \delta_\epsilon K_a & = - \bar{\nu}_a \Gamma^I [X_I, \psi] + i \frac{\Omega}{4} \bar{\nu}_a \Gamma^{1 2 3} \psi \,. \label{eq:SUSY_32}
}
}
We then impose $\{ \delta_{\epsilon_1}, \delta_{\epsilon_2} \} (\mathrm{fields}) = \delta_B (\mathrm{fields})$ on $X^I$, $\psi_\alpha$ and $K_a$ where $\delta_B$ is some combination of $SU(N) \times SO(3) \times SO(7)$ generators. 

Before diving in, let us introduce some useful identities. We will use  Majorana flip relations, see e.g. \cite[Sec.~3.2.1]{Freedman:2012zz},
\begin{equation}
     (\mathcal{C} \Gamma^A)^\top = - t_{r_A} (\mathcal{C} \Gamma^A) \,,
\end{equation}
where $\Gamma^A = \Gamma^{\mu_1,..., \mu_{r_A}}$ is a rank-$r_A$ gamma matrix, and in our convention $t_0 = t_3 = 1$ and $t_1 = t_2 = -1$. Recalling that we consider $\epsilon_i$ to be Grassmann-even (for Grassmann-odd fermions, invert the signs), this implies
\begin{equation}
    \bar{\epsilon}_1 \Gamma^{A} \epsilon_2 = - \bar{\epsilon}_2 \Gamma^{A} \epsilon_1\,, \qquad r_A = 0,3,4,7,8 \label{eq:Rank_Vanishing}
\end{equation}
\begin{equation}
    \bar{\epsilon}_1 \Gamma^{A} \epsilon_2 = \bar{\epsilon}_2 \Gamma^{A} \epsilon_1\,. \qquad r_A = 1,2,5,6,9,10 
\end{equation}
Note also that for left-handed Weyl spinors $\epsilon_1$, $\epsilon_2$,
\begin{equation}
    \bar \epsilon_1 \Gamma^A \epsilon_2 = 0\,, \qquad r_A \text{ even} \label{eq:Rank_Vanishing2}
\end{equation}
which follows from $\bar \epsilon_1 \Gamma^A \epsilon_2 = \bar \epsilon_1 \Gamma^A \Gamma^* \epsilon_2 = (-1)^{r_A} \bar \epsilon_1  \Gamma^* \Gamma^A \epsilon_2 = (-1)^{r_A + 1} \bar \epsilon_1  \Gamma^A \epsilon_2$. When considering anticommutators $\{ \delta_{\epsilon_1}, \delta_{\epsilon_2} \}$, we will symmetrize $\delta_{\epsilon_1} \delta_{\epsilon_2}$ in $(\epsilon_1, \nu_a^1) \leftrightarrow (\epsilon_2, \nu_a^2)$, where $\nu_a^i$ is the ``auxiliary'' spinor that will be constrained in terms of $\epsilon_i$. Thus, 
it is useful to note that the only non-vanishing symmetrized expressions are of the form
\begin{equation}
    \bar \epsilon_1 \Gamma^A \epsilon_2 + (1 \leftrightarrow 2) \neq 0 \iff r_A = 1,5,9 \,. \label{eq:Rank_Vanishing3}
\end{equation}
All other gamma matrices vanish when symmetrized over $\epsilon_1, \epsilon_2$.

\paragraph{On $X^I$: } We obtain
\al{
\spl{
\{ \delta_{\epsilon_1}, \delta_{\epsilon_2} \} X^I & = \frac{i}{2} \bar \epsilon_2 \Gamma^I \Gamma^{J K} \epsilon_1 [X_J, X_K] + \bar{\epsilon}_2 \Gamma^I \nu_a^1 K_a \\ & \quad + \frac{3 \Omega}{8} \bar \epsilon_2 \Gamma^I \Gamma^{1 2 3} \Gamma^{j} \epsilon_1 X_j + \frac{\Omega}{8} \bar \epsilon_2 \Gamma^I \Gamma^{1 2 3} \Gamma^{q} \epsilon_1 X_q + \{\epsilon_1, \nu_a^1\} \leftrightarrow \{\epsilon_2, \nu_a^2\} \,.
}
}
It is easy to notice that the auxiliary matrices $K_a$ cannot appear in order to have algebra closure. Thus, we obtain the first constraint
\begin{equation}
    \bar{\epsilon}_2 \Gamma^I \nu_a^1 = \bar{\epsilon}_1 \Gamma^I \nu_a^2 = 0 \,. \label{eq:Constraint1}
\end{equation}

Let us now consider $I = i = 1,2,3$ and $I=p=4,5,...,10$ separately.
Using that $\Gamma^{I} \Gamma^{J K} = \Gamma^{I J K} + \delta^{I J} \Gamma^K - \delta^{I K} \Gamma^J$, where $\Gamma^{I J K}$ do not contribute according to \eqref{eq:Rank_Vanishing3} because they are rank-3 gamma matrices, and that $\Gamma^{i} \Gamma^{1 2 3} \Gamma^{q}$, $\Gamma^p \Gamma^{1 2 3} \Gamma^j$ are rank-3 gamma matrices which do not contribute as well, we have
\begin{equation}
    \{ \delta_{\epsilon_1}, \delta_{\epsilon_2} \} X^i = -2 i \bar{\epsilon}_1 \Gamma^J \epsilon_2 [X^J, X^i] + \frac{3 \Omega}{4} \bar{\epsilon}_1 \Gamma^{1 2 3} \Gamma^{i j} \epsilon_2 X^j \,, \label{eq:deltasquare_Xi}
\end{equation}
\begin{equation}
    \{ \delta_{\epsilon_1}, \delta_{\epsilon_2} \} X^p = -2 i \bar{\epsilon}_1 \Gamma^J \epsilon_2 [X^J, X^p] - \frac{ \Omega}{4} \bar{\epsilon}_1 \Gamma^{1 2 3} \Gamma^{p q} \epsilon_2 X^q \,, \label{eq:deltasquare_Xp}
\end{equation}
where we recognize a sum of gauge generator and $SO(3) \times SO(7)$ rotations acting on $X^I$.

\paragraph{On $\psi_\alpha$: } Using \eqref{eq:SUSY_32}, we have
\al{
\spl{
\{ \delta_{\epsilon_1}, \delta_{\epsilon_2} \} \psi & = - i \Gamma^{J I} \epsilon_2 [\bar \epsilon_1 \Gamma^I \psi, X^J] - i \nu_a^2 \bar \nu_a^1 \Gamma^I [X_I, \psi] \\ & + \frac{3 \Omega}{8} \Gamma^{1 2 3} \Gamma^i \epsilon_2 (\bar \epsilon_1 \Gamma^i \psi) + \frac{\Omega}{8} \Gamma^{1 2 3} \Gamma^p \epsilon_2 (\bar \epsilon_1 \Gamma^p \psi) - \frac{\Omega}{4} \nu_a^2 \bar \nu_a^1 \Gamma^{1 2 3} \psi + \{\epsilon_1, \nu_a^1\} \leftrightarrow \{\epsilon_2, \nu_a^2\} \,. \label{eq:Fermion Closure}
}
}
Let us start by looking at the term independent of $\Omega$. First, we separate $\Gamma^{J I} = \Gamma^J \Gamma^I - \delta^{I J}$. Then, we use on the term containing $\Gamma^J \Gamma^I$ the Fierz identity \cite{Freedman:2012zz}
\al{
\spl{
\Gamma^I_{\alpha \beta} \Gamma^I_{\gamma \delta} & = \frac{1}{2^5} \sum_A v_{r_A} \Gamma^{A}_{\alpha \delta} (\Gamma_A)_{\gamma \beta} \\ & \equiv \frac{1}{2^5} \left( v_0 \d_{\a \d} \d_{\g \b} + v_1 \G^I_{\a \d} \G^I_{\g \b} + \frac{v_2}{2!} \G^{I J}_{\a \d} \G^{J I}_{\g \b} + \dots + \frac{v_{10}}{10!} \G^{I_1 ... I_{10}}_{\a \d} \G^{I_{10} ... I_1}_{\g \b}  \right)\,,
}
}
where
\begin{equation}
    v_{r_A} \equiv (-1)^{r_A} (10 - 2 r_A) \,.
\end{equation}
Nicely, this vanishes for $r_A = 5$. Note that from condition \eqref{eq:Rank_Vanishing3}, we only have contributions from $r_A = 1$ and its hodge dual $r_A = 9$. For Weyl spinors, the two give the same contribution. Altogether, we obtain
\begin{equation}
     - i \Gamma^{J} \Gamma^I \epsilon_2 [\bar \epsilon_1 \Gamma^I \psi, X^J] + (1 \leftrightarrow 2) = \frac{i}{2} \bar \epsilon_1 \Gamma^I \epsilon_2 [\Gamma^J \Gamma^I \psi, X_J] + (1 \leftrightarrow 2) \,.
\end{equation}
This identity allows to write the $\Omega = 0$ part of \eqref{eq:Fermion Closure} as
\al{
\spl{
\{ \delta_{\epsilon_1}, \delta_{\epsilon_2} \} \psi_\a|_{\Omega = 0} & = -i \bar \epsilon_1 \Gamma^I \epsilon_2 [X^I, \psi_\a] - \frac{i}{2} (\bar \epsilon_1 \Gamma^I \epsilon_2) \Gamma^I_{\alpha \beta} [(\Gamma^J \psi)_\beta, X^J] \\ & + i (\epsilon_2)_\alpha (\bar \epsilon_1)_\b [(\G^I \psi)_\beta,X^I] + i (\nu_a^2)_\alpha (\bar \nu_a^1)_\b [(\G^I \psi)_\beta,X^I] + \{\epsilon_1, \nu_a^1\} \leftrightarrow \{\epsilon_2, \nu_a^2 \} \,.
}
}
Only the first term is compatible with the algebra closure. This gives the constraint
\begin{equation}
    \frac{1}{2} (\bar{\epsilon}_1 \Gamma^I \epsilon_2) \Gamma^I_{\alpha \beta} = (\epsilon_2)_\alpha (\bar{\epsilon}_1)_\beta + (\nu_a^2)_\alpha (\bar{\nu}_a^1)_{\beta} \,.\label{eq:Constraint2}
\end{equation}

Treating the $\Omega$ term in \eqref{eq:Fermion Closure} is now straightforward by solving the $\nu$ dependence using the above constraint. We obtain
\begin{equation}
    \{ \delta_{\epsilon_1} , \delta_{\epsilon_2} \} \psi = - 2 i \bar{\epsilon}_1 \Gamma^I \epsilon_2 [X^I,\psi] + \frac{3 \Omega}{16} \bar{\epsilon}_1 \Gamma^{1 2 3} \Gamma^{i j} \epsilon_2 \Gamma^{i j} \psi - \frac{\Omega}{16} \bar{\epsilon}_1 \Gamma^{1 2 3} \Gamma^{p q} \epsilon_2 \Gamma^{p q} \psi \,. \label{eq:deltasquare_psi}
\end{equation}

\paragraph{On $K_a$: } 
The anticommutator $\{ \delta_{\epsilon_1}, \delta_{\epsilon_2} \} K_a$ has three pieces, one of order 1, one of order $\Omega$, and one of order $\Omega^2$. Let us separate
\begin{equation}
\{ \delta_{\epsilon_1}, \delta_{\epsilon_2} \} K_a = \{ \delta_{\epsilon_1}, \delta_{\epsilon_2} \} K_a|_{\Omega = 0} + \{ \delta_{\epsilon_1}, \delta_{\epsilon_2} \} K_a|_{\Omega} + \{ \delta_{\epsilon_1}, \delta_{\epsilon_2} \} K_a|_{\Omega^2} \,.
\end{equation}
We start with the $\Omega$-independent piece,
\al{
\spl{
\{ \delta_{\epsilon_1}, \delta_{\epsilon_2} \} K_a|_{\Omega = 0} & = - i \bar \nu_a^2 \Gamma^I \nu_b^1 [X^I, K_b] + i (\bar \nu_a^2 \G^I \psi) (\bar \epsilon_1 \G^I \psi) + i (\bar \epsilon_1 \G^I \psi) (\bar \nu_a^2 \G^I \psi) \\ & - \frac{i}{2} \bar{\nu}_a^2 \Gamma^I \Gamma^{J K} \epsilon_1 [X^I,[X^J, X^K]] + \{\epsilon_1, \nu_a^1\} \leftrightarrow \{\epsilon_2, \nu_a^2\} \,.
}
}
The first term (+ $\nu_a^1 \leftrightarrow \nu_a^2$) is exactly the gauge generator $-2 i (\bar \epsilon_1 \Gamma^I \epsilon_2) [X^I, \cdot]$ that we obtained in \eqref{eq:deltasquare_Xi}, \eqref{eq:deltasquare_Xp}, \eqref{eq:deltasquare_psi} provided that we impose
\begin{equation}
    \bar \nu_a^1 \Gamma^I \nu_b^2 = \delta_{a b} \bar \epsilon_1 \G^I \epsilon_2 \,, \label{eq:Constraint3}
\end{equation}
whereas the rest of the first line is 0 because $\psi$ are Grassmann-odd, and the second line vanishes thanks to the decomposition $\Gamma^I \Gamma^{J K} = \Gamma^{I J K} + \delta^{I J} \Gamma^K - \delta^{I K} \Gamma^J$, and the use of Jacobi identity on $\Gamma^{I J K}$, and the condition $\bar \nu_a \Gamma^I \epsilon = 0$ on $\Gamma^K$, $\Gamma^J$.

Let us now look at the $\Omega$-dependent piece of $\{ \delta_{\epsilon_1}, \delta_{\epsilon_2} \} K_a$. We obtain
\al{
\spl{
\{ \delta_{\epsilon_1}, \delta_{\epsilon_2} \} K_a|_{\Omega} & = - \frac{3 \Omega}{8} \bar \nu_a^2 \Gamma^I [X^I, \G^{1 2 3} \G^i \epsilon_1 X_i] - \frac{\Omega}{8} \bar \nu_a^2 \Gamma^I [X^I, \G^{1 2 3} \G^p \epsilon_1 X_p] \\ & \quad - \frac{\Omega}{8} \bar \nu_a^2 \Gamma^{1 2 3} \Gamma^{I J} \epsilon_1 [X_I, X_J] + (\epsilon_1, \nu_a^1 \leftrightarrow \epsilon_2, \nu_a^2)  = 0 \,.
}
}
where in order to obtain the last equality, we separated the contributions from $I=i$, $I=p$ and $J = j$, $J = q$, anticommuted the matrices around and we used the constraint $\bar \nu \Gamma^I \epsilon = 0$.
Finally,
\al{
\spl{
\{ \delta_{\epsilon_1}, \delta_{\epsilon_2} \} K_a|_{\Omega^2} & = - i \frac{3\Omega^2}{32} \bar \nu_a^2\Gamma^i \epsilon_1 X_i - i \frac{\Omega^2}{32} \bar \nu_a^2 \Gamma^i \epsilon_1 X_i + (\epsilon_1, \nu_a^1 \leftrightarrow \epsilon_2, \nu_a^2) = 0 \,,
}
}
where we used $(\Gamma^{1 2 3})^2 = - \mathbb{1}_{32}$ and the constraint $\bar \nu \Gamma^I \epsilon = 0$.
Thus, we only obtain the gauge generator without $SO(3) \times SO(7)$ transformation, i.e.
\begin{equation}
    \{ \delta_{\epsilon_1} , \delta_{\epsilon_2} \} K_a = - 2 i \bar{\epsilon}_1 \Gamma^I \epsilon_2 [X^I,K_a] \,.
    \label{eq:deltasquare_Ka}
\end{equation}

\paragraph{Identifying the symmetries: } Let us define $SO(10)$ rotations by an antisymmetrized parameter $\omega_{I J}$ (from which $SO(3)$ and $SO(7)$ are a subgroup) as
\begin{equation}
    \delta_{SO(10),\omega} X^I = i \omega_{K L} (M_{K L})_{I J} X^J \,, \qquad \delta_{SO(10),\omega} \psi = \frac{i}{2} \omega_{K L}\Gamma^{K L} \psi \,,
\end{equation}
where
\begin{equation}
    (M_{KL})_{I J} = i (\delta_{I K} \delta_{J L} - \delta_{K J} \delta_{I L}) \,.
\end{equation}
The $SO(3)$ and $SO(7)$ generators are a simple subgroup of the above with parameters $\omega_{i j}$ and $\omega_{p q}$ respectively. We also define the $SU(N)$ generator with parameter $\alpha$ (Hermitian matrix) as
\begin{equation}
    \delta_{SU(N), \alpha} = i [\alpha, \cdot] \,.
\end{equation}
Then, from \eqref{eq:deltasquare_Xi}, \eqref{eq:deltasquare_Xp}, \eqref{eq:deltasquare_psi}, \eqref{eq:deltasquare_Ka}, we derived\footnote{We also see that $\delta_{SO(3)} K_a = \delta_{SO(7)} K_a = 0$}
\begin{equation}
 \{ \delta_{\epsilon_1}, \delta_{\epsilon_2} \} = \delta_{SU(N),-2 \bar \epsilon_1 \Gamma^J \epsilon_2 X^J} + \frac{3 \Omega}{8} \delta_{SO(3), -i \bar \epsilon_1 \Gamma^{1 2 3} \G^{i j} \epsilon_2} - \frac{ \Omega}{8} \delta_{SO(7), -i \bar \epsilon_1 \Gamma^{1 2 3} \G^{p q} \epsilon_2}  \,. \label{eq:SUSYCommutatorResult}
\end{equation}

\paragraph{Summary:} We derived the constraints \eqref{eq:Constraint1}, \eqref{eq:Constraint2} and \eqref{eq:Constraint3}, which are equivalent to \eqref{eq:nu_constraints} once written in 16-component notation. We also obtained along the way the action of the SUSY anticommutator on all fields \eqref{eq:deltasquare_Xi}, \eqref{eq:deltasquare_Xp}, \eqref{eq:deltasquare_psi}, \eqref{eq:deltasquare_Ka} which can be written as
\begin{equation}
    \{ \delta_{\epsilon_1}, \delta_{\epsilon_2} \} = \delta_{SU(N)} + \delta_{SO(3) \times SO(7)}\,,
\end{equation}
as shown with explicit parameters in \eqref{eq:SUSYCommutatorResult}.

\subsection{Solving the auxiliary constraints}
\label{app:constraintsFromAuxiliaries}
We now solve the constraints on $\nu_a$ such that $\delta_s^2$ closes off-shell. Here we will use the notation 
\begin{equation}
    M = 1,2,4,5,6,...,9
\end{equation}
Note that we have a gamma matrix representation where $\bar \gamma^3$ and $\bar \gamma^{10}$ are proportional to the identity on both 8-component subspaces, but also the other $\bar \gamma^{I \neq 3,10} = \bar \gamma^M$ matrices have $8 \times 8$ off-diagonal blocks \eqref{eq:SO9Matrices}, \eqref{eq:Weyl_matrices}. In particular, 
\begin{equation}
    - \bar \gamma^3 \epsilon = -i \bar \gamma^{10} \epsilon = \epsilon = \begin{pmatrix}
        \eta \\ 0
    \end{pmatrix}\,, \qquad \qquad \bar \gamma^{M} \epsilon  = \begin{pmatrix}
       0 \\ \tilde \eta_{M}
    \end{pmatrix}\,,
\end{equation}
where we recall that $M \neq 3,10$. Note also that the eight $\tilde \eta_{M}$ form a basis of the 8-component spinor subspace since they are orthogonal to each other. Thus, it is easy to notice that the first constraint in \eqref{eq:nu_constraints} is satisfied by a set of spinors 
\begin{equation}
    \nu_{a = 1,...,7} = \begin{pmatrix}
        \tilde \nu_{a} \\ 0
    \end{pmatrix} \,, \qquad \qquad \tilde \nu_a^\top \eta = 0 \,.
\end{equation}
Let us now look at the third constraint. Since $\bar \gamma^{M}$ are off-diagonal, this constraint is trivial for $I = M \neq 3,10$. For $I=3,10$, $\bar \gamma^I$ is proportional to the identity on $\tilde \eta_a$ and $\epsilon$, and it takes the same value on the left and right-hand-side. The constraint thus reduces to
\begin{equation}
    \nu_a^\top \nu_b = \delta_{a b} \,.
\end{equation}
This implies that $\nu_a$ have to be orthogonal to each other on the 8-component subspace, as well as being orthogonal to $\epsilon$. In particular, this can only be satisfied with $a = 1,...,7$.
Finally, the second constraint in \eqref{eq:nu_constraints} can be derived from the first and third constraints. This has been proven in \cite{Pestun:2007rz}. Thus, it is unnecessary to discuss it, but it is not hard to see that it is satisfied by the above solution.

\subsection{Constructing the BRST cohomology}
\label{app:BRST_cohomology}
In this section, we construct naturally the BRST cohomology described in the main text, justifying along the way why we need all the ghosts appearing in \eqref{eq:ghost action}. This section only serves as some motivation for the construction \eqref{eq:action of Q}. 

Let us start by the naive gauge fixing action, with ghosts $(\tilde C, C, b)$, which has zero modes in general
\begin{equation}
    S_\mathrm{gh} = \mathrm{Tr} (i b F + i \tilde C [L_j, [X_j, C]]) \,.
\end{equation}
It is known from the BRST formalism that this can be written as
\begin{equation}
    S_\mathrm{gh} = \delta_b V_\mathrm{gh} \,, \quad V_\mathrm{gh} = \mathrm{Tr} \tilde C F \,,
\end{equation}
where
\begin{gather}
    \delta_b \text{boson} = [\text{boson}, C] \,, \quad \delta_b \text{fermion} = - \{\text{fermion}, C \} \,, \\ \delta_b \tilde C = i b \,, \quad \delta_b C = - C^2 \,, \quad \delta_b b = 0 \,,
\end{gather}
where ``boson'' and ``fermion'' are bosonic and fermionic physical fields. These transformations are constrained from requiring $\delta_b V_\mathrm{gh} = S_\mathrm{gh}$, $\delta_b^2 = 0$ and $\delta_b \mathrm{Tr} (\text{physical fields}) = 0$.

However, we also want to get rid of the zero modes as discussed in section \ref{subsec:ghost action}. Thus, we want the action with three new ghost zero-modes $(C_0, \tilde C_0, b_0)$ which commute with $L_i$,
\begin{equation}
    S_\mathrm{gh} \to S_\mathrm{gh} = \mathrm{Tr} (i b (F + b_0) + i \tilde C [L_j, [X_j, C]]) + i \tilde C C_0 + i C \tilde C_0 ) \,.
\end{equation}
This action is a valid gauge fixing action. We could stop here. However, we want to have a BRST formalism to treat this action on the same footing as the localization deformation. Let us try to write it as $Q V_\mathrm{gh}$ = $(\delta_s + \delta_b) V_\mathrm{gh}$. 

\paragraph{Step 1: Add a ghost} First, the new term $\mathrm{Tr} i b b_0$ is obtained from shifting $F \to F + b_0$ in $V_\mathrm{gh}$. By defining $\delta_b b_0 = - i C_0$, this also creates the term $\tilde C C_0$ along the way, namely
\begin{equation}
    \mathrm{Tr} (i b (F + b_0) + i \tilde C [L_j,[X_j,C]] + i \tilde C C_0) = \delta_b \mathrm{Tr} \tilde C (F + b_0) \,,
\end{equation}
with
\begin{gather}
\delta_b \text{boson} = [\text{boson}, C] \,, \quad \delta_b \text{fermion} = - \{\text{fermion}, C \} \,, \\ \delta_b \tilde C = i b\,, \quad \delta_b C = - C^2\,, \quad \delta_b b = 0\,, \quad \delta_b b_0 = - i C_0 \,.
\end{gather}
To create the third term $i C \tilde C_0$, we need a zero-mode bosonic field such that 
$\delta_b ( ...) = \tilde C_0$, but we already used $b_0$. This forces us to introduce a \textbf{new ghost}, which we denote by $\tilde a_0$ such that 
\begin{equation}
    \delta_b \tilde a_0 = - i \tilde C_0 \, .
\end{equation} 
Then, $i C \tilde C_0 \subset \delta_b (C \tilde a_0)$. We thus need to deform $V_\mathrm{gh}$ to
\begin{equation}
    V_\mathrm{gh} \to V_\mathrm{gh} = \mathrm{Tr}( \tilde C (F + b_0) + C \tilde a_0) \,,
\end{equation}
which is now such that
\begin{equation}
    \delta_b V_\mathrm{gh} = \mathrm{Tr} (i b (F + b_0) + i \tilde C [L_j, [X_j, C]] + i \tilde C C_0 + i C \tilde C_0 - C^2 \tilde a_0) \,.
\end{equation}
\paragraph{Step 2: Add another ghost} The above is not sufficient since $\tilde a_0$ doesn't appear quadratically multiplied to other bosons. The integral over $e^{- S_\mathrm{gh}}$ would be divergent. This is why we need to introduce yet \textbf{another bosonic ghost} $a_0'$ whose job in life is to multiply $\tilde a_0$.\footnote{$a_0'$ is not yet $a_0$, because we will perform a shift as described in the last step.} Looking at $V_\mathrm{gh}$, a minimal way is to modify the BRST transformations by adding $ia_0'\subset\delta_b C $ so that it comes multiplied to $\tilde a_0$. Thus, with the deformed BRST transformations,
\begin{gather}
\delta_b \text{boson} = [\text{boson}, C] \,, \quad \delta_b \text{fermion} = - \{\text{fermion}, C \} \,, \nonumber \\ \delta_b \tilde C = i b \,, \quad \delta_b C = i a_0' - C^2 \,, \quad \delta_b b = 0 \,, \quad \delta_b b_0 = - i C_0 \,, \quad \delta_b \tilde a_0 = - i \tilde C_0 \, , \label{eq:IntermediateBRSTalgebra}
\end{gather}
we obtain
\al{
\spl{
\delta_b V_\mathrm{gh} & = \mathrm{Tr} (i b (F + b_0) + i \tilde C [L_j, [X_j, C]] + i \tilde C C_0 + i C \tilde C_0 \\ & \qquad \qquad + i (a_0' + i C^2) \tilde a_0) \,.
}
}
This is exactly the $\delta_b V_\mathrm{gh}$ part of \eqref{eq:ghost action}. It is now, again, a valid gauge fixing action, because the integral over $\tilde a_0$ and $a_0'$ would reduce to a number. However, we also want to incorporate supersymmetry which is the next step. Before doing that, let us complete the algebra of $\delta_b$. We did not yet write $\delta_b \tilde C_0$, $\delta_b C_0$ and $\delta_b a_0'$. Note that, from \eqref{eq:IntermediateBRSTalgebra},
\begin{equation}
    \delta_b^2 \text{(physical field)} = i [\text{(physical field)}, a_0'] \,.
\end{equation}
Thus, we require $\delta_b^2 = i [\cdot, a_0']$. This constrains $\delta_b \tilde C_0$, $\delta_b C_0$ to involve commutators with $a_0'$ and also modifies $\d_b b$, yielding
\al{
\begin{gathered}
    \delta_b \text{boson} = [\text{boson}, C] \,, \quad \delta_b \text{fermion} = - \{\text{fermion}, C \} \,, \\
    \delta_b C = i a_0'  - C^2 \,, \\
    \delta_b \tilde{C} = i b \,, \qquad  \delta_b b = [\tilde{C}, a_0'] \,, \\
    \delta_b \tilde{a}_0 = - i \tilde{C}_0 \,, \qquad  \delta_b \tilde{C}_0 = - [\tilde{a}_0, a_0'] \,, \\ \delta_b b_0 = - i C_0 \,, \qquad \qquad \delta_b C_0 = - [b_0, a_0'] \,, \\ \delta_b a_0' = 0 \,.
\end{gathered}
\label{eq:BRST trans in app}}
This gives precisely \eqref{eq:BRST_algebra}. 
\paragraph{Step 3: Include SUSY} We now include $\delta_s$ by promoting
\begin{equation}
    \delta_b V_\mathrm{gh} \to Q V_\mathrm{gh} \equiv (\delta_s + \delta_b) V_\mathrm{gh} \,,
\end{equation}
to treat $V_\mathrm{gh}$ on the same footing as $V_0$. The operation of $\delta_s$ on the ghosts must be such that $Q^2$ acts on the same way on all matrices, reducing to some global + gauge symmetry generator. Assuming that $\delta_s (\text{all ghosts}) = 0$ is inconsistent, because this would give $Q^2 \text{ghosts} = i[\text{ghosts}, a_0']$ whereas, for example $Q^2 X_{I'} = i [X_{I'}, a_0'] - i \Omega (\delta_\phi + \delta_{U(1)}) X_{I'}$.
A solution is to impose that
\begin{equation}
    \delta_s C = \phi \,, \qquad \qquad \delta_s (\text{other ghosts}) = 0 \,.
\end{equation}
The resulting cohomology is consistent, as described in the main text, and is such that
\al{
\spl{
Q V_\mathrm{gh} & = \mathrm{Tr} \Bigl( i b (F[X] + b_0) + i \tilde{C} [L_j, [X_j, C]] + i \tilde{C} C_0 + i C \tilde{C}_0 \\ & \qquad + i \left(a_0' - i \phi + i C^2\right) \tilde{a}_0  + \tilde{C} [L_j, \psi_j] \Bigr)\,,
}
}
\begin{equation}
    Q^2 = - i \Omega \delta_{U(1)} + i [\cdot, a_0'] \,.
\end{equation}

\paragraph{Step 4: Shift $a_0'$} Finally, we choose to define $a_0' \equiv a_0 + i \frac{3 \Omega}{8} L_3$ where $[a_0, L_i] = 0$.\footnote{Using \eqref{eq:BRST trans in app}, $[L_i,b_0]=0$ and the Jacobi identity, one can show $[b_0,[a_0',L_i]]=0$. Therefore, this choice is consistent with the BRST transformation \eqref{eq:BRST trans in app}.} To understand why we do this shift, note that $a_0'$ and the fluctuations of $X_{10}$ (the components of $X_{10}$ which are orthogonal to $M$) appear in our localization computation through the variable
\begin{equation}
    \tilde \phi = a_0' - i \phi + i C^2 \,.
\end{equation}
At the saddle point, $\phi = \frac{3 \Omega}{8} L_3 - i M$ and the integral of $\tilde{a}_0$ sets $\tilde \phi = 0$. It is thus important that $a_0' = M + \frac{3 \Omega}{8} L_3$ at the saddle point. This is only consistent with the shift we mentioned above.\footnote{Also note that this choice is important for the property \eqref{eq:RZequalsRZ0} to hold. If, at the saddle point, $a_0' = M$, this property would be spoiled and \eqref{eq:Deformation_NiceForm} would not be quadratic in the fluctuations.}

\subsection{Explicit formulae for the localization deformation}
\label{app:explicit_formulae}
Here, we write explicit formulae for $V_0$ and $V$, which are useful in order to understand the arguments underlying the claimed form of the deformation \eqref{eq:Deformation_NiceForm}.

In the spinor basis \eqref{eq:spinor_basis} the supersymmetry acts as \eqref{eq:SUSY_spinorbasis}, implying
\begin{equation}
    (\delta_s \psi_{I'})^\dagger = i [\bar \phi, X_{I'}] + \Omega \delta_{U(1)} X_{I'} \,, \quad \quad (\delta_s \chi_a)^\dagger = - i H_a^\dagger \,,
\end{equation}
where $\bar \phi = X_3 + i X_{10}$. Thus,
\begin{equation}
    V_0 = \mathrm{Tr}( \psi_{I'} (\delta_s \psi_{I'})^\dagger + \chi_a (\delta_s \chi_a)^\dagger) = \mathrm{Tr}\Bigl(\psi_{I'} (i [\bar \phi, X_{I'}] + \Omega \delta_{U(1)} X_{I'}) - i H_a^\dagger \chi_a\Bigr) \,.
\end{equation}
In fact,
\begin{equation}
    H_a^\dagger = H_a + 2 i s_a \,.
\end{equation}
Performing the change of variables \eqref{eq:change_of_var} and introducing the ghost potential $V_\mathrm{gh}$, we find that
\al{
\spl{
V = V_0 + V_\mathrm{gh} & = \mathrm{Tr} \Biggl((\tilde \psi_{I'}  -i [X_{I'}, C])(i [\bar \phi, X_{I'}] + \Omega \delta_{U(1)} X_{I'}) \\ & - i \tilde H_a \chi_a - \{ \chi_a, C \} \chi_a + 2 s_a \chi_a - i \tilde C [L_i, X_i] + \tilde c b_0 + c \tilde a_0 \Biggr) \,.
}
}
When expanding around the saddle, we need to make sure that there are no constant or linear piece in the fluctuation as claimed in arguing for \eqref{eq:Deformation_NiceForm}. There are of course no constant terms because of the presence of fermions (which are all set to 0 at the saddle point). To make sure that there is no linear term we need to make sure that every term multiplying fermions does vanish at the saddle point. It is straight-forward to verify this. We indeed obtain that at the saddle, $X_i = \frac{3 \Omega}{8} L_i$, $\tilde H_a = \tilde a_0 = b_0 = 0$, $\bar \phi = \frac{3 \Omega}{8} L_3 + i M$,
\begin{equation}
    i [\bar \phi, X_{i}] + \Omega \delta_{U(1)} X_{i} |_\mathrm{saddle} = 0 \, ,
\end{equation}
\begin{equation}
    s_a|_\mathrm{saddle} = 0 \, .
\end{equation}

Finally, let us verify the claim $R Z = R_0 \tilde Z + \mathrm{subleading}$, where $\tilde Z$ is a notation for all the fluctuations and we remind the reader that
\begin{equation}
    R = - i \Omega \delta_{U(1)} + i \left[\cdot, a_0 + i \frac{3 \Omega}{8} L_3\right], \quad \quad R_0 = - i \Omega \delta_{U(1)} + i \left[\cdot, M + i \frac{3 \Omega}{8} L_3\right] \, .
\end{equation}
This claim is not obvious for matrices which are expanded around a non-trivial constant. This is only the case for $X_i$ which are expanded around $\frac{3 \Omega}{8} L_i$, that could generate a constant term. However, we have that
\al{
\spl{
R X_i & = - i \left(\frac{3 \Omega}{8}\right)^2 \epsilon_{3 i j} L_j + i \left[\frac{3 \Omega}{8}L_i, M + i \frac{3 \Omega}{8} L_3\right] + R_0 \tilde X_i + \mathrm{subleading} \\ & = - i \left( \frac{3 \Omega}{8} \right)^2\epsilon_{3 i j} L_j - \left( \frac{3 \Omega}{8} \right)^2 [L_i, L_3] + R_0 \tilde X_i + \mathrm{subleading}  = R_0 \tilde X_i + \mathrm{subleading} \,. \label{eq:RZequalsRZ0}
}
}

\subsection{Eigenvalues and the 1-loop determinant}
\label{app:eigenvalues}
In this section, we compute the 1-loop determinant\footnote{We reserve the notation $Z_\mathrm{1-loop}$ for a nicely rescaled expression.}
\begin{equation}
    \tilde Z_\mathrm{1-loop} = \Delta \left(\frac{\det_{\tilde{Z}_f} R_0}{\det_{\tilde{Z}_b} R_0}\right)^{1/2} \,,
\end{equation}
where $\Delta$ is the Vandermonde determinant \eqref{eq:VDM_factor} appearing from the diagonalization of $M$. As discussed in the main section, computing these determinants amounts to compute the eigenvalues of $R_0$ on the vector spaces spanned by the matrices $\tilde{Z}_b$ and $\tilde{Z}_f$, where we remind the reader that
\begin{equation}
    Z_b \equiv (X_{I'}, \tilde{a}_0, b_0) \,, \quad Z_f \equiv (\chi_a, C, \tilde{C}) \,.
\end{equation}
To do it, we first define
\begin{gather}
    B_1 \equiv X_1 - i X_2 \,, \quad \quad \quad B_2 \equiv X_4 - i X_5 \,, \nonumber \\  B_3 \equiv X_6 - i X_7 \,,  \quad \quad \quad B_4 \equiv X_8 - i X_9 \,, \\
\xi_1 \equiv \chi_1 + i \chi_6 \,, \quad \quad \xi_2 \equiv  \chi_2 - i \chi_5 \,, \quad \quad \xi_3 \equiv \chi_3 + i \chi_4 \,. \nonumber
\end{gather}
We group the matrices in the following way,
\al{
\spl{
\mathrm{(1) \ : \ } & (B_2,B_3,B_4, B_2^\dagger, B_3^\dagger, B_4^\dagger), (\xi_1, \xi_2, \xi_3, \xi_1^\dagger, \xi_2^\dagger, \xi_3^\dagger) \,, \\
\mathrm{(2) \ : \ } & (B_1, B_1^\dagger), (C, \tilde{C}) \,, \\
\mathrm{(3) \ : \ } & (X_3), (\chi_7) \,, \\
\mathrm{(4) \ : \ } & (\tilde{a}_0, b_0) \,.
}
}
These are just a rewriting of the $9+2$ bosonic fields in $Z_b$ and the $7+2$ fermions of $Z_f$. With this rewriting,
\begin{equation}
    - i \Omega \delta_{U (1)} B_1 = \frac{3 \Omega}{8} B_1 \,,  \quad - i \Omega \delta_{U (1)} B_{i = 2, 3, 4} = \frac{\Omega}{8} B_i \,, \quad - i \Omega \delta_{U(1)} \xi_i = \frac{\Omega}{4} \xi_i \,,
\end{equation}
\begin{equation}
    - i \Omega \delta_{U(1)} (\mathrm{other \ matrices}) = 0 \,.
\end{equation}
We recall that
\begin{equation}
    R_0 \equiv - i \Omega \delta_{U(1)} - i \left[ M + i \frac{3 \Omega}{8} L_3 , \cdot \right] \,.
\end{equation}
To diagonalize $[L_3, \cdot]$, we use the fuzzy sphere harmonics expansion (see e.g. \cite[Sec.4.3]{Ishiki:2006yr}) for all our (non-zero mode) matrices. More precisely, we split the matrices $\Phi$ into rectangular blocks $\Phi^{(s,t)}$, and expand each block as
\begin{equation}
    \Phi^{(s,t)} = \sum_{J = |j_s - j_t|}^{j_s + j_t} \sum_{m = -J}^J \Phi^{(s,t)}_{J m} \otimes Y_{Jm (j_s,j_t)} \,.
\end{equation}
In terms of matrix dimension we have ($N_{s/t} = 2j_{s/t}+1$, no sum in $s,t$)
\al{\spl{
\F^{(s,t)}:\  n_s N_s\times n_t N_t\,,\qquad 
\F^{(s,t)}_{Jm}:\  n_s\times n_t\,,
\qquad
Y_{Jm(j_s,j_t)}:\  N_s\times N_t\,.
}}
The fuzzy sphere harmonics $Y_{Jm(j_s,j_t)}$ satisfy 
\al{\spl{
&\hat{L}_\pm Y_{Jm(j_s,j_t)} = \sqrt{(J\mp m)(J\pm m+1)} Y_{Jm(j_s,j_t)}\,,
\\
&\hat{L}_3 Y_{Jm(j_s,j_t)} = m Y_{Jm(j_s,j_t)}\,,
}}
where $\hat L$'s denote operators rather than matrices (see \cite{Ishiki:2006yr} for more details). It follows that under the action of $[L_3, \cdot]$, $\Phi_{J m}^{(s,t)} \to m \Phi_{J m}^{(s,t)}$. Also note that for diagonal $M = \mathrm{diag}(m_{s i},s=1, ..., q, i = 1, ..., n_{s})$, 
\begin{equation}
    ([M, \Phi]_{J m}^{(s,t)})_{i j} = (m_{s i} - m_{t j}) (\Phi^{(s,t)}_{J m})_{i j} \,.
\end{equation}
Finally, we wrote matrices in a diagonal form with respect to $- i \Omega \delta_{U(1)}$. Let us write
\begin{equation}
    - i \Omega \delta_{U(1)}\Phi \equiv \Omega r_{U(1)}^{(\Phi)} \Phi \,.
\end{equation}
Thus, $R_0$ acts diagonally on $\Phi_{J m}^{(s,t)}$, and we obtain
\begin{equation}
    (R_0 \Phi)^{(s,t)} =  \sum_{J = |j_s - j_t|}^{j_s + j_t} \sum_{m = -J}^J \left[ \Omega \left(r_{U(1)}^{(\Phi)} + \frac{3}{8} m\right)\Phi^{(s,t)}_{J m} - i [M, \Phi]_{J m}^{(s,t)} \right]\otimes Y_{Jm (j_s,j_t)} \,.
\end{equation}
Thus, all the components $\left(\Phi^{(s,t)}_{J m}\right)_{i j}$ are eigenvectors with eigenvalue
$$ \Omega \left(r_{U(1)}^{(\Phi)} + \frac{3}{8} m\right) - i (m_{s i} - m_{t j})\,. $$
Let us now consider $B_{i=2,...,4}, B^\dagger_{i=2,...,4}$. They have $r_{U(1)} = \frac{\Omega}{8}$ and $r_{U(1)} = - \frac{\Omega}{8}$ respectively. Thus, they contribute
\al{
\spl{
& \mathrm{det}_{B_{i = 2,3,4},B^\dagger_{i = 2,3,4}} R_0 \\ \quad & = \prod_{s,t} \prod_J \prod_m \prod_{i, j} \left(\Omega \left(\frac{1}{8} + \frac{3}{8} m\right) - i (m_{s i} - m_{t j})\right)^3 \left(\Omega \left(-\frac{1}{8} + \frac{3}{8} m\right) - i (m_{s i} - m_{t j})\right)^3 \,.
}
}
For the 6 fermions $\xi_{i=1,2,3}$, $\xi_{i=1,2,3}^\dagger$, since $r_{U(1)} = \pm \frac{\Omega}{4}$, we obtain
\al{
\spl{
& \mathrm{det}_{\xi_{i = 1,2,3},\xi^\dagger_{i = 1,2,3}} R_0 \\ \quad & =  \left(\Omega \left(-\frac{1}{4} + \frac{3}{8} m\right) - i (m_{s i} - m_{t j})\right)^3 \left(\Omega \left(\frac{1}{4} + \frac{3}{8} m\right) - i (m_{s i} - m_{t j})\right)^3 \,.
}
}
Some nice cancellations of the $m$ product happens when dividing the fermionic part by the bosonic, namely
\al{
\spl{
& \prod_{m = - J}^J  \frac{\left(\Omega \left(-\frac{1}{4} + \frac{3}{8} m\right) - i (m_{s i} - m_{t j})\right)^3 \left(\Omega \left(\frac{1}{4} + \frac{3}{8} m\right) - i (m_{s i} - m_{t j})\right)^3}{\left(\Omega \left(\frac{1}{8} + \frac{3}{8} m\right) - i (m_{s i} - m_{t j})\right)^3 \left(\Omega \left(-\frac{1}{8} + \frac{3}{8} m\right) - i (m_{s i} - m_{t j})\right)^3} \\ & = \prod_{m = - J}^J  \frac{\left(\Omega \left(\frac{1}{8} + \frac{3}{8} (m-1)\right) - i (m_{s i} - m_{t j})\right)^3 \left(\Omega \left(-\frac{1}{8} + \frac{3}{8} (m+1)\right) - i (m_{s i} - m_{t j})\right)^3}{\left(\Omega \left(\frac{1}{8} + \frac{3}{8} m\right) - i (m_{s i} - m_{t j})\right)^3 \left(\Omega \left(-\frac{1}{8} + \frac{3}{8} m\right) - i (m_{s i} - m_{t j})\right)^3} \,.
}
}
Note that the numerator is just a shift of the denominator, thus leading to many cancellations. The first numerator term and the first denominator term contribute respectively when $m=-J$ and $m=J$, other terms cancelling each other. Similarily, the second numerator term and the second denominator term contribute respectively when $m = J$ and $m = - J$.
Thus, the first group contributes
\begin{equation}
    \mathrm{det}^{(1)} \equiv \frac{\det_{\xi_{i = 1,2,3},\xi^\dagger_{i = 1,2,3}} R_0}{\det_{B_{i = 2,3,4},B^\dagger_{i = 2,3,4}} R_0} = \prod_{s,t} \prod_{J = |j_s - j_t|}^{j_s + j_t} \prod_{i, j} \left(\frac{\left( \frac{\Omega}{8} \right)^2 (2 + 3 J)^2 + (m_{s i} - m_{t j})^2 }{\left( \frac{\Omega}{8} \right)^2 (1 + 3 J)^2 + (m_{s i} - m_{t j})^2}\right)^3 \,.
\end{equation}
For the second group,
\al{
\spl{
& \mathrm{det}_{B_1,B^\dagger_1} R_0 \\ \quad & = \prod_{s,t} \prod_J \prod_m \prod_{i, j} \left(\Omega \left(\frac{3}{8} + \frac{3}{8} m\right) - i (m_{s i} - m_{t j})\right) \left(\Omega \left(-\frac{3}{8} + \frac{3}{8} m\right) - i (m_{s i} - m_{t j})\right) \,,
}
}
\al{
\spl{
& \mathrm{det}_{C,\tilde{C}} R_0  = \prod_{s,t} \prod_J \prod_m \prod_{i, j} \left(\frac{3 \Omega}{8} m - i (m_{s i} - m_{t j})\right) \left(\frac{3 \Omega}{8} m - i (m_{s i} - m_{t j})\right) \,.
}
}
Again, we see that they only differ by $m \to m \pm 1$, leading to cancellations. We obtain
\begin{equation}
    \mathrm{det}^{(2)} \equiv \frac{\mathrm{det}_{C,\tilde{C}} R_0}{\mathrm{det}_{B_1,B^\dagger_1} R_0} = \prod_{s,t} \prod_{J = |j_s - j_t|}^{j_s + j_t} \prod_{i, j} \left(\frac{\left( \frac{\Omega}{8} \right)^2 (3 J)^2 + (m_{s i} - m_{t j})^2 }{\left( \frac{\Omega}{8} \right)^2 (3 + 3 J)^2 + (m_{s i} - m_{t j})^2}\right) \,.
\end{equation}
For the last two groups the story is even simpler. Since $X_3$ and $\chi_7$ do not transform under the $U(1)$, they have the same $R_0$ eigenvalues, thus the determinants cancel each other.
\begin{equation}
    \mathrm{det}^{(3)} \equiv \frac{\mathrm{det}_{\chi_7} R_0}{\mathrm{det}_{X_3} R_0} = 1 \,.
\end{equation}
Finally, the group (4) is a group of bosonic ghost zero modes. They need to commute with $L_i$, thus they have the same form as $M$ and have components $(\tilde{a}_0)_{s; i j}$, $(b_0)_{s; i j}$ in the diagonal block $(s,s)$, and their $R_0$ value is
\begin{equation}
    (R_0 \tilde{a}_0)_{s; i j} = - i (m_{s i} - m_{s j}) (\tilde{a}_0)_{s; i j} \,, \quad \quad (R_0 b_0)_{s; i j} = - i (m_{s i} - m_{s j}) (b_0)_{s; i j} \,.
\end{equation}
Thus,
\begin{equation}
    \mathrm{det}^{(4)} \equiv \frac{1}{\mathrm{det}_{\tilde{a}_0, b_0} R_0} = (-1)^{\sum_s n_s^2} \prod_{s} \prod_{i, j} \frac{1}{(m_{s i} - m_{s j})^2} \,.
\end{equation}
This precisely cancels the zero-mode $J = 0$ ($s = t$) term in the numerator of $\mathrm{det}^{(2)}$.\footnote{One could avoid these 0/0's, by not integrating over the $J=0$ $i=j$ components of $C$, $\tilde C$, $\tilde a_0$ and $b$.} We also need the Vandermonde determinant from the diagonalization of $M$. It reads
\begin{equation}
    \Delta = \left(\prod_{s} \prod_{i \neq j} (m_{s i} - m_{s j})^2 \right)^{1/2} \,.
    \label{eq:VDM_factor}
\end{equation}
This allows again $J= 0$ in the numerator of $\mathrm{det}^{(2)}$, provided $i \neq j$.
Putting everything together,
\al{
\spl{
\tilde Z_\mathrm{1-loop} & = \Delta \ \mathrm{det}^{(1)} \mathrm{det}^{(2)} \mathrm{det}^{(3)} \mathrm{det}^{(4)} \\ & = (-1)^{\sum_s n_s^2} \left( \frac{64}{3 \Omega} \right)^{\sum_s n_s} \prod_{s,t = 1}^{q} \prod_{J = |j_s - j_t|}^{j_s + j_t} \prod_{\substack{i, j = 1\\ J = 0 : i \neq j}}^{n_s, n_t}  \\ &  \quad \quad \quad \quad \left[ \frac{\left[\left( \frac{\Omega}{8} \right)^2 (2 + 3 J)^2 + (m_{s i} - m_{t j})^2 \right]^3 \left[ \left( \frac{\Omega}{8} \right)^2 (3 J)^2 + (m_{s i} - m_{t j})^2  \right]}{\left[\left( \frac{\Omega}{8} \right)^2 (1 + 3 J)^2 + (m_{s i} - m_{t j})^2 \right]^3 \left[ \left( \frac{\Omega}{8} \right)^2 (3 + 3 J)^2 + (m_{s i} - m_{t j})^2  \right]} \right]^{1/2}  \,. \label{eq:tilde_1-loop}
}
}
where the product is not taken over $i = j$ when $J=0$. Separating $i = j$ and $i \neq j$ contributions,
\al{
\spl{
\tilde Z_{\mathrm{1-loop}} & = (-1)^{\sum_s n_s^2} \left(\frac{64}{3 \Omega} \right)^{\sum_s n_s} \prod_{s = 1}^q \left(\frac{1}{N_s} \prod_{J = 1}^{N_s -1} \frac{(2 + 3 J)^3}{(1 + 3 J)^3}\right)^{n_s} \prod_{s,t = 1}^q \prod_{\substack{i,j = 1 \\ i \neq j}}^{n_s, n_t} \prod_{J = |j_s-j_t|}^{j_s + j_t} \\ & \quad \quad \quad \quad \left[ \frac{\left[\left( \frac{\Omega}{8} \right)^2 (2 + 3 J)^2 + (m_{s i} - m_{t j})^2 \right]^3 \left[ \left( \frac{\Omega}{8} \right)^2 (3 J)^2 + (m_{s i} - m_{t j})^2  \right]}{\left[\left( \frac{\Omega}{8} \right)^2 (1 + 3 J)^2 + (m_{s i} - m_{t j})^2 \right]^3 \left[ \left( \frac{\Omega}{8} \right)^2 (3 + 3 J)^2 + (m_{s i} - m_{t j})^2  \right]} \right]^{1/2}  \,.
}
}

The partition function around the saddle $\mathcal{R}$ is thus given by
\begin{equation}
    Z_\mathcal{R} \propto e^{\frac{\Omega^4}{g_\mathrm{YM}^2} \frac{9}{8192} \mathrm{Tr} L_i^2} \int \prod_{s = 1}^{q} \prod_{i = 1}^{n_s} d m_{s i} \tilde Z_{\mathrm{1-loop}} \ \mathrm{exp} \left( - \frac{3 \Omega^2}{2^7} \sum_{s = 1}^{q} \sum_{i = 1}^{n_s} (2j_s + 1) m_{s i}^2 \right) \,.
\end{equation}
Note that we can get rid of the $\Omega$-dependence in $\tilde Z_{\mathrm{1-loop}}$ by the transformation $m_{s i} \to \Omega m_{s i}$. 
The full partition function is thus given by
\begin{equation}
    Z = \sum_{\mathcal{R}} C_\mathcal{R} e^{ \frac{9 \Omega^4}{2^{13}} \mathrm{Tr} L_i^2} \int \prod_s \prod_{i = 1}^{n_s} d m_{s i} Z_{\mathrm{1-loop}} \ \mathrm{exp} \left( - \frac{3 \Omega^4}{2^7} \sum_s \sum_{i = 1}^{n_s} N_s m_{s i}^2 \right) \,,
\end{equation}
where
\al{
\spl{
& Z_{\mathrm{1-loop}}  = \prod_{s,t = 1}^{q} \prod_{\substack{i, j = 1\\ i \neq j}}^{n_s, n_t} \prod_{J = |j_s - j_t|}^{j_s + j_t}  \\ &  \quad \quad \quad \quad \left[ \frac{\left[\left( \frac{1}{8} \right)^2 (2 + 3 J)^2 + (m_{s i} - m_{t j})^2 \right]^3 \left[ \left( \frac{1}{8} \right)^2 (3 J)^2 + (m_{s i} - m_{t j})^2  \right]}{\left[\left( \frac{1}{8} \right)^2 (1 + 3 J)^2 + (m_{s i} - m_{t j})^2 \right]^3 \left[ \left( \frac{1}{8} \right)^2 (3 + 3 J)^2 + (m_{s i} - m_{t j})^2  \right]} \right]^{1/2} \,.
}
}
The proportionality constant $C_\mathcal{R}$ is determined in the following appendix \ref{app:Normalization_factors}, where we have absorbed the factor $(-1)^{\sum_s n_s^2} \left(\frac{64}{3 \Omega} \right)^{\sum_s n_s} \prod_{s = 1}^q \left(\frac{1}{N_s} \prod_{J = 1}^{N_s -1} \frac{(2 + 3 J)^3}{(1 + 3 J)^3}\right)^{n_s}$.

\subsection{Normalization factors}
\label{app:Normalization_factors}
For bosonic $X$ and fermionic $\psi$ matrices we use the measures
\al{
\ga{
[d X] \equiv \prod_{i,j} d  \tilde X_{i j} = 2^{\frac{N^2 - N}{2}} \prod_{i \leq j} d \mathrm{Re} X_{i j} \prod_{i < j} d \mathrm{Im} X_{i j} \,, \\
[d \psi] \equiv \prod_{i,j} d  \tilde \psi_{i j} = 2^{-\frac{N^2 - N}{2}} \prod_{i \leq j} d \mathrm{Re} \psi_{i j} \prod_{i < j} d \mathrm{Im} \psi_{i j} \,, \label{eq:measure}
}
}
where $\tilde{X}$ and $\tilde{\psi}$ are matrices with $N^2$ real entries which contain all the information about the Hermitian matrices.
\al{
\ga{
\tilde X_{i j} = \mathrm{Re} X_{i j} + \mathrm{Im} X_{i j} \,, \\
\tilde \psi_{i j} = \mathrm{Re} \psi_{i j} + \mathrm{Im} \psi_{i j} \,.
}
}
Indeed, since $\mathrm{Re} X_{i j}$ is symmetric and $\mathrm{Im} X_{i j}$ is antisymmetric, they can be obtained from $\tilde X_{i j}$ by taking its symmetric and antisymmetric parts.
This measure satisfies the nice property that
\begin{equation}
    \int [d X] e^{- \lambda \mathrm{Tr} X^2} = \int \prod_{i,j} d \tilde X_{i j} e^{- \lambda \sum_{i, j} \tilde X_{i j}^2} = \left(\frac{\pi}{\lambda}\right)^{N^2/2} \,, \label{eq:BosonicGaussian}
\end{equation}
\begin{equation}
    \int [d \eta] [d \psi]  e^{- \mathrm{Tr} \eta \psi} = 1 \,. \label{eq:FermionicToyGaussian}
\end{equation}
Note that it is the same as the $U(N)$ measure, namely, if we expand the matrices $X$ and $\psi$ in $U(N)$ generator components $X = X^A T^A$, $\psi = \psi^A T^A$ such that $\mathrm{Tr} T^A T^B = \delta^{A B}$, the measure is 
\begin{equation}
    [d X] = \prod_{A = 1}^{N^2} d X^A \,, \qquad \qquad [d \psi] = \prod_{A = 1}^{N^2} d \psi^A \,.
\end{equation}
Indeed, this measure has the same Gaussian integrals \eqref{eq:BosonicGaussian}, \eqref{eq:FermionicToyGaussian}.
Let us consider a more complicated example of Grassmann integrals, where we have an interpolating matrix $M_{a b}^{i j}$ which is antisymmetric in the exchange $(a,i) \leftrightarrow (b,j)$. One obtains
\al{
\spl{
\int \prod_{i = 1}^{n_f} [d \psi^i] e^{- \sum_{a, b, i, j} \psi_{a b}^i M_{a b}^{i j} \psi_{b a}^j} = 2^{\frac{n_f}{2} N^2} \prod_{a,b} (\det M_{a b})^{1/2} \,, \label{eq:fermionic_gaussian}
}
}
where the determinant of $M_{a b}^{i j}$ is taken over the $i,j$ indices with $a,b$ fixed. Note that we also had zero-mode ghosts $a_0, b_0,...$. They are all expressed in terms of $n_s$ by $n_s$ Hermitian matrices. We thus define the measure similarily for each $s$. The result is the same by replacing $N^2 \to \sum_s n_s^2$. Now we get the following factors. 

First, we find that
\al{
\spl{
\int & [d C] [d \tilde{C}] [d b] [d C_0] [d \tilde C_0] [d b_0] [d a_0] [d \tilde{a}_0] \exp \Biggl\{ \mathrm{Tr} \Bigl( i b (F + b_0) + i \tilde{C} [L_i, [X_i, C]] \\ & \quad \quad \quad \quad + i \tilde{C} C_0 + i C \tilde{C}_0 + i \left(a_0 + i \frac{3 \mu}{8} L_3 - i \phi + i C^2\right) \tilde{a}_0  \Bigr) \Biggr\} \\ & = (-1)^{\sum_s n_s^2} (2 \pi)^{N^2} (2 \pi)^{\sum_s n_s^2} \mathrm{det}' \left(\frac{\delta F}{\delta \alpha} \right) \Biggl|_{\alpha = 0} \delta'(F) \,.
}
}
The factor $(-1)^{\sum_s n_s^2}$ comes from integration over $C_0,\tilde C_0$. The factor $(2 \pi)^{N^2}$ comes from the integral over $b$ and $b_0$, whereas the last factor comes from $\int [d a_0] [d \tilde a_0] e^{i \mathrm{Tr} (a_0 + ...) \tilde a_0}$.
Thus, when using the Fadeev-Popov trick \eqref{eq:FaddeevPopovTrick2}, one should divide by these factors
\begin{equation}
    \mathrm{det}'\left(\frac{\delta F}{\delta \alpha}\right) \delta'(F) 
 = \frac{(-1)^{\sum_s n_s^2}}{(2 \pi)^{N^2} (2 \pi)^{\sum_s n_s^2}} \int [d \mathrm{ghosts}] e^{- t S_\mathrm{{g h}}} \label{eq:factor1} \,.
\end{equation}
Secondly, we find using \eqref{eq:BosonicGaussian}, \eqref{eq:fermionic_gaussian} that
\begin{equation}
\int [d Z_b] [d Z_b'] [d Z_f] [d Z_f'] e^{- t Q (V_0 + V_\mathrm{gh})} = \pi^{9 N^2} \pi^{\sum_s n_s^2} 2^{9 N^2} 2^{\sum_s n_s^2} \left(\frac{\det_{Z_f} R_0}{\det_{Z_b} R_0}\right)^{1/2} \,.
\label{eq:factor2}
\end{equation}
Thirdly, we have two factors from ``gauge fixing''. First, since $\mathcal{G}_\mathcal{R} = U(N) / \otimes_s U(n_s)$, we get the volume factor appearing in the gauge fixing procedure
\begin{equation}
    \mathrm{Vol} \ \mathcal{G}_\mathcal{R} = \frac{\mathrm{Vol} \ U(N)}{\prod_s \mathrm{Vol} \ U(n_s)} \,,
\end{equation}
where
\begin{equation}
    \mathrm{Vol}(U(N)) = \frac{(2 \pi)^{(N^2 + N)/2}}{\prod_{k=1}^{N-1} k!} \,.
\end{equation}
We are also diagonalizing $M$. Thus, we are basically gauge fixing each of its blocks $s = 1,...,q$ using $SU(n_s)$, up to $U(1)^{n_s-1}$ remaining gauge redundancy. However, we don't gauge fix the permutation symmetry. This would correspond to gauge fixing, for example, $m_{s i} < m_{s j} \ \forall i < j$. This means we need to divide by $n_s!$. Thus, the diagonalization of $M$ reads
\begin{equation}
    \int [d M] = \prod_s \left( \frac{1}{n_s !}  \frac{\mathrm{Vol} \ U(n_s)}{\mathrm{Vol} \  U(1)^{n_s}} \right) \int \prod_{s,i} d m_{s i} \Delta(m) \,.
\end{equation}
Combining these two volume factors we obtain
\begin{equation}
    \frac{(2 \pi)^{(N^2 + N)/2 - \sum_s n_s}}{\prod_{k = 1}^{N-1} k! \prod_s n_s !} \,. \label{eq:factor3} 
\end{equation}
Now note that this is the path integral including the auxiliary fields $K_a$. The auxiliary fields contribute $(2 \pi)^{7 N^2/2}$ to the path integral. Thus,
\begin{equation}
    \int [d X_a] [d \psi_\alpha] e^{- S} = (2 \pi)^{- 7 N^2 / 2} \int [d X_a] [d \psi_\alpha] [d K_a] e^{- S|_\text{with auxiliaries}}  \,. \label{eq:factor4}
\end{equation}
Combining all these factors \eqref{eq:factor1}, \eqref{eq:factor2}, \eqref{eq:factor3}, \eqref{eq:factor4}, as well as absorbing the factor $(-1)^{\sum_s n_s^2} (64/3)^{\sum_s n_s} \prod_{s = 1}^q \left(\frac{1}{N_s} \prod_{J = 1}^{N_s -1} \frac{(2 + 3 J)^3}{(1 + 3 J)^3}\right)^{n_s}$ in \eqref{eq:tilde_1-loop} we obtain
\al{
\spl{
C_\mathcal{R} & = \frac{(2 \pi)^{9 N^2/2 + (N^2 + N)/2 - \sum_s n_s} }{\prod_{k = 1}^{N-1} k! \prod_s n_s !} \left(\frac{64}{3}\right)^{\sum_s n_s} \prod_{s = 1}^q \left(\frac{1}{N_s} \prod_{J = 1}^{N_s -1} \frac{(2 + 3 J)^3}{(1 + 3 J)^3}\right)^{n_s} \\ & = \frac{ (2 \pi)^
    {5 N^2 +  N/2 }}{\prod_{k=1}^{N-1} k! \prod_s n_s !} \left(
    \frac{32}{3\pi } 
    \right)^{\sum_s n_s} \prod_{s = 1}^q \left(\frac{1}{N_s} \prod_{J = 1}^{N_s -1} \frac{(2 + 3 J)^3}{(1 + 3 J)^3}\right)^{n_s} \,, \label{eq:normalizationPartFunction}
}
}
as claimed in the introduction \eqref{eq:normalization}.

\section{Approximate solutions to the electrostatic problem}
\label{app:approxsollutions}
In this appendix we solve the electrostatic problem \eqref{eq:SaddlePointEigenvalues} and \eqref{eq:ConstraintsEigenDensities} (or equivalently \eqref{eq:LinearCharge} and \eqref{eq:ConditionsLinearDensity}) with a single disk in the limits $z_s/R_s \gg 1$  and $z_s/R_s \ll 1$.
We also solve the general problem with one disk numerically. 

\subsection{Analytics}

As usual in matrix models, it is convenient to introduce resolvents
\begin{align}
    G_s(z) \equiv \int_{-R_s}^{R_s} dr \frac{f_s(r)}{z-r} = \frac{Q_s}{z} +O\left(\frac{1}{z^3}\right)\,.
\end{align}
Notice that $G_s(-z)=-G_s(z)$ because $f_s(r)=f_s(-r)$.
Equation \eqref{eq:LinearCharge} can be written as
\begin{align}
    \tilde{V}_s &= V_b3 r^2 z_s - \sum_{t + \mathrm{images}} \frac{1}{8\pi^2} \mathcal{R} \hspace{-.4cm}\int_{- R_t}^{R_t} dr' f_t(r')
    \left[ \frac{1}{(r'-r-i(z_s - z_t) )^2}+\frac{1}{(r'-r+i(z_s - z_t) )^2}
    \right]\nonumber\\
    &= V_b3 r^2 z_s   - \sum_{t + \mathrm{images}} \frac{1}{8\pi^2} \oint \frac{dr'}{2\pi i} G_t(r')
    \left[ \frac{1}{(r'-r-i(z_s - z_t) )^2}+\frac{1}{(r'-r+i(z_s - z_t) )^2}
    \right]\nonumber\\
    &= V_b3 r^2 z_s   + \sum_{t + \mathrm{images}} \frac{1}{8\pi^2} 
    \left[ G_t'(r+i(z_s - z_t)) +G_t'(r-i(z_s - z_t)) 
    \right] \,, \label{Resolventinz}
\end{align}
where $ \tilde{V}_s = V_s+V_bz_s^3$, $\oint$ denotes the contour around the cut of $G_t$ and in the last line we opened the contour and picked up the poles of the integrand.

The last equation tell us that 
$G_s'(z+i\epsilon) +G_s'(z-i\epsilon)$ is analytic in a region around the branch point of $G_s(z)$ at $z=R_s$. Therefore, we must have an expansion in half integers powers, i.e.  
$G_s(z)=\sum_{n=0}^\infty  c_n(z-R_s)^{\frac{n}{2}}$ with $c_n \in \mathbb{R}$.
It is then convenient to change variable to remove the branch points at $z=\pm R_s$.
More precisely, let us map the $z$ complex plane minus the cut $[-R_s, R_s]$ to the interior of the unit disk using the Zhukovsky map\footnote{We choose the solution $w=z/R_s-\sqrt{(z/R_s)^2-1}$ so that $w$ is inside the unit disk.}
\begin{align}
  \frac{2z}{R_s} = w+\frac{1}{w} \,,\qquad\qquad
   H_s(w)\equiv G'_s(z)\,.
\end{align}
By construction, $H_s(w)$ is analytic for $|w|<1$. In addition, it is an even function,  obeying $H_s(w)=H_s(-w)=-4Q_s/R_s^2 w^2 +O(w^4)$
Equation \eqref{Resolventinz}, can then be written as
\begin{align}
    \tilde{V}_s = V_b\frac{3}{4} R_s^2 z_s \left(w+\frac{1}{w}\right)^2  + \sum_{t + \mathrm{images}} \frac{1}{8\pi^2} 
    \left[ H_t(w_t^+) +H_t(w_t^-) 
    \right]\,,\label{ResolventH(w)}
\end{align}
where 
\begin{align}
 2 \frac{z}{R_s} = w+\frac{1}{w} \,,\qquad\qquad
  2\frac{z\pm i(z_s - z_t)}{R_t} = w_t^\pm+\frac{1}{w_t^\pm}\,.
\end{align}
For $|w|=1$ we choose the  solutions obeying $|w_t^\pm|<1$. For the special case $t=s$, we choose $w_s^+=w$ and $w_s^-=1/w$. Equation \eqref{ResolventH(w)} is valid for $|w|=1$ but since it is  analytic in $w$ we can analytically continue it. Moving $w$ slightly outside the unit disk, we can express $H_s(w)$ in terms of analytic functions and a linear combination of $H_t(w_t^\pm)$ evaluated at $|w_t^\pm|<1$. This allows us to expand the region of analyticity of $H_s(w)$ from the unit disk to a strictly bigger region. Therefore, the series expansion  of $H_s(w)$ around $w=0$ has a radius of convergence larger than 1.

Let us consider the special case of only one ball. Then
\begin{align}
    \tilde{V}_s = V_b\frac{3}{4} R_s^2 z_s \left(w+\frac{1}{w}\right)^2  + \frac{1}{8\pi^2} 
    \left[ 
    H_s(w) +H_s(1/w) 
     - H_s(w_{\tilde{s}}^+) - H_s(w_{\tilde{s}}^-) 
    \right]
    \,,\label{ResolventH(w)1ball}
\end{align}
with 
\begin{align}
w_{\tilde{s}}^\pm+\frac{1}{w_{\tilde{s}}^\pm}= w+\frac{1}{w} \pm i
  \frac{4z_s}{R_s}  \,.
\end{align}
We don't know how to solve this equation analytically in full generality but the limits of $z_s/R_s \gg 1$ and $z_s/R_s \ll 1$ are tractable. 
Let us start with the simpler case $z_s/R_s \gg 1$. Then, $w_{\tilde{s}}^\pm = \mp i \frac{R_s}{4z_s} +O(R_s/z_s)^2 \ll 1$ and we can neglect the effect of the image ball. In this limit, we find
\begin{align}
    \tilde{V}_s = V_b\frac{3}{4} R_s^2 z_s \left(w+\frac{1}{w}\right)^2  + \frac{1}{8\pi^2} 
    \left[ 
    H_s(w) +H_s(1/w) 
    \right] \,,
\end{align}
whose solution is simply
\begin{align}
    H(w) = - 6\pi^2 V_b R_s^2 z_s w^2\,,\qquad
    \qquad \tilde{V}_s = V_b\frac{3}{2} R_s^2 z_s\,\qquad \qquad Q_s= \frac{3}{2}\pi^2 V_b R_s^4 z_s\,.
    \label{eq:zs_to_infinity}
\end{align}
In this approximation we can also compute the octopole 
\begin{equation}
    q_{3,s} =2z_s^3 Q_s-6z_s \int_{-R_s}^{R_s}drr^2f_s(r) \approx 2 Q_s z_s^3.
    \label{eq: q3 large z}
\end{equation}



Let us now consider the limit $z_s/R_s \ll 1$.
In this limit, we expect $G_s$ to vary slowly on the length scale $z_s$.
Therefore, we can approximate the exact equation
\begin{align}
    \tilde{V}_s = V_b3 r^2 z_s  +  \frac{1}{8\pi^2} 
    \left[ G_s'(r+i\epsilon) +G_s'(r-i\epsilon) -
    G_s'(r+2iz_s) -G_s'(r-2iz_s) 
    \right]\,,\nonumber
\end{align}
with
\begin{align}
    \tilde{V}_s = V_b3 r^2 z_s  -  \frac{2iz_s}{8\pi^2} 
    \left[ G_s''(r+i\epsilon) -G_s''(r-i\epsilon) +O(z_s) 
    \right]\,,
    \label{eq:1ballapprox}
\end{align}
which gives us the discontinuity of $G_s''$.
From the definition, we find that
\begin{align}
    G_s''(z) = \int_{-R_s}^{R_s} drf_s(r) \frac{d^2}{dz^2}\frac{1}{z-r} = \int_{-R_s}^{R_s} drf_s(r) \frac{d^2}{dr^2}\frac{1}{z-r} = 
    \int_{-R_s}^{R_s} drf_s''(r) \frac{1}{z-r} \,, 
\end{align}
where we used that $f_s(\pm R_s) = f_s'(\pm R_s) = 0$.
Therefore, equation \eqref{eq:1ballapprox} gives 
\begin{align}
  f_s''(r)\approx   2\pi  \left(V_b 3 r^2-\frac{\tilde{V}_s}{z_s}  \right) 
  \,,
\end{align}
which leads to
\begin{align}
  f_s(r)&\approx    -\pi \frac{\tilde{V}_s}{z_s}  r^2  +\frac{\pi}{2}V_b r^4 +const.\\
 &=\frac{\pi}{2}  V_b
 \left( R_s^2 -r^2
 \right)^2 \,,
 \label{eq:sigma_0}
\end{align}
where we imposed $f_s(\pm R_s) = f_s'(\pm R_s) = 0$ to determine the integration constant  and $\tilde{V}_s = V_b z_s R_s^2$.
Notice that this approximation fails for $r$ close to $R_s$ because it does not give the expected behavior $f_s(r) \sim (R_s-r)^\frac{3}{2}$. However, this is a small effect at the tip that does not affect the leading behavior of the macroscopic quantities.
The total charge is then
\begin{align}
    Q_s= \int_{-R_s}^{R_s}drf_s(r) = \frac{8\pi}{15}V_b R_s^5\,.
    \label{eq:total charge sigma0}
\end{align}
The octopole is given by
\begin{align}
    q_{3,s}=2z_s^3 Q_s-6z_s \int_{-R_s}^{R_s}drr^2f_s(r) \approx -\frac{16\pi}{35}V_b z_s R_s^7\,.
    \label{eq: q3 sigma0}
\end{align}
The on-shell action is then
\begin{align}
    S_\mathrm{on-shell} = \frac{1}{3  \pi^4 g_s^2\alpha'^4 \mu^5}\sum_s \left(  \frac{2}{V_b}  Q_s V_s -  q_{3,s} \right)= \frac{5}{7 \ 2^7}\left(\frac{15\pi}{8N_s}\right)^{2/5}\Omega^{12/5} N^{7/5} \,,
\end{align}
where we used $V_b=\mu^5/2^7$, $N=N_s n_s$, $\mu= \Omega \sqrt{g_{\text{YM}}}$, $g_s=(2\pi)^3 \alpha'^2 g_{\text{YM}}^2$ and the relations \eqref{eq:zs and Qs in terms of Ns and ns}. 

\subsection{Numerics}
We want to solve the equation 
\begin{equation}
    V_{\mathrm{ball},s}(r,z_s) + V_{\mathrm{image},s}(r,z_s) + V_{bg}(r,z_s) = V_s, \qquad 0 \leq r \leq R_s.
    \label{eq:EOM_numerics}
\end{equation}
The potential due to the ball reads 
\begin{equation}
V_{\mathrm{ball},s}(r,z) = \frac{1}{4\pi r} \int_0^{R_s} du u \sigma_s(u) \operatorname{log}\left(1+ \frac{4 r u}{(r-u)^2+(z-z_s)^2} \right).
\end{equation}
We can combine the contributions from the ball and its image, giving 
\begin{equation}
V_{\mathrm{ball},s}(r,z_s)+V_{\mathrm{image},s}(r,z_s) = \frac{1}{4\pi r} \operatorname{p.v.}\int_0^{R_s} du u \sigma_s(u) \operatorname{log}\left( \frac{(r+u)^2}{(r-u)^2 \left( 1+ \frac{4 r u}{(r-u)^2+4 z_s^2} \right)}\right).
\end{equation}
To solve the equation numerically we can write $\sigma(r)$ as an expansion into a complete basis of functions $f_n(r)$ on $[0,R_s ]$. A convenient choice are the shifted Legendre polynomials.
We can then expand 
\begin{equation}
    \sigma_s(r) = \sum_{n=0}^{N_{max}} c_n P_n\left(2\frac{r}{R_s}-1 \right).
\end{equation}
From the discussion of the previous section (see also the near-tip expansion in \cite{Lin:2005nh}) we expect the charge density to behave as $\sqrt{R_s^2-r^2}$ close to the tip. This is hard to reproduce with a finite number of polynomials, so to improve the convergence we consider 
\begin{equation}
    \sigma_s(r) = \sum_{n=0}^{N_{max}} c_n P_n\left(2\frac{r}{R_s}-1 \right)+ \tilde{c} \sqrt{R_s^2-r^2},
\end{equation}
where $\tilde{c}$ is a new free parameter.
To impose the equations of motion \eqref{eq:EOM_numerics} we take a uniform grid with $N_{grid}$ points on the segment $[0,R_s]$. We call $r_i$ the points of that grid. We construct the constraint function 
\begin{equation}
    \operatorname{constr}^{(EOM)} = \sum_{i=1}^{N_{grid}} \left(\operatorname{EOM}(r_i)\right)^2,
\end{equation}
which is always positive and vanishes on the physical solution. Our strategy is to do this minimization numerically.
In addition we want to impose that $\sigma$ vanishes at $r=R_s$ and has zero derivative at $r=0$. We can easily implement this by modifying the constraint function. The numerical problem we solve is then\footnote{One could worry that the result can be unphysical if the minimum is not zero. However in all our results we find that the minimum is zero (up to numerical accuracy).}
\begin{equation}
\underset{\{c_i, \tilde{c}, V_s \}}{\text{Minimize}} \left[\sum_{i=1}^{N_{grid}} \left(\operatorname{EOM}(r_i)\right)^2+ \sigma_s(R_s)^2 + \sigma'_s(0)^2 \right]
\label{eq:numerical_problem}
\end{equation}
For the value of $N_{grid}$ we choose $2 N_{max}$ so that we always have more constraints than free parameters. We also set $V_b=1$.

\begin{figure}[h]
    \centering
\includegraphics[scale=0.6]{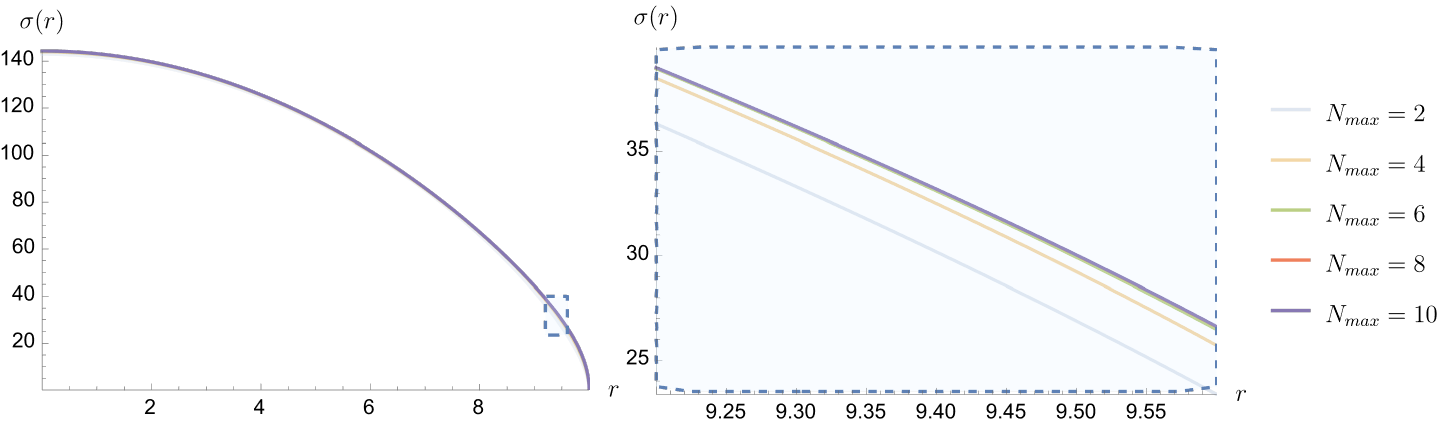}
    \caption{Convergence study of the numerical problem \eqref{eq:numerical_problem} with $N_{max}$. We solved for the charge density with $z_s=1$ and $R_s=10$.}
    \label{fig:convergence}
\end{figure}

On figure \ref{fig:convergence} we study the convergence of the result with $N_{max}$. We see that the convergence is really fast.

\begin{figure}[h]
    \centering
\includegraphics[scale=0.8]{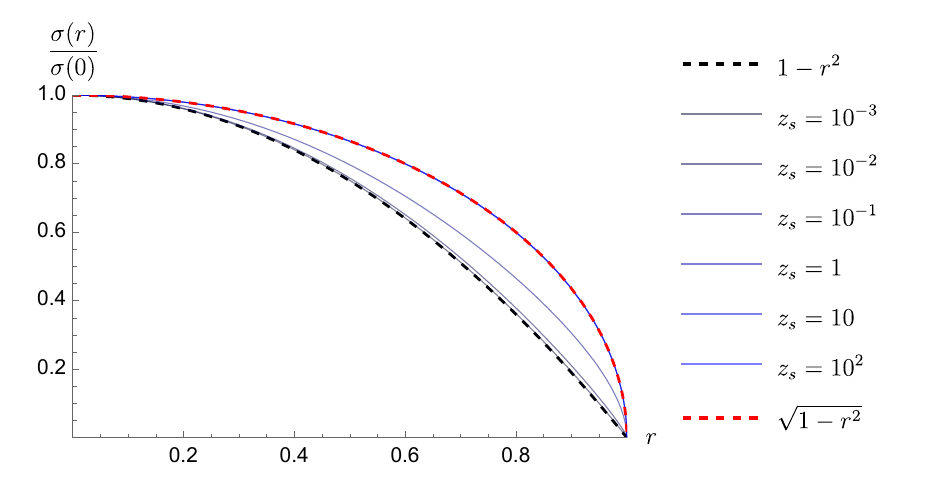}
    \caption{Numerical results for the (normalized) charge density for different heights $z_s$ when $R_s=1$. We used $N_{max}=4$. The black and red dashed lines are the analytic approximations when $z_s \ll R_s, z_s \gg R_s$ given by \eqref{eq:sigma_0} and \eqref{eq:zs_to_infinity}.}
    \label{fig:shape}
\end{figure}

On figure \ref{fig:shape} we study the shape of the charge density in different regimes of the dimensionless ratio $z_s/R_s$. The analytic solutions in both limits $z_s/R_s \to 0$ and  $z_s/R_s \to \infty$ are derived in  \eqref{eq:sigma_0} and the appendix of \cite{Komatsu:2024bop} respectively. We see that the numerical solution nicely interpolates between those results.

\begin{figure}[h]
    \centering
\includegraphics[scale=0.6]{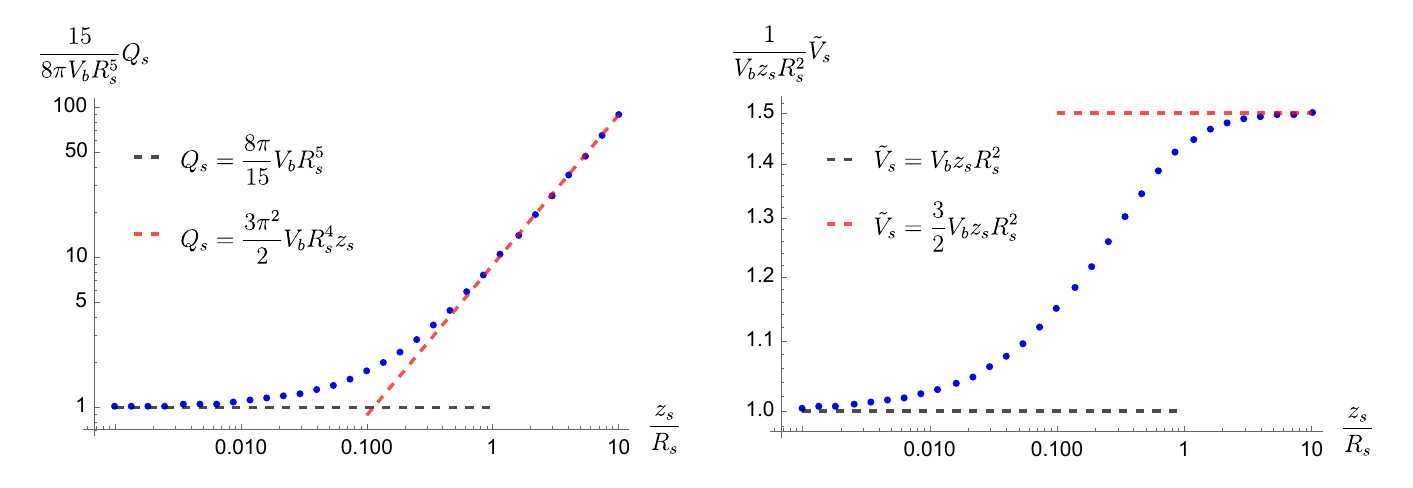}
    \caption{Numerical results (blue dots) for the total charge $Q_s$ and the potential $V_s$. We used $N_{max}=4$. The black and red dashed lines are the analytic approximations when $z_s \ll R_s, z_s \gg R_s$ given by \eqref{eq:total charge sigma0} and \eqref{eq:zs_to_infinity}.}
    \label{fig:QVzs}
\end{figure}

\begin{figure}[h]
    \centering
\includegraphics[scale=0.75]{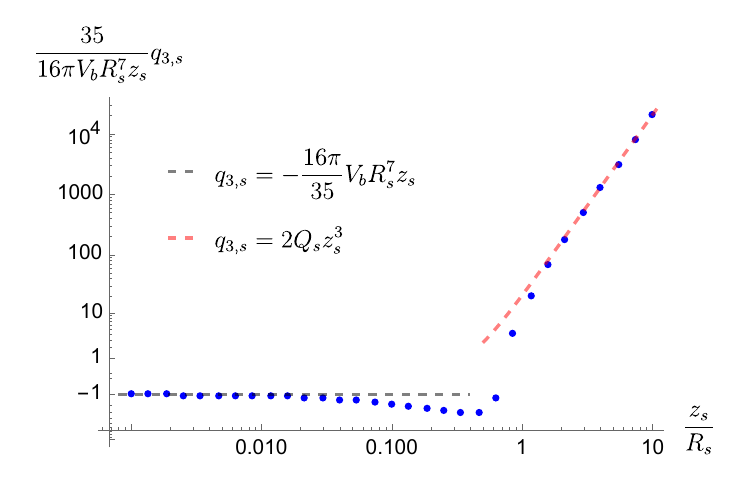}
    \caption{Numerical results (blue dots) for the octopole $q_{3,s}$. We used $N_{max}=4$. The black and red dashed lines are the analytic approximations when $z_s \ll R_s, z_s \gg R_s$ given by \eqref{eq: q3 sigma0} and \eqref{eq: q3 large z}.}
    \label{fig:q3}
\end{figure}

We also study the total charge, the potential and to octopole, shown on figures \ref{fig:QVzs} and \ref{fig:q3}. We also observe a clean transition from the asymptotics $z_s/R_s \to 0,\infty$ where the numerics agree with the analytical results.

\begin{figure}[h]
    \centering
\includegraphics[scale=0.75]{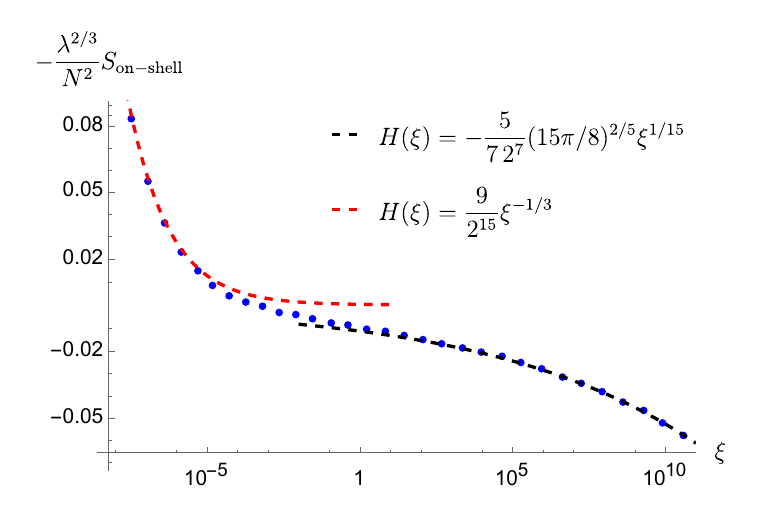}
    \caption{Numerical results (blue dots) for the on shell action \eqref{eq:Hxi}. We used $N_{max}=4$. The black and red dashed lines are the analytic approximations when $\xi \ll 1$ and $\xi \gg 1$ respectively.}
    \label{fig:S_onshell}
\end{figure}

Finally we can measure numerically the function $H(\xi)$ defined in \eqref{eq:Hxi} by plotting the on-shell action as function of the parameter $\xi \equiv n_s \Omega^{-4} N_s^{-5}$. In terms of the electrostatic variables we have 
\begin{equation}
    \xi = \frac{3^5}{2^{20}\pi^2} \frac{Q_s}{z_s^5 \mu^5},
\end{equation}
which can be computed with our numerics.
The result for $H(\xi) = -S_\mathrm{on-shell} \lambda^{2/3}/N^2$ is shown on figure \ref{fig:S_onshell}. 

\newpage

\bibliographystyle{utphys} 
\bibliography{refs}

\end{document}